\documentclass[11pt]{article}
\usepackage{latexsym,amssymb,amstext,amsmath}
\usepackage{hyperref}
\renewcommand{\baselinestretch}{1.2}
\allowdisplaybreaks
\def\Slash#1{\rlap{\hbox{$\mskip 3 mu /$}}#1}      % " upper
\def\oneone{\rlap 1\mkern4mu{\rm l}} % unit matrix
\newcommand{\ft}[2]{{\textstyle\frac{#1}{#2}}}
\usepackage{color}
\definecolor{MyDarkBlue}{rgb}{0,0.08,0.45}
\definecolor{cadgreen}{rgb}{0.0, 0.42, 0.24}
\hypersetup{
colorlinks=true,
citecolor=cadgreen,
linkcolor=MyDarkBlue,
urlcolor=MyDarkBlue,
pdfauthor={N. Banerjee, B. de Wit and S. Katmadas },
pdftitle={The off-shell c-map},
pdfsubject={hep-th}
}
\begin{document}
%%%%%%%%%%%%%%%%%%%%%%%%%%%%%%%%%%%%%%%%%%%%%%%%%%%%%%%%%%%%%%
%
\begin{titlepage}
\begin{flushright} \small
 Nikhef-2015-046 \\~
\end{flushright}
\bigskip

\begin{center}
 {\LARGE\bfseries  The off-shell c-map}
\\[10mm]
\textbf{Nabamita Banerjee$^a$, Bernard de Wit$^{b,c}$ and Stefanos
  Katmadas$^{d,e}$ }\\[5mm]
\vskip 4mm $^a${\em IISER Pune, Department of Physics, Homi Bhaba
  Road, Pashan, Pune,  India}\\ 
$^b${\em Nikhef Theory Group, Science Park 105, 1098 XG Amsterdam, The
  Netherlands}\\
$^c${\em Institute for Theoretical Physics, Utrecht
  University,} \\
{\em Leuvenlaan 4, 3584 CE Utrecht, The Netherlands}\\
$^d${\em Dipartimento di Fisica, Universit\'a di Milano-Bicocca,}\\
  {\em Piazza della Scienza 3, I-20126 Milano, Italy} \\
$^e${\em INFN, Sezione di Milano-Bicocca,}\\
  {\em Piazza della Scienza 3, I-20126 Milano, Italy}\\ [3mm]
{\tt nabamita@iiserpune.ac.in}\,,\;{\tt B.deWit@uu.nl}\,,\;{\tt
  stefanos.katmadas@unimib.it}
\end{center}

\vspace{3ex}

\begin{center}
{\bfseries Abstract}
\end{center}
\begin{quotation} \noindent The off-shell version of the c-map is
  presented, based on a systematic off-shell reduction from four to
  three space-time dimensions for supergravity theories with eight
  supercharges. In the reduction, the R-symmetry group is enhanced to
  local $[\mathrm{SU}(2)\times \mathrm{SU}(2)]/\mathbb{Z}_2 \cong
  \mathrm{SO}(4)$ and the c-map is effected by a parity transformation
  in the internal space that interchanges the two $\mathrm{SU}(2)$
  factors. Vector and tensor supermultiplets are each others conjugate
  under the c-map and both can be dualized in three dimensions to
  (on-shell) hypermultiplets. \\
  As shown in this paper the off-shell formulation indeed leads to a
  clarification of many of the intricate issues that play a role in
  the c-map. The results for off-shell Lagrangians quadratic in
  space-time derivatives are analyzed in detail and compared to the
  literature. The underlying reasons are identified why not all of the
  four-dimensional tensor multiplet Lagrangians can be in the image of
  the c-map. The advantage of the off-shell approach is, that it also
  enables a systematic analysis of theories with higher-derivative
  couplings. This is demonstrated for a particular subclass of such
  theories, which, under certain conditions, are consistent under the
  c-map. In principle, explicit results for realistic four-dimensional
  type-II string compactifications can be explored in this way.
\end{quotation}

\vfill

%%%%%%%
%\flushleft{\today}
%%%%%%
\end{titlepage}

%%%%%%%%%%%%%%%%%%%%%%%%%%%%%%%%%%%%%%%%%%%%%%%%%%%%%%%%%%%%%%%
%%%%%%%%%%%%%%%%%%%%%%%%%%%%%%%%%%%%%%%%%%%%%%%%%%%%%%%%%%%%%%%
\section{Introduction}
\label{sec:introduction}
\setcounter{equation}{0}
%%%%%%%%%%%%%%%%%%%%%%%%%%%%%%%%%%%%%%%%%%%%%%%%%%%%%%%%%%%%%%%
Dimensional reduction of supersymmetric theories is usually performed
in the context of {\it on-shell} field representations. For theories
with a large number of supercharges this is unavoidable, as off-shell
representations are usually not available. For theories based on
off-shell representations it is often not worthwhile to define a full
and consistent off-shell dimensional reduction scheme, because the
extra auxiliary fields contained in the off-shell configuration can be
removed by solving their corresponding (algebraic) field equations. In
the presence of higher-derivative couplings, however, these field
equations are no longer algebraic. In their on-shell form these
couplings will therefore take the form of an iterative expansion in
ever increasing powers of space-time derivatives, which will
completely obscure their underlying structure. In this case an
off-shell reduction scheme is indispensable, as one obtains a
supercovariant dictionary expressing the higher-dimensional fields into the
lower-dimensional ones, so that different invariants can be reduced on
a case by case basis.

The dimensional reduction of $4D$ $N=2$ supergravity theories to $3D$
dimensions is special and is relevant for the so-called c-map
\cite{Cecotti:1988qn}. Because the number of supersymmetries remains
the same, four-dimensional theories with $N=2$ supersymmetry yield
three-dimensional theories with $N=4$ supersymmetry. Dimensional
reduction is usually applied to Lagrangians that are at most quadratic
in space-time derivatives and the c-map has mainly been studied at the
on-shell level
\cite{Cecotti:1988qn,Ferrara:1989ik,deWit:1992wf,DeJaegher:1997ka}. In
its original form it maps vector multiplets into hypermultiplets. But
in its off-shell form it maps vector into tensor supermultiplets and
vice versa \cite{deWit:2006gn}. Both these types of multiplets can be
converted to hypermultiplets in $3D$ by vector-scalar duality.

The c-map is related to T-duality for type-II string theories with
one spatial dimension compactified on a circle
\cite{Dine:1989vu,Dai:1989ua}. In the compactification of type-IIA
string theory the spectrum of 1/2-BPS states consists of the massless
states described by $9D$ supergravity, coupled to momentum and winding states
associated with the circle. Denoting the circumference of the circle
by $L$, the momentum states have masses of order $1/L$, while the
winding modes have masses of order $L$. In the limit $L\to\infty$ the
momentum states become massless and the theory decompactifies with
massless states described by type-IIA supergravity. Obviously a second
decompactification limit exists for $L\to 0$, where the winding states
become massless. In the latter case the massless states are those described
by type-IIB supergravity.  The momentum and winding modes belong to
different representations associated with different central charges of
the $9D$ supersymmetry algebra \cite{AbouZeid:1999fv}. This is then
consistent with the fact that the massless spectra of IIA and IIB
string theory are different.

The inequivalent representations of the massless states in type-IIA
and type-IIB string theory have also direct consequences for massless
states when compactifying on a Calabi-Yau three-fold. For a Calabi-Yau
manifold with Hodge numbers $h_{11}$ an $h_{12}$ the massless states
of the $N=2$ four-dimensional effective field theory on the IIA side
correspond to the states of $N=2$ supergravity with $h_{11}$ vector
supermultiplets and $h_{12}+1$ hypermultiplets. Likewise, the massless
states on the IIB side correspond to those of $N=2$ supergravity, but
now with $h_{12}$ vector supermultiplets and $h_{11}+1$
hypermultiplets. Those are the two configurations that emerge in the
circle decompactification limits of the type-II string theories when
compactified on a Calabi-Yau space times a circle. We should mention
that an additional intriguing feature of Calabi-Yau three-folds, which
will not be directly relevant for this paper, is that they appear in
pairs which are topologically different and related by the fact that
$h_{12}$ and $h_{11}$, which define the number of complex structure
moduli and of K\"ahler form moduli, respectively, are
interchanged. This surprising phenomenon is known as mirror symmetry,
and it can be combined with T-duality to obtain important results for
string effective actions (for an early reference, see
e.g. \cite{Candelas:1990qd}).

Some time ago it was demonstrated how to carry out the dimensional
reduction of $5D$ {\it off-shell} supergravity field configurations
with eight supercharges to the corresponding $4D$ ones, based on a
corresponding reduction of the off-shell supersymmetry algebra
\cite{Banerjee:2011ts} and its representations. This reduction can be
performed systematically on separate supersymmetric invariants and in
particular on actions containing higher-derivative couplings. To
accomplish the reduction one maps a supermultiplet in higher dimension
to a corresponding, not necessarily irreducible, supermultiplet in
lower dimension, possibly in a certain conformal supergravity
background. When considering the supersymmetry algebra in the context
of a lower-dimensional space-time, the dimension of the automorphism
group of the algebra (the R-symmetry group) usually increases, and
this has to be taken into account when casting the resulting
supermultiplet in a form that is appropriate for the lower-dimensional
theory.

In three dimensions, the massless matter states can be characterized
in terms of vector and tensor supermultiplets (or of on-shell
hypermultiplets). As is to be expected, the $4D$ R-symmetry group
$\big(\mathrm{SU}(2)\times \mathrm{U}(1)\big)/\mathbb{Z}_2$ is
enhanced to $(\mathrm{SU}(2)\!\times\mathrm{SU}(2))/\mathbb{Z}_2
\cong\mathrm{SO}(4)$ in three dimensions. Under the c-map the two
factors of the $3D$ R-symmetry group will be interchanged and so are
the vector and tensor supermultiplets. In addition the matter fields
of the Weyl multiplet, two scalars and one spinor, will change sign. A
similar phenomenon takes place for hypermultiplets, as their scalar
fields parametrize a local product of two quaternion-K\"ahler spaces,
each of them associated with one of the $\mathrm{SU}(2)$ factors of
the R-symmetry group.\footnote{ %%%%%%%%%
  Our attention will, however, not be focused on the conversion to
  hypermultiplets
  \cite{deWit:2001dj,Rocek:2005ij,Rocek:2006xb,Antoniadis:1993ze}.
} %%%%%%%%%%%%%%%%%%%%%%%%%%%
Some of the final results of the dimensional reduction procedure can
be compared to existing results in the literature on $N=4$ (conformal)
supergravity theories in three dimensions (see e.g.
\cite{Howe:1995zm,Bergshoeff:2010ui,Kuzenko:2011xg,
  Gran:2012mg,Butter:2013goa,Butter:2013rba} where further references
can be found). We will discuss the details in due course.

As in \cite{Banerjee:2011ts}, the off-shell reduction scheme is
subtle, especially in view of the fact that the $4D$ Weyl multiplet
decomposes into a $3D$ Weyl multiplet and an additional (Kaluza-Klein)
vector multiplet. Both in four and in three dimensions, the matter
multiplets are defined in a superconformal background consisting only
of the $4D$ or the $3D$ Weyl multiplet fields, respectively. To fully
establish this fact requires to also consider the transformation rules
beyond the linearised approximation. The fact that the R-symmetry
group is enhanced upon dimensional reduction requires a conversion of
the spinor basis.  Furthermore, to realize the extended R-symmetry
locally it is necessary to introduce an $\mathrm{SU}(2)/\mathrm{U}(1)$
local phase factor that ensures that the $4D$ and $3D$ local
R-symmetries can coexist. The central result of this paper is then to
express the $4D$ off-shell fields in terms of the $3D$ ones. This
leads to a covariant dictionary which enables us to write any $4D$
supersymmetric action in terms of its $3D$ counterparts by direct
substitution. While this is relatively straightforward for
hypermultiplet and tensor multiplet Lagrangians quadratic in
derivatives, it is much more subtle for the vector multiplet
Lagrangians. The reason is that the number of vector multiplets is
increased in the reduction by the addition a Kaluza-Klein vector
supermultiplet that originates from the $4D$ Weyl multiplet. Therefore
the resulting $3D$ Lagrangian has to be completely reformulated to
match the form of the generic $3D$ tensor multiplet Lagrangians. In
doing so, one establishes that the $3D$ vector Lagrangians, although
identical in structure to the $3D$ tensor Lagrangians, belong to a
restricted class. This can be inferred from the fact that they are
{\it manifestly} invariant under both the vector gauge symmetry and
under local R-symmetry. Moreover they are invariant under a group of
rigid transformations that are characteristic for the dimensional
reduction of $4D$ vector multiplets (for an extensive classification,
see \cite{deWit:1992wf}). None of these features are generically
present in the $3D$ tensor Lagrangians. Therefore not all the tensor
multiplets can belong to the image of the c-map. The corresponding
phenomenon for hypermultiplets has been noted long ago
\cite{Cecotti:1988qn}.

The supercovariant dictionary can straightforwardly be applied to any $4D$
off-shell supersymmetric Lagrangian including the ones with
higher-derivative couplings. We present a few examples of
higher-derivative Lagrangians and discuss their implication for the
c-map. In principle these results are relevant for explicit
four-dimensional type-II string compactifications, such as given in
\cite{Berkovits:1995cb}. This last topic definitely warrants further
study, but this is outside the scope of the present paper.

This paper is organized as follows. Section
\ref{sec:off-shell-dim-red-Weyl} presents the off-shell reduction to
three space-time dimensions of the $4D$ Weyl multiplet. After a first
discussion of its reduction we establish the resulting decomposition
into the $3D$ Weyl multiplet and a separate Kaluza-Klein vector
supermultiplet.  The necessary conversion of $4D$ into $3D$ spinors is
introduced in subsection \ref{sec:g-compensator-KKvector}. The
resulting $3D$ Weyl multiplet corresponds to $N=4$ conformal
supergravity and is considered in detail in section
\ref{sec:3d-weyl-multiplet}. Its characteristic features, in
particular those related to the c-map, are discussed and compared to
the literature.  Section \ref{sec:summary-4d-matter} analyzes the
reduction of the supersymmetry transformations for the various $4D$
matter supermultiplets: the vector supermultiplet, the tensor
supermultiplet and the hypermultiplet, by expressing all the $4D$
fields into $3D$ fields. All the results are then expressed in the
form of a supercovariant dictionary, which expresses all the $4D$ fields
into the $3D$ fields. This is done in section
\ref{sec:five-four-dimens-fields-lagr}, where we also apply the
dictionary to the$4D$ supersymmetric actions with at most two
derivatives. In a third subsection we then describe the conditions
upon which a $3D$ Lagrangian can be uplifted to two inequivalent $4D$
Lagrangians with a different field content by making use of the
c-map. A more novel application concerns the reduction of
higher-derivative couplings. This is the topic of section
\ref{sec:high-deriv-coupl} where we present a few examples and discuss
their properties in relation to the c-map. Finally there are two
appendices. Appendix \ref{App:4-3D-Riemann-curv} discusses the
relation between $4D$ and $3D$ Riemann curvatures, while the more
technical aspects of $4D$ to $3D$ spinor conversion are presented in
appendix \ref{sec:conv-spin-basis}.

%%%%%%%%%%%%%%%%%%%%%%%%%%%%%%%%%%%%%%%%%%%%%%%%%%%%%%%%%%%%%%%%
\section{Off-shell dimensional reduction; the Weyl multiplet}
\label{sec:off-shell-dim-red-Weyl}
\setcounter{equation}{0}
%%%%%%%%%%%%%%%%%%%%%%%%%%%%%%%%%%%%%%%%%%%%%%%%%%%%%%%%%%%%%%%%
Starting from the super conformal transformations for $4D$
supermultiplets we compactify one spatial dimension on a circle which
will be shrunk to zero size, so that the space-time dimension is
reduced to $3D$. Subsequently we reinterpret the results in terms of
$3D$ super conformal transformations. The first multiplet to
consider is the Weyl multiplet, because it acts as a background for
the other supermultiplets: the vector and tensor multiplet 
and the hypermultiplet. A second reason why the Weyl multiplet
deserves priority, is that it becomes reducible upon the reduction,
unlike the other (matter) supermultiplets. The $N=2$ Weyl multiplet
in $D=4$ comprises $24+24$ bosonic and fermionic degrees of freedom,
which, in the reduction to $D=3$ dimensions will decompose into the
Weyl multiplet comprising $16+16$ degrees of freedom, and a vector
multiplet comprising $8+8$ degrees of freedom. As we shall see, this
decomposition takes a subtle form off-shell.

The independent fields of the Weyl multiplet of four-dimensional $N=2$
conformal supergravity consist of the vierbein $e_M{}^A$, the
gravitino fields $\psi_M{}^i$, the dilatational gauge field $b_M$, the
R-symmetry gauge fields $\mathcal{V}_{M i}{}^j$ (which is an
anti-hermitian, traceless matrix in the $\mathrm{SU}(2)$ indices
$i,j$) and $A_M$, an anti-selfdual tensor field $T_{AB}{}^{ij}$, a
scalar field $D$ and a spinor field $\chi^i$. All spinor fields are
Majorana spinors which have been decomposed into chiral
components. Our $4D$ conventions are as in \cite{deWit:2010za}. The
three gauge fields $\omega_M{}^{AB}$, $f_M{}^A$ and $\phi_M{}^i$,
associated with local Lorentz transformations, conformal boosts and
S-supersymmetry, respectively, are not independent as will be
discussed later. The infinitesimal Q, S and K transformations of the
independent fields, parametrized by spinors $\epsilon^i$ and $\eta^i$
and a vector $\Lambda_\mathrm{K}{}^A$, respectively, are as
follows,\footnote{%%%%%%%%%%%%%%%%%%%%%%%%%%%%%%%%%%%%%%%%%%%%
  In four dimensions we consistently use world indices $M,N,\ldots$
  and tangent space indices $A,B,\ldots$. For fields that do not carry
  such indices the distinction between $4D$ and $3D$ fields may not
  always be manifest, but it will be specified in the text whenever
  necessary. We use Pauli-K\"all\'en conventions with hermitian
  gamma matrices and label the coordinates by $x^M =
  (x^4,x^1,x^2,x^3)$, where $x^\mu=(x^1,x^2,x^3)$ with
  $x^3=\mathrm{i}x^0$. Consistency with the four-dimensional results
  that we will use requires that $\varepsilon_{4123}=1$,
  $\gamma^1\gamma^2\gamma^3= \gamma^4\gamma^5$ and
  $\varepsilon_{123}=1$. From subsection
  \ref{sec:g-compensator-KKvector} we will employ proper $3D$ gamma
  matrices, which are defined in appendix
  \ref{sec:conv-spin-basis}} %%%%%%%%%%%%%%%%%%%%%%%%%%%%%%%%%%%
\begin{align}
  \label{eq:weyl-multiplet}
  \delta e_M{}^A  =&\, \bar{\epsilon}^i \, \gamma^A \psi_{ M i} +
  \bar{\epsilon}_i \, \gamma^A \psi_{ M}{}^i \, , \nonumber\\[1mm]
  \delta \psi_{M}{}^{i} =&\, 2 \,\mathcal{D}_M \epsilon^i - \tfrac{1}{8}
  T_{AB}{}^{ij} \gamma^{AB}\gamma_M \epsilon_j - \gamma_M \eta^i
  \, \nonumber \\[1mm]
  \delta b_M =&\, \tfrac{1}{2} \bar{\epsilon}^i \phi_{M i} -
  \tfrac{3}{4} \bar{\epsilon}^i \gamma_M \chi_i - \tfrac{1}{2}
  \bar{\eta}^i \psi_{M i} + \mbox{h.c.} + \Lambda_\mathrm{K}{}^A\, e_{M A} \, ,
  \nonumber \\[1mm]
  \delta A_{M} =&\, \tfrac{1}{2} \mathrm{i} \bar{\epsilon}^i \phi_{M i} +
  \tfrac{3}{4} \mathrm{i} \bar{\epsilon}^i \gamma_M \, \chi_i +
  \tfrac{1}{2} \mathrm{i}
  \bar{\eta}^i \psi_{M i} + \mbox{h.c.} \, , \nonumber\\[1mm]
  \delta \mathcal{V}_M{}^{i}{}_j =&\, 2\, \bar{\epsilon}_j
  \phi_M{}^i - 3
  \bar{\epsilon}_j \gamma_M \, \chi^i + 2 \bar{\eta}_j \, \psi_{M}{}^i
  - (\mbox{h.c. ; traceless}) \, , \nonumber \\[1mm]
  \delta T_{AB}{}^{ij} =&\, 8 \,\bar{\epsilon}^{[i} R(Q)_{AB}{}^{j]} \,
  , \nonumber \\[1mm]
  \delta \chi^i =&\, - \tfrac{1}{12} \gamma^{AB} \, \Slash{D} T_{AB}{}^{ij}
  \, \epsilon_j + \tfrac{1}{6} R(\mathcal{V})_{MN}{}^i{}_j
  \gamma^{MN} \epsilon^j -
  \tfrac{1}{3} \mathrm{i} R_{MN}(A) \gamma^{MN} \epsilon^i\nonumber\\
  &\, + D \, \epsilon^i +
  \tfrac{1}{12} \gamma_{AB} T^{AB ij} \eta_j \, , \nonumber \\[1mm]
  \delta D =&\, \bar{\epsilon}^i \,  \Slash{D} \chi_i +
  \bar{\epsilon}_i \,\Slash{D}\chi^i \, .
\end{align}
The above supersymmetry variations and also the conventional
constraints that we have to deal with in due time, depend on a number
of supercovariant curvature tensors, which will be defined shortly.
The full superconformally covariant derivative is denoted by $D_M$,
while $\mathcal{D}_M$ denotes a covariant derivative with respect to
Lorentz, dilatation, and chiral $\mathrm{SU}(2)\times \mathrm{U}(1)$
transformations, e.g.
\begin{equation}
  \label{eq:D-epslon}
  \mathcal{D}_{M} \epsilon^i = \big(\partial_M - \tfrac{1}{4}
    \omega_M{}^{AB} \, \gamma_{AB} + \tfrac1{2} \, b_M +
    \tfrac{1}{2}\mathrm{i} \, A_M  \big) \epsilon^i + \tfrac1{2} \,
  \mathcal{V}_{M}{}^i{}_j \, \epsilon^j  \,.
\end{equation}
Under local scale and $\mathrm{U}(1)$ transformations the various
fields and transformation parameters transform as indicated in table
\ref{table:weyl}.

%%%%%%%%%%%%%%%%%%%%%%%%%%%%%%%%%%%%%%%%%%%%%%%%%%%%%%%%%%
%
\begin{table}[t]
\begin{tabular*}{\textwidth}{@{\extracolsep{\fill}} |c||cccccccc|ccc||ccc| }
\hline
 $4D$& &\multicolumn{9}{c}{Weyl multiplet} & &
 \multicolumn{2}{c}{parameters} & \\[1mm]  \hline \hline
 field & $e_M{}^{A}$ & $\psi_M{}^i$ & $b_M$ & $A_M$ &
 $\mathcal{V}_M{}^i{}_j$ & $T_{AB}{}^{ij} $ &
 $ \chi^i $ & $D$ & $\omega_M^{AB}$ & $f_M{}^A$ & $\phi_M{}^i$ &
 $\epsilon^i$ & $\eta^i$
 & \\[.5mm] \hline
$w$  & $-1$ & $-\tfrac12 $ & 0 &  0 & 0 & 1 & $\tfrac{3}{2}$ & 2 & 0 &
1 & $\tfrac12 $ & $ -\tfrac12 $  & $ \tfrac12  $ & \\[.5mm] \hline
$c$  & $0$ & $-\tfrac12 $ & 0 &  0 & 0 & $-1$ & $-\tfrac{1}{2}$ & 0 &
0 & 0 & $-\tfrac12 $ & $ -\tfrac12 $  & $ -\tfrac12  $ & \\[.5mm] \hline
 $\gamma_5$   &  & + &   &    &   &   & + &  &  &  & $-$ & $ + $  & $
 -  $ & \\ \hline
\end{tabular*}
\vskip 2mm
\renewcommand{\baselinestretch}{1}
\parbox[c]{\textwidth}{\caption{\label{table:weyl}{\footnotesize
Weyl and chiral weights ($w$ and $c$) and fermion
chirality $(\gamma_5)$ of the Weyl multiplet component fields and the
supersymmetry transformation parameters in four space-time dimensions.}}}
\end{table}
%%%%%%%%%%%%%%%%%%%%%%%%%%%%%%%%%%%%%%%%%%%%%%%%%%%%%%%%%%

The gauge fields associated with local Lorentz transformations,
S-supersymmetry and special conformal boosts, $\omega_{M}{}^{AB}$,
$\phi_M{}^i$ and $f_{M}{}^A$, respectively, are composite and
determined by conventional constraints. In this case these constraints
are S-supersymmetry invariant and they take the following form,
\begin{align}
  \label{eq:conv-constraints}
  &R(P)_{M N}{}^A =  0 \, , \nonumber \\[1mm]
  &\gamma^M R(Q)_{M N}{}^i + \tfrac32 \gamma_{N}
  \chi^i = 0 \, , \nonumber\\[1mm]
  &
  e^{N}{}_B \,R(M)_{M N A}{}^B - \mathrm{i} \tilde{R}(A)_{M A} +
  \tfrac1{8} T_{ABij} T_M{}^{Bij} -\tfrac{3}{2} D \,e_{M A} = 0
  \,.
\end{align}
The curvatures appearing in \eqref{eq:conv-constraints} take the
following form,
\begin{align}
  \label{eq:curvatures-4}
  R(P)_{M N}{}^A  = & \, 2 \, \partial_{[M} \, e_{N]}{}^A + 2 \,
  b_{[M} \, e_{N]}{}^A -2 \, \omega_{[M}{}^{AB} \, e_{N]B} -
  \tfrac1{2} ( \bar\psi_{[M}{}^i \gamma^A \psi_{N]i} +
  \mbox{h.c.} ) \, , \nonumber\\[.2ex]
  R(Q)_{M N}{}^i = & \, 2 \, \mathcal{D}_{[M} \psi_{N]}{}^i -
  \gamma_{[M}   \phi_{N]}{}^i - \tfrac{1}{8} \, T^{ABij} \,
  \gamma_{AB} \, \gamma_{[M} \psi_{N]j} \, , \nonumber\\[.2ex]
  R(M)_{M N}{}^{AB} = & \,
  \, 2 \,\partial_{[M} \omega_{N]}{}^{AB} - 2\, \omega_{[M}{}^{AC}
  \omega_{N]C}{}^B
  - 4 f_{[M}{}^{[A} e_{N]}{}^{B]}
  + \tfrac12 (\bar{\psi}_{[M}{}^i \, \gamma^{AB} \,
  \phi_{N]i} + \mbox{h.c.} ) \nonumber\\
& \, + ( \tfrac14 \bar{\psi}_{M}{}^i   \,
  \psi_{N}{}^j  \, T^{AB}{}_{ij}
  - \tfrac{3}{4} \bar{\psi}_{[M}{}^i \, \gamma_{N]} \, \gamma^{AB}
  \chi_i
  - \bar{\psi}_{[M}{}^i \, \gamma_{N]} \,R(Q)^{AB}{}_i
  + \mbox{h.c.} ) \, , \nonumber\\[.2ex]
  R(A)_{M N} = & \, 2 \, \partial_{[M} A_{N ]} - \mathrm{i}
  \left( \tfrac12
    \bar{\psi}_{[M}{}^i \phi_{N]i} + \tfrac{3}{4} \bar{\psi}_{[M}{}^i
    \gamma_{N ]} \chi_i - \mbox{h.c.} \right) 
  \,.
\end{align}
%%%%%%%

%%%%%%%%%%%%%%%%%%%%%%%%%%%%%%%%%%%%%%%%%%%%%%%%%%%%%%%%%%%%
\subsection{Reduction ans\"atze}
\label{sec:ansatze}
%%%%%%%%%%%%%%%%%%%%%%%%%%%%%%%%%%%%%%%%%%%%%%%%%%%%%%%%%%%%
The reduction to three space-time dimensions is effected by first
carrying out the standard Kaluza-Klein decompositions on the various
fields, to ensure that the resulting $3D$ fields will transform
consistently under four-dimensional diffeomorphisms. The space-time
coordinates are decomposed into $x^M\to (x^4,x^\mu)$, where $x^4$
denotes the (spatial) coordinate that will be suppressed in the
reduction. Subsequently the vielbein field and the dilatational gauge
field are then written in special form, by means of an appropriate
local Lorentz transformation and a conformal boost, respectively. In
obvious notation,
\begin{equation}
  \label{eq:kk-ansatz}
  e_M{}^A= \begin{pmatrix} e_\mu{}^a &  B_\mu\phi^{-1} \\[4mm]
    0 & \phi^{-1}
    \end{pmatrix} \;,\qquad
    e_A{}^M= \begin{pmatrix} e_a{}^\mu & - e_a{}^\nu B_\nu \\[4mm]
    0 & \phi
    \end{pmatrix}\;,\qquad
    b_M = \begin{pmatrix} b_\mu \\[4mm]  0
  \end{pmatrix} \,.
\end{equation}
On the right-hand side of these decompositions, we exclusively used
three-dimensional notation, with world and tangent-space indices,
$\mu,\nu,\ldots$ and $a,b,\ldots$, taking three values. Observe that
the scaling weights for $e_M{}^A$ and $e_\mu{}^a$ are equal to $w=-1$,
while for $\phi$ we have $w=1$. The fields $b_M$ and $b_\mu$ have
weight $w=0$. The above formulae suffice to express the $4D$ Riemann
curvature tensor in terms of the $3D$ Riemann tensor and the fields
$\phi$ and $B_\mu$. The corresponding equations are collected in
appendix \ref{App:4-3D-Riemann-curv} and will be needed later on.

We now turn to the supersymmetry transformations. Since we have
imposed gauge choices on the vielbein field and the dilatational gauge
field, one has to include compensating Lorentz and special conformal
transformations when deriving the $3D$ Q-supersymmetry transformations
to ensure that the gauge conditions are preserved. Only the parameter
of the Lorentz transformation is relevant, and it is given by,
\begin{equation}
  \label{eq:comp-Lor}
  \varepsilon^{a4} = -\varepsilon^{4a} = - \phi\,
  \big(\bar\epsilon_i\gamma^a\psi^i +\bar\epsilon^i\gamma^a\psi_i)
  \,,
\end{equation}
where we assumed the standard Kaluza-Klein decomposition on the
gravitino fields,
\begin{equation}
  \label{eq:gravitino-KK}
  \psi_M{}^i =  \begin{pmatrix}\psi_\mu{}^i+ B_\mu \psi^i\\[4mm]
    \psi^i \end{pmatrix}\;,
\end{equation}
which ensures that $\psi_\mu{}^i$ on the right-hand side transforms as
a $3D$ vector.  Upon including the extra term \eqref{eq:comp-Lor}, one
can write down the Q- and S-supersymmetry transformations on the $3D$
fields defined above. As a result of this, the $3D$ and $4D$
supersymmetry transformation will be different. For instance, the
supersymmetry transformations of the $3D$ fields $e_\mu{}^a$, $\phi$
and $B_\mu$ read,
\begin{align}
  \label{eq:susy-e-B-phi}
  \delta e_\mu{}^a =&\,  \bar\epsilon_i\gamma^a\psi_\mu{}^i+
\bar\epsilon^i\gamma^a\psi_\mu{}_i
  \,, \nonumber\\[.2ex]
%%%
  \delta\phi =&\, - \phi^2\,(\bar\epsilon_i\gamma_4\psi^i
  +\bar\epsilon^i \gamma_4\psi_i)\,,
  \nonumber \\[.2ex]
%%%
  \delta B_\mu=&\,  \phi^2 \,(\bar\epsilon_i\gamma_\mu\psi^i+
\bar\epsilon^i\gamma_\mu\psi_i) +
  \phi \,(\bar\epsilon_i \gamma_4\psi_\mu{}^i+\bar\epsilon^i
  \gamma_4\psi_\mu{}_i) \,, 
\end{align}
where the first term in $\delta B_\mu$ originates from the
compensating transformation \eqref{eq:comp-Lor}. Consequently the
supercovariant field strength of $B_\mu$ contains a term that is not
contained in the supercovariant four-dimensional curvature
$R(P)_{MN}{}^A$. Therefore the $4D$ spin-connection components are not
supercovariant with respect to $3D$ supersymmetry, as is exhibited
below,
\begin{align}
  \label{eq:spin-connection}
  \omega_M{}^{ab} =&\, \begin{pmatrix} \omega_\mu{}^{ab} \\[4mm]
    0 \end{pmatrix} + \tfrac12 \phi^{-2} \hat F(B)^{ab} \,
  \begin{pmatrix} B_\mu \\[4mm] 1\end{pmatrix} \;, \nonumber \\[.8ex]
%%%%%%%
  \omega_M{}^{a4} =&\, -\tfrac12 \begin{pmatrix} \phi^{-1}
    \hat F(B)_\mu{}^a +\phi\,(\bar\psi_{\mu
      i}\gamma^a\psi^i+\bar\psi_{\mu}{}^i\gamma^a\psi_i)   \\[4mm]
    0 \end{pmatrix} -\phi^{-2} D^{a}\phi  \,
  \begin{pmatrix} B_\mu \\[4mm] 1\end{pmatrix} \;.
\end{align}
Here we introduced the supercovariant field strength and derivative
(with respect to $3D$ supersymmetry),
\begin{align}
  \label{eq:supercov-FB-Dphi}
  \hat F(B)_{\mu\nu} =&\, 2\,\partial_{[\mu} B_{\nu]} - \phi^2\,
  (\bar\psi_{[\mu i}\gamma_{\nu]}
  \psi^i+\bar\psi_{[\mu}{}^i\gamma_{\nu]} \psi_i)  -
  \phi\,\bar\psi_{[\mu}{}^i \gamma_4 \psi_{\nu] i} \,,\nonumber\\
  D_\mu\phi =&\, (\partial_\mu -b_\mu) \phi + \tfrac12 \phi^2
  \,(\bar\psi_{\mu i} \gamma_4\psi^i -\bar\psi_{\mu}{}^i \gamma_4 \psi_i) \,.
\end{align}

Subsequently we write down corresponding Kaluza-Klein decompositions
for some of the other fields of the Weyl multiplet, which do not
require special gauge choices,
\begin{equation}
  \label{eq:1Weyl-KK}
    \mathcal{V}_M{}^i{}_j=
    \begin{pmatrix}\mathcal{V}_\mu{}^i{}_j+ B_\mu
      \mathcal{V}^{i}{}_j\\[4mm]
      \mathcal{V}^{i}{}_j \end{pmatrix}\;,\qquad
    {A}_{M }=
    \begin{pmatrix}{A}_{\mu}+ B_\mu A \\[4mm]
      {A} \end{pmatrix}\;,\qquad
    \phi_M{}^i =  \begin{pmatrix}\phi_\mu{}^i+ B_\mu \phi^i\\[4mm]
      \phi^i
    \end{pmatrix}\;.
\end{equation}
Furthermore we define two complex $3D$ target-space vectors $A_a^\pm$
such that
\begin{equation}
  \label{eq:def-T}
  \begin{array}{rcl}
    T_{a4}{}^{ij}& =& A^-_a\,\varepsilon^{ij} \,, \\
    T_{a4\,ij}& =& A^+_{a}\, \varepsilon_{ij}\,,
    \end{array}
\qquad
  \begin{array}{rcl}
    T_{ab}{}^{ij}& =& \varepsilon_{abc} \,A^{- c}\,\varepsilon^{ij} \,, \\
    T_{ab\,ij}& =& - \varepsilon_{abc} \,A^{+c} \,\varepsilon_{ij}\,,
    \end{array}
\end{equation}
where $A^+_a$ and $A^-_a$ are related by complex conjugation. Here one
has to bear in mind that we are using Pauli-K\"all\'en notation, so
that $A^+_a+A^-_a$ is real when $a$ denotes a spatial component, and
imaginary when $a$ denotes the time component. This is reflected in
the different sign in the last two terms of \eqref{eq:def-T}.  All
gamma matrices are hermitian. We recall that the convention for the
Levi-Civita tensors is $\varepsilon^{4abc} = \varepsilon^{abc}$, and
$\varepsilon^{123}=1$. Correspondingly for the gamma matrices we have
the conventions that $\gamma_5=\gamma_4\gamma_1\gamma_2\gamma_3$ and
we note the useful relation $\gamma^{ab}= \varepsilon^{abc}\,\gamma_c
\gamma_4\gamma_5$, so that e.g. $T_{AB}{}^{ij}\,\gamma^{AB}= 2\,
A^-_a\,\gamma^a\gamma_4(1+\gamma_5)\,\varepsilon^{ij}$.

%%%%%%%%%%%%%%%%%%%%%%%%%%%%%%%%%%%%%%%%%%%%%%%%%%%%%%%%%%%
\subsection{Decomposition of the 4D Weyl multiplet}
\label{sec:decomp-4D-Weyl}
%%%%%%%%%%%%%%%%%%%%%%%%%%%%%%%%%%%%%%%%%%%%%%%%%%%%%%%%%%%
Hence we are now ready to consider the Q- and S-supersymmetry
transformations of the spinor fields originating from the $4D$
gravitino fields. Up to possible higher-order spinor terms, one
derives from \eqref{eq:weyl-multiplet},
\begin{align}
  \label{eq:susy-W-gravitino}
  \delta\psi^i =&\, \big[-\tfrac12
  \phi^{-2} \hat F(B)_{ab} \gamma^{ab} + \phi^{-2}  \Slash{D} \phi \gamma_4
  +\mathrm{i} A \big] \epsilon^i  + \mathcal{V}^i{}_j
  \,\epsilon^j  -  \phi^{-1} \Slash{A}^- \varepsilon^{ij}  \epsilon_j
  \nonumber\\
  &\,  -  \phi^{-1}\gamma_4 \big( \eta^i
  +\tfrac12 \Slash{A}^- \gamma_4 \varepsilon^{ij} \epsilon_j -\tfrac14
  \phi^{-1}\hat
  F(B)_{ab}\gamma^{ab}\gamma_4 \epsilon^i  \big) \,, \nonumber\\
  \nonumber\\[.1ex]
 %%%
  \delta\psi_\mu{}^i =&\,2\,\big(\partial_\mu
  -\tfrac14\omega_\mu{}^{ab}\gamma_{ab}+\tfrac12 b_\mu
  +\tfrac12\mathrm{i} A_\mu -\tfrac18 \mathrm{i}\phi^{-1}
  \hat F(B)_\mu  \big)\epsilon^i + {\cal V}_{\mu}{}^i{}_j
  \epsilon^j + A^-_\mu \,\varepsilon^{ij}\gamma_4 \epsilon_j
  \nonumber\\
  &\, -\gamma_{\mu}\big( \eta^i +\tfrac12 \Slash{A}^-
  \gamma_4 \varepsilon^{ij}\epsilon_j -\tfrac14 \phi^{-1}\hat
  F(B)_{ab}\gamma^{ab}\gamma_4 \,\epsilon^i \big) \,,
\end{align}
where $\hat F(B)_\mu = \mathrm{i} e\varepsilon_{\mu\nu\rho} \hat
F(B)^{\nu\rho}$. Although the results \eqref{eq:susy-e-B-phi} and
\eqref{eq:susy-W-gravitino} are still incomplete, they already exhibit
some of the systematic features that will turn out to be
universal. Therefore let us first have a brief perusal of these
initial results.

The fields whose transformations we have determined will belong to two
$3D$ supermultiplets, namely the Weyl and the Kaluza-Klein vector
multiplet. Clearly, the fields $e_\mu{}^a$ and $\psi_\mu{}^i$ belong
to the Weyl multiplet, whereas $\phi$, $B_\mu$ and $\psi^i$ belong to
the vector multiplet. An obvious puzzle is the fact that we have
identified only one real scalar, whereas the $3D$ vector multiplet
contains three scalars. This is related to a generic feature of
dimensional reduction, namely that lower-dimensional results are often
obtained in a gauge-fixed version of the (local) R-symmetry group.
Another aspect of this phenomenon is that the vector fields
$A^{\pm}{}_a$ seem to play the role of a complex gauge field, because
they appear to covariantize the derivatives on $\phi$ and $\epsilon^i$
in \eqref{eq:susy-W-gravitino}, in spite of the fact that the
$A^{\pm}{}_a$ are actually auxiliary fields in $D=4$. As we shall see
shortly, $A^\pm{}_a$ combined with $A_\mu$ will provide the
$\mathrm{SU}(2)$ gauge fields associated with the enhancement of the
$\mathrm{U}(1)$ factor of the $4D$ R-symmetry group. This additional
$\mathrm{SU}(2)$ group emerges in the reduction, in addition to the
manifest $\mathrm{SU}(2)$ R-symmetry group of the $4D$
theory. Hence the full $3D$ R-symmetry group equals
$(\mathrm{SU}(2)\times\mathrm{SU}(2))/\mathbb{Z}_2 \cong
\mathrm{SO}(4)$. This situation is in close analogy to what was
encountered in five dimensions \cite{Banerjee:2011ts}, where the
$\mathrm{SU}(2)$ R-symmetry group was enhanced to
$\mathrm{SU}(2)\times\mathrm{U}(1)/\mathbb{Z}_2$. Observe that in both
cases the fermions remain irreducible under the extended R-symmetry
group.

Just as in \cite{Banerjee:2011ts}, we will discover that the
higher-dimensional supersymmetry transformations yield the
lower-dimensional ones, but with parameters that involve additional
field-dependent terms. These field-dependent terms can be dropped
eventually. We see this already in the uniform field-dependent
additions to the S-supersymmetry transformations in
\eqref{eq:susy-W-gravitino} and we will discover similar modifications of
the R-symmetry transformations in due course. Some of those can be
interpreted as compensating transformations related to the fact that
the formulation that we obtain is gauge-fixed. This gauge-fixing will
be removed at the end by introducing a local
$\mathrm{SU}(2)/\mathrm{U}(1)$ phase factor, which provides the
missing two fields to the Kaluza-Klein vector multiplet.

To demonstrate some of this in more detail, let us present the higher-order
completion of \eqref{eq:susy-W-gravitino}. The pattern that we find is
repeated in the results for the matter supermultiplets that will be
presented in section \ref{sec:summary-4d-matter}. Explicit calculation
leads to the following results,
\begin{align}
  \label{eq:susy-W-grav-complete}
  \delta\big(\varepsilon_{ij}\phi^2\psi^j\big) =&\,-\tfrac12 \hat
  F(B)_{ab} \gamma^{ab}\,\varepsilon_{ij}\epsilon^j +
  \gamma^\mu\big({D}_\mu \phi\, \gamma_4 \,\varepsilon_{ij}\epsilon^j
  + \mathcal{A}^-_\mu \phi \,\epsilon_i \big)
  +\mathrm{i} C \,\phi\, \varepsilon_{ij} \epsilon^j
  +  Y_i{}^{j}{}^0\varepsilon_{jk} \,\epsilon^k
  \nonumber\\
  &\, - \phi\,\gamma_4 \varepsilon_{ij} \big(\eta^j +\tilde\eta^j\big)
  +\tfrac12 \tilde\Lambda_i{}^j (\varepsilon_{jk} \phi^2 \psi^k)
  +\tfrac12 \tilde \Sigma^-\,\varepsilon_{ij} \gamma_4
  (\varepsilon^{jk} \,\phi^2 \psi_k) \,, \nonumber\\
  \nonumber\\[.1ex]
 %%%
  \delta\psi_\mu{}^i =&\,2\,\big(\partial_\mu
  -\tfrac14\omega_\mu{}^{ab}\gamma_{ab}+\tfrac12 b_\mu
  +\tfrac12\mathrm{i} \mathcal{A}_\mu{}^0\big) \epsilon^i + {\cal
    V}_{\mu}{}^i{}_j \,\epsilon^j + \mathcal{A}^-_\mu \,\varepsilon^{ij}\gamma_4
  \epsilon_j
  \nonumber\\
  &\, -\gamma_{\mu}\big( \eta^i +\tilde\eta^i\big)  +\tfrac12
  \tilde\Lambda^i{}_j \,\psi_\mu{}^j
  -\tfrac12 \tilde \Sigma^-\,\varepsilon^{ij} \gamma_4\,
   \psi_{\mu j} \,.
\end{align}
Here we have used the definition,
\begin{align}
  \label{eq:def-two-fields}
  C =&\, \phi\,A
  -\tfrac{1}2\,\mathrm{i}\,\phi^2\,\bar\psi^k\gamma_4\psi_k\,,
  \nonumber\\[.1ex]
  Y^{i}{}_{j}{}^0=&\, \phi^2\, \big( \mathcal{V}^i{}_j +
  3\,\phi\,\bar\psi^i\gamma_4\psi_j -
  \tfrac32\,\phi\,\bar\psi^l\gamma_4\psi_l\, \delta^i{}_j \big)\,.
\end{align}
Obviously the field $Y^{0\,i}{}_{j}$ will correspond to the auxiliary
field of the Kaluza-Klein vector multiplet. The field $C$ will turn
out to belong to the $3D$ Weyl multiplet.

Furthermore we have introduced six vector fields which are related to
the two sets of SU(2) gauge fields associated with the $3D$ R-symmetry
group,
\begin{align}
  \label{eq:def-weyl-connections}
  \mathcal{A}_\mu{}^0 = &\,A_\mu -\tfrac14 \phi^{-1} \hat F(B)_\mu
  -\tfrac12 \mathrm{i} \phi^2\bar\psi^k\gamma_\mu\psi_k \,,
  \nonumber\\[.1ex]
  \mathcal{A}_\mu{}^- = &\, e_\mu{}^a\,A^-_a +\phi\,
  \varepsilon_{ij}\,\bar\psi_\mu{}^i\psi^j
  +\tfrac12\,\phi^2\,\varepsilon_{ij}\,\bar\psi^{i}
  \gamma_\mu\gamma_4\psi^{j} \,,
  \nonumber\\[.1ex]
  \mathcal{A}_\mu{}^+ = &\, e_\mu{}^a\,A^+_a + \phi\,
  \varepsilon^{ij}\,\bar\psi_{\mu i}\psi_{j}
  +\tfrac12\,\phi^2\,\varepsilon^{ij}\,\bar\psi_{i}
  \gamma_\mu\gamma_4\psi_{j}
  \,,   \nonumber\\[.1ex]
  \mathcal{V}_\mu{}^i{}_j =&\, \mathcal{V}_\mu{}^i{}_j\big\vert_{4D} +
  \phi \big(\bar\psi_\mu{}^i\gamma_4\psi_j + \bar\psi^i \gamma^4
  \psi_{\mu j} - \tfrac12\delta^i{}_j (\bar\psi_\mu{}^k\gamma_4\psi_k
  +
  \bar\psi^k \gamma^4 \psi_{\mu k}) \big)\nonumber\\
  &\,+\phi^2\, \big(\bar\psi^i\gamma_\mu\psi_j - \tfrac12
  \delta^i{}_j\, \bar\psi^k\gamma_\mu\psi_k \big) \,.
\end{align}
The remaining quantities are given by
\begin{align}
  \label{eq:field-dep-S-R}
  \tilde\eta^i=&\,\tfrac12 \Slash{A}^- \gamma_4 \varepsilon^{ij}
  \epsilon_j -\tfrac14 \phi^{-1}\hat F(B)_{ab}\gamma^{ab}\gamma_4
  \epsilon^i  \nonumber \\
  &\, +\tfrac1{2} \phi^{2} \big(\bar\psi^{(i} \psi^{j)}\,\epsilon_j
  +\bar\psi^i\gamma^a \psi_j\,\gamma_a\epsilon^j
  +\bar\psi^i\gamma_4\psi_j\, \gamma_4 \epsilon^j \big)\,,  \nonumber\\[.1ex]
%%%%%%%%
  \tilde\Lambda_i{}^j  =&\, 2\phi\,\big( \bar\epsilon_i
  \gamma_4 \psi^j + \bar\psi_i \gamma_4\epsilon^j  - \tfrac12
  \delta_i{}^j\,( \bar\epsilon_k \gamma_4 \psi^k
  +\bar\psi_k\gamma_4\epsilon^k) \big) \,, \nonumber
  \\[.1ex]
%%%%%%%%
  \tilde\Sigma^-=&\,(\tilde\Sigma^+)^\ast = 2\, \phi\,\varepsilon_{ij}
  \bar\epsilon^i\psi^j \,,
\end{align}
and are related to the various field-dependent transformations
mentioned above. They will appear universally for all fields and
define the decomposition of the $4D$ Q-supersymmetry variations,
in terms of the $3D$ Q-supersymmetry
variations combined with a field-dependent S-supersymmetry
transformation, a field-dependent $\mathrm{SU}(2)$ R-symmetry
transformation, and a field-dependent $\mathrm{SU}(2)/\mathrm{U}(1)$ chiral
transformation. The latter should be regarded as compensating
transformations associated with the fact that the reduction leads to a
gauge-fixed formulation with respect to the new (local) R-symmetry
transformations,
\begin{equation}
  \label{eq:D4-D3-decomp}
  \delta_\mathrm{Q}(\epsilon)\big|^\mathrm{reduced}_{4D} \Psi =
  \delta_\mathrm{Q}(\epsilon)\big|_{3D} \Psi +
  \delta_\mathrm{S}(\tilde\eta)\big|_{3D} \Psi +
  \delta_{\mathrm{SU}(2)}(\tilde\Lambda)\big|_{3D}\Psi  +
  \delta_{\mathrm{SU}(2)/\mathrm{U}(1)}(\tilde\Sigma)\big|_{3D}\Psi\,.
\end{equation}
To give a meaning to the right-hand side one has to identify fields
$\Psi$ that transform covariantly in the $3D$ setting, so that all
transformations in the above decomposition are clearly defined. The
identification of these fields is done iteratively. Here one has to
realize that the $4D$ transformations for the Weyl multiplet are
defined in a background consisting of the $4D$ Weyl multiplet, whereas
the $3D$ transformations of the matter multiplets are defined in the
$3D$ background. But the field-dependent parameters in
\eqref{eq:D4-D3-decomp} still depend on a variety of the $4D$ Weyl
multiplet fields. When these parameters are associated with proper
$3D$ symmetries they can be safely suppressed and this is what we will
do henceforth. Obviously this concerns the parameters $\tilde \eta^i$
and $\tilde\Lambda_i{}^j$, but not $\tilde\Sigma^{\pm}$. The fate of
$\tilde\Sigma^\pm$ will be become clear shortly in the next subsection
\ref{sec:g-compensator-KKvector}.

Let us examine some further properties of the newly defined fields
\eqref{eq:def-two-fields} and \eqref{eq:def-weyl-connections} before
proceeding. First of all, an explicit calculation reveals the
following transformations under S-supersymmetry,
\begin{align}
  \label{eq:s-susy-extra fields}
  \delta_\mathrm{S}C =&\, 0\,,  \nonumber\\[.1ex]
  \delta_\mathrm{S}Y^i{}_{j}{}^0=&\, \phi^2 \big(\bar\psi^i \eta_j -
  \psi_j \eta^i\big)
  -\tfrac12 \delta^i{\!}_j\, \phi^2 \big( \bar\psi^{k} \eta_k - \bar\psi_k \eta^k\big)
  \,, \nonumber\\[.1ex] 
  \delta_\mathrm{S}\mathcal{A}_\mu{}^0 = &\, \tfrac12\mathrm{i}
  \big(\bar\psi_{\mu{i}} \eta^i -\bar \psi_\mu{}^i\eta_i\big) \,,
  \nonumber\\[.1ex]
  \delta_\mathrm{S}\mathcal{A}_\mu{}^- = &\, -\varepsilon_{ij}\,\bar
  \psi_\mu{}^i\gamma_4\,\eta^j \,, \nonumber\\[.1ex]
  \delta_\mathrm{S}\mathcal{A}_\mu{}^+ = &\, -\varepsilon^{ij}\,\bar
  \psi_{\mu {i}}\gamma_4\,\eta_j\,,   \nonumber\\[.1ex]
  \delta_\mathrm{S}\mathcal{V}_\mu{}^i{}_j =&\, \bar\psi_\mu{}^i
  \eta_j - \bar\psi_{\mu{j}} \,\eta^i - \tfrac12 \delta^i{\!}_j
  (\bar\psi_\mu{}^k \eta_k - \bar\psi_{\mu{k}} \,\eta^k)\,.
\end{align}
Note that the S-supersymmetry transformations of the fields
$\mathcal{A}_\mu{}^0$, $\mathcal{A}_\mu^\pm$ and
$\mathcal{V}_\mu{}^i{}_j$ are very similar, which confirms that they
will indeed provide the connections associated with the
$\mathrm{SU}(2)\times\mathrm{SU}(2)$ R-symmetry group. Note also that
the Q- and S-supersymmetry transformations of the gravitini in
\eqref{eq:susy-W-grav-complete} no longer contain any auxiliary
fields, but only the connections associated with the local Lorentz
group, dilatations, and R-symmetry.

The structure of the $3D$ Weyl supermultiplet is almost covered
completely at this stage, except for the auxiliary spinor $\chi^i$ and
the scalar $D$. To see what they will represent in the $3D$ theory,
let us consider the variation of the S-invariant scalar $C$, defined
in \eqref{eq:def-two-fields}. Under Q-supersymmetry it transforms as
\begin{equation}
  \label{eq:delta-C}
   \delta C = \tfrac{1}2\,\mathrm{i}\,\bar{\epsilon}^i\,\gamma_4\,
   \breve\chi_i  + \mbox{h.c.}\,,
\end{equation}
where $\breve\chi_i$ equals
\begin{align}
  \label{eq:def-chi}
  \breve\chi_i=&\, \tfrac52\, \gamma_4\,\chi_i\big\vert_{4D}\nonumber\\
  &\, +\tfrac23 \phi\gamma_4\Slash{D}\psi_i
  +\tfrac23\,\phi\,\Slash{A}^+\,\varepsilon_{ij}\psi^j
  + \mathrm{i}\,\phi^2\,A\,\psi_i
  - \Slash{D}\phi\,\gamma_4\,\psi_i + \phi^2\,\mathcal{V}_i{}^j\psi_j
  - \tfrac1{12}\,\hat F(B)_{ab} \gamma^{ab} \, \psi_{i}\,,
\end{align}
where the right-hand side is expressed in terms of the original $4D$
fields and covariant derivatives.  In this result we used that the components
of the $4D$ S-supersymmetry gauge field are given by (up to terms
cubic in fermions)
\begin{align}
  \label{eq:phi-4}
   \phi_{\hat 4}{}^i=&\, \phi^{-1} \gamma_4 \chi^i\big|_{4D}
   +\tfrac{2}3\Slash{D}\psi^i
 +\tfrac{1}{6}\,\phi^{-1} \gamma^{ab} F(B)_{ab}\gamma_4\psi^i
 -\tfrac16\,\Slash{A}^-\gamma_4\varepsilon^{ij}\psi_j\,,
\nonumber\\[.2ex]
 \phi_\mu{}^i=&\, \phi_\mu{}^i\big\vert_{3D}
 + \tfrac12 \Slash{A}^- \gamma_4 \varepsilon^{ij} \psi_\mu{}_j
 -\tfrac14 \phi^{-1}\hat F(B)_{ab}\gamma^{ab}\gamma_4 \psi_\mu{}^i
 +\tfrac{1}{2}\, \hat F(B)_{ab}\gamma^{ab} \gamma_\mu \psi^i
 +\tfrac12\, \gamma_\mu \chi^i\big\vert_{4D}
\nonumber\\[.2ex]
 &-\tfrac{2}{3}\,\phi\gamma_4\left( e^a_\mu + \gamma^a\gamma_\mu \right)
    \left( D_a\psi^i+\tfrac1{4} \phi^{-1}\hat F(B)_{ab}\gamma^{b} \gamma_4\psi^i
   + \tfrac14\,\gamma_4 \Slash{A}^- \gamma_{a} \varepsilon^{ij} \psi_j
 \right)\,.
\end{align}
The definition of the $3D$ S-supersymmetry gauge field $\phi_\mu{}^i$
will be discussed in section \ref{sec:3d-weyl-multiplet}. 
The scalar field of the Weyl multiplet related to the $4D$ scalar $D$
can be identified by analyzing the supersymmetry transformations on the
spinor $\breve\chi_i$ defined in \eqref{eq:def-chi},
\begin{align}
   \label{eq:delta-chi}
   \delta \breve\chi^i =&\, 2\,\mathrm{i}\, \gamma_4\,\Slash{D}C \,\epsilon^i
   + D\big\vert_{3D}\, \epsilon^i
   +\tfrac12\,\gamma_4\,\gamma_c\,\varepsilon^{abc}
   R(\mathcal{V})_{ab}{}^i{}_j \epsilon^j
   \nonumber\\
   &\,
   - \tfrac12\,\gamma_4\,\gamma_c\,\varepsilon^{abc} \,
     e_a{}^\mu e_b{}^\nu \big[\big(2\,\partial_{\mu}\mathcal{A}_\nu{}^-
      +\mathrm{i}\mathcal{A}_{\mu}{}^0 \mathcal{A}_\nu{}^- \big)
   \gamma_{4}\varepsilon^{ij} \epsilon_j
   + \mathrm{i}\big(R_{\mu\nu}(\mathcal{A}^0)
   + \mathrm{i}\, \mathcal{A}_\mu{}^-\mathcal{A}_\nu{}^+\big) \epsilon^i \big]
    \nonumber\\
    &\,
    +2\mathrm{i}\,C\,\gamma_4\, \eta^i 
    -\tfrac12 \tilde \Sigma^-\,\varepsilon^{ij} \gamma_4\,\breve\chi_{j} \,,
\end{align}
where the scalar $D\vert_{3D}$ equals 
\begin{equation}
  \label{eq:def-D}
  D= 2\, D\big\vert_{4D} - \tfrac23 \phi\,D_aD^a \phi^{-1} +\phi^2
  (D_a\phi^{-1})^2  +\tfrac16\phi^{-2}\,F(B)_{ab}{}^2 
  +C^2 + \tfrac12\,\phi^{-2}\,Y^{0\,i}{}_{j}\, Y^{0\,j}{}_{i} \,.
\end{equation}
Neither in \eqref{eq:delta-chi} nor in \eqref{eq:def-D} did we include
higher-order fermionic terms. Observe that the bracket in the
second line of \eqref{eq:delta-chi} will lead to the field strengths
associated with the new $\mathrm{SU}(2)_\mathrm{R}$ symmetry. Here and
in the formulae below we will include the $3D$ conformal gauge field
$f_\mu{}^a$ in the second-order covariant derivatives, $D_\mu D^a\phi=
\mathcal{D}_\mu \mathcal{D}^a \phi + w\, f_\mu{}^a\phi$, where $w=1$
is the Weyl weight of $\phi$. The gauge field $f_\mu{}^a$ will be
defined explicitly in section \ref{sec:3d-weyl-multiplet}.

We now give the expressions for the components of the $4D$ conformal
gauge fields $f_M{}^A$. The first one, $f_\mu{}^a$ is defined in terms
of the $3D$ fields by 
\begin{align}
  \label{eq:f-33}
   f_\mu{}^a=&\,  f_\mu{}^a\big\vert_{3D}
  +\tfrac14 \phi^{-2} \, \left[ F(B)^{ac}F(B)_{\mu c}-
    \tfrac1{12}\,e^a_\mu\,F(B)_{bc}{}^2\right]
   + \tfrac1{2}\,\phi \,\left[D_\mu D^a\phi^{-1}
             -\tfrac13\,e_\mu{}^a \,D_b D^b\phi^{-1}\right]\nonumber\\[.2ex]
   &\,  -\tfrac14\,\mathrm{i} \varepsilon_{\mu}{}^{ab} D_b(C\phi^{-1} )
   +\tfrac18\, \big[ A^{+}{}_\mu A^{-\,a} + A^{-}{}_\mu A^{+\,a}
      - e_\mu{}^a\,A^{+}{}_b A^{-\,b} \big] -\tfrac14 e_\mu{}^a\, D\big\vert_{4D}\,,
\end{align}
up to fermionic terms. The remaining components are $f_{\hat 4}{}^a$,
$f_\mu{}^4$ and $f_{\hat 4}{}^4$, and are given by (we recall that $4$
denotes the tangent space index and $\hat4$ the world index associated
with the compactified coordinate)
\begin{align}
  \label{eq:f-4}
  f_{\hat 4}{}^a=&\, -\tfrac1{4}\phi^{-1} \, D_b(\omega)F(B)^{ab}
  +\tfrac34 \phi^{-2}\,F(B)^{ab}\, D_{b}\phi
  -\tfrac14\,\mathrm{i} \varepsilon^{a b c}\,[R(A)_{b c} + 
  C\phi^{-1}\,F(B)_{bc}] \nonumber\\
  &\, +\tfrac18\, \varepsilon^{a b c}\,A^{-}{}_{b} A^{+}{}_{c}
  \,, \nonumber\\[.2ex]
  f_\mu{}^4 =&\,  -\tfrac1{4}\phi^{-1} \, D^\nu(\omega)F(B)_{\mu\nu}
  +\tfrac34 \phi^{-2}\,F(B)_{\mu\nu}\, D^{\nu}\phi
  +\tfrac14\,\mathrm{i} \varepsilon_{\mu a b}\,[R(A)^{ab}
  +C\phi^{-1}\,F(B)^{ab}] \nonumber\\
  &\,+\tfrac18\, \varepsilon_{\mu a b}\,A^{-\,a} A^{+\,b} \,,
  \nonumber\\[.2ex]
  f_{\hat 4}{}^4=&\,\tfrac13 D_aD^a\phi^{-1}
  - \tfrac7{48}\phi^{-3} F(B)_{ab}{}^2
  - \tfrac1{4}\phi^{-1}\,D\big\vert_{4D}
  + \tfrac1{8}\phi^{-1} A_a^-\,A^{+\,a}\,.
\end{align}
With the exception of the last equation in \eqref{eq:f-4}, all the
linear combinations of $D_aD_b\phi$ and $(D_a\phi)^2$ appearing in
equations \eqref{eq:def-D}, \eqref{eq:f-33} and \eqref{eq:f-4} are
conformally invariant.

%%%%%%%%%%%%%%%%%%%%%%%%%%%%%%%%%%%%%%%%%%%%%%%%%%%%%%%%%%%%%%%%
\subsection{Gauge compensator and the Kaluza-Klein vector
  supermultiplet}
\label{sec:g-compensator-KKvector}
%%%%%%%%%%%%%%%%%%%%%%%%%%%%%%%%%%%%%%%%%%%%%%%%%%%%%%%%%%%%%%%%
At several occasions it was already pointed out that the $4D$
R-symmetry group is enhanced to a larger symmetry group upon
dimensional reduction. More specifically the $\mathrm{U}(1)$ factor of
the $4D$ R-symmetry group is extended to the group
$\mathrm{SU}(2)$. Hence in $3D$ one is dealing with two
$\mathrm{SU}(2)$ factors in the R-symmetry group, one that was
originally present in $4D$ and another one that emerges in the
reduction. Therefore $3D$ spinors will carry two indices, namely one
index denoted by $i,j,\ldots=1,2$ that is carried already by the $4D$
fields, and an additional index denoted by $p,q,\ldots=+,-$, that
indicates the $4D$ $\mathrm{U}(1)$ charge to be equal to $+\tfrac12$
or $-\tfrac12$, respectively. Every $3D$ spinor can thus be written as
$\Psi^{i\, p}$. It satisfies a Majorana constraint, so that it
comprises eight components, just as in the $N=2$, $4D$ setting. Here
it is crucial that spinors in $3D$ Minkowski space are real
two-component spinors.  The expressions for the $3D$ spinors in terms
of $4D$ ones involve an arbitrary phase factor and the relative phase
factors between the spinors belonging to different $3D$
supermultiplets will eventually follow from insisting on uniformity of
the R-symmetry assignments in various supersymmetry
transformations. The group-theoretical aspects of all this is
described in detail in appendix \ref{sec:conv-spin-basis}, where we
also present the relation between the $4D$ and $3D$ gamma matrices.

However, at this stage the new R-symmetry transformations are not
realized locally, whereas the ones originating from $4D$ are, as is
standard in the superconformal formulation. This phenomenon is well known
and was, for instance, also observed in the dimensional reduction from
five to four dimensions \cite{Banerjee:2011ts}. As it turns out the
resulting lower-dimensional theory is always obtained in a gauge where
all the new gauge degrees of freedom are put to zero. In the case at
hand, this can be avoided by simply re-introducing
the missing gauge degrees of freedom. This is done by introducing a
new field $\Phi^p{}_q$, which is an element of $\mathrm{SU}(2)$,
\begin{equation}
  \label{eq:Phi-prop}
  \Phi\in\mathrm{SU}(2) \Longrightarrow \Phi \,\Phi^\dagger =
  \oneone_2   \,,\quad  \det \Phi =1 \,.
\end{equation}
and which is assigned the following transformation under the new local
$\mathrm{SU}(2)$ and the original local $\mathrm{U}(1)$,
\begin{equation}
  \label{eq:local-Phi}
  \Phi \to V\,\Phi \,\begin{pmatrix} 
    \mathrm{e}^{-\mathrm{i}\Lambda_\mathrm{A}/2}&0 \\[.2ex]
    0& \mathrm{e}^{\mathrm{i}\Lambda_\mathrm{A}/2} 
    \end{pmatrix}\,,
\end{equation}
were $V$ denotes the new (local) $\mathrm{SU}(2)_\mathrm{R}$
transformation and $\Lambda_\mathrm{A}$ denotes the parameter of the
original $\mathrm{U}(1)_\mathrm{R}$ group. Obviously, when fixing
$\Phi$ to the identity, there is only one gauge transformation that is
left unaffected, corresponding to the diagonal $\mathrm{U}(1)$
subgroup. Subsequently we require that $\Phi$ transforms as follows
under Q-supersymmetry,
\begin{equation}
  \label{eq:susy-Phi}
  \Phi^{-1} \delta\Phi = \tfrac12  \begin{pmatrix} 
    0 & -\tilde\Sigma^+ \\[.3ex]
    \tilde\Sigma^- & 0 
    \end{pmatrix}\,,
\end{equation}
where $\tilde\Sigma^\pm$ is defined in \eqref{eq:field-dep-S-R}. It is
important to observe that, when proceeding to the special gauge
$\Phi=\oneone$, one will induce compensating $\mathrm{SU}(2)$
transformations proportional to $\tilde\Sigma^\pm$ in the
supersymmetry transformation. This implies that in the fully local
version of the extended R-symmetry, those terms will cancel. In due
course we see that this is indeed the case. We will subsequently
redefine all spinors by multiplying them with $\Phi$, so that they
will transform {\it locally} under $\mathrm{SU}(2)\times
\mathrm{SU}(2)$. Before doing so we have to specify the correct $3D$
spinor basis for the various fields. For the fields corresponding to
the $4D$ fields $\psi_\mu{}^i$ and $\psi^i$, the spinor parameters of
Q- and S-supersymmetry $\epsilon^i$ and $\eta^i$, their conjugate
spinors, the S-supersymmetry spinor gauge fields $\phi_\mu{}^i$ and
$\phi_{\mu i}$, and the matter spinors of the Weyl multiplet, denoted
by $\breve\chi^i$, the required expressions follow from appendix
\ref{sec:conv-spin-basis},
\begin{equation}
  \label{eq:4D-Weyl-fermion-def}
  %%%%%%%%%%%%%%%%%%
  \begin{array}{rcl}
    \psi_\mu{}^{i\,p}&\!\!\!= \!\!\!&
    \begin{pmatrix}
      \varepsilon^{ij}\,\gamma_4\,\psi_{\mu\,j} \\  \psi_\mu{}^i
    \end{pmatrix}\,, \vspace{2mm}\\
  %%%%%%%%
    \epsilon^{i\, p} &\!\!\!=\!\!\! &
    \begin{pmatrix}
      \varepsilon^{ij}\,\gamma_4\,\epsilon_j \\  \epsilon^i
    \end{pmatrix}\,, \vspace{2mm}\\
    %%%%%%%%%%%%%%%%%%% 
    \bar\epsilon_{i\, p} &\!\!\!=\!\!\!&
    \begin{pmatrix}
      \mathrm{i}\varepsilon_{ij}\,\bar\epsilon^j \\ 
      -\mathrm{i}\bar\epsilon_i\gamma_4
    \end{pmatrix} \,, \vspace{2mm} \\
         \phi_\mu{}^{i\,p} &\!\!\!=\!\!\!& 
    \begin{pmatrix} -\mathrm{i}\varepsilon^{ij} \phi_{\mu j} \\  
      \mathrm{i} \gamma_4 \phi_\mu{}^i
    \end{pmatrix} \,, 
  \end{array}
  \qquad %%%%%%%%%%%%%%%%%%%%%%%%%%%%%
    \begin{array}{rcl} 
  \psi^{i\,p}&\!\!\!=\!\!\! & \phi^2\,
  \begin{pmatrix}
    \varepsilon^{ij}\,\gamma_4\,\psi_j \\  \psi^i
  \end{pmatrix}\,,   \vspace{2mm}\\
  %%%%%%%%%%%%
    \eta^{i\,p}&\!\!\!= \!\!\!& 
    \begin{pmatrix} -\mathrm{i}\varepsilon^{ij} \eta_j\\  
      \mathrm{i} \gamma_4 \eta^i
    \end{pmatrix} \,,  \vspace{2mm} \\
    %%%%%%%%%%
    \bar\eta_{i\, p} &\!\!\!=\!\!\! &
    \begin{pmatrix}
        \varepsilon_{ij}\,\bar\eta^j \\ 
      \bar\eta_i\gamma_4
    \end{pmatrix} \,. \vspace{2mm} \\
   \chi^{i\, p} &\!\!\!=\!\!\! &
 \begin{pmatrix}
  \varepsilon^{ij}\,\gamma_4\, \breve\chi_j\\ \breve\chi^i
  \end{pmatrix}\,.
    \end{array} 
\end{equation}
Redefining the spinors will also affect the bosonic expressions that
emerge upon applying supersymmetry and there will be extra terms in
the supersymmetry transformations proportional to $\tilde\Sigma^\pm$
as a result of \eqref{eq:susy-Phi}, which cancel corresponding terms
in the Q-supersymmetry transformation rules
\eqref{eq:susy-W-grav-complete} and \eqref{eq:delta-chi} originating
from the $4D$ transformations. This cancellation is a non-trivial
check on the correctness of our strategy.  A first example of the
modification of the bosonic fields concerns the gauge fields
$\mathcal{A}_\mu{}^p{}_q$ associated with the new
$\mathrm{SU}(2)_\mathrm{R}$, which follow from the second equation in
\eqref{eq:susy-W-grav-complete}, and take the form,
\begin{equation}
  \label{eq:new-R-conn}
  \Phi^{-1}\big(\oneone \,\partial_\mu  +
  \tfrac12\mathcal{A}_\mu\big)\Phi 
  = \tfrac12  
  \begin{pmatrix} -\mathrm{i} \mathcal{A}_\mu{}^0
    & -\mathcal{A}_\mu{}^+ \\[2mm]  \mathcal{A}_\mu{}^- &\mathrm{i}
    \mathcal{A}_\mu{}^0
  \end{pmatrix} \;,
\end{equation}
where the quantities on the right-hand side are the ones obtained
previously from the $4D$ theory, which were listed in
\eqref{eq:def-weyl-connections}. From the above result one can
directly derive an equation for the field strengths associated with
the new $\mathrm{SU}(2)_\mathrm{R}$ gauge fields,
\begin{align}
  \label{eq:new-su(2)-field-strengths}
  &\Phi^{-1}\big(\partial_{[\mu} \mathcal{A}_{\nu]} +\tfrac12\mathcal{A}_{[\mu}\,,
  \mathcal{A}_{\nu]}  \big) \Phi   \nonumber\\[.8ex]
  &= \begin{pmatrix}
    -\mathrm{i} \partial_{[\mu} \mathcal{A}_{\nu]}{}^0 
    & -\partial_{\mu} \mathcal{A}_{\nu]}{}^+
    \\[2mm]  \partial_{[\mu}\mathcal{A}_{\nu]}{}^- &\mathrm{i}\partial_{[\mu}
    \mathcal{A}_{\nu]}{}^0
  \end{pmatrix}  
  + \tfrac12 \begin{pmatrix} -\mathrm{i} 
    \mathcal{A}_{[\mu}{}^0 & -\mathcal{A}_{[\mu}{}^+ \\[2mm]
    \mathcal{A}_{[\mu}{}^- &\mathrm{i} \mathcal{A}_{[\mu}{}^0
  \end{pmatrix}\,
\begin{pmatrix}
    -\mathrm{i} \mathcal{A}_{\nu]}{}^0 
    & -\mathcal{A}_{\nu]}{}^+
    \\[2mm]  \mathcal{A}_{\nu]}{}^- &\mathrm{i}
    \mathcal{A}_{\nu]}{}^0
  \end{pmatrix}
\end{align}
Obviously the gauge fields $\mathcal{A}_\mu$ transform under local
$\mathrm{SU}(2)$ transformations as 
\begin{equation}
  \label{eq:su2-A}
  \mathcal{A}_\mu \to V\,\mathcal{A}_\mu V^{-1} -2\, \partial_\mu V
  V^{-1} \,. 
\end{equation}

Likewise, the scalar field $\phi$ originating from the $4D$ metric
will now be extended to a triplet of scalar fields encoded in an
anti-hermitian matrix $L^{p}{}_q{\!}^0$ that transforms under the new
$\mathrm{SU}(2)$ R-symmetry.  Subsequently we use the phase factor
$\Phi$ to define $L^{0\,p}{}_q$,
\begin{equation}
  \label{eq:scalar-KKphoton}
  L^{p}{}_q{}^0 = \Phi\, \begin{pmatrix} -\mathrm{i} \phi&0\\[.2ex] 0& \mathrm{i}
    \phi\end{pmatrix} \Phi^{-1} \,,\qquad L^0\to(L^0)'= V\,L^0\,V^{-1}\,, 
\end{equation}
which now transforms consistently under $\mathrm{SU}(2)$ and is
invariant under the $4D$ $\mathrm{U}(1)$ R-symmetry.  Let us first
decompose the triplet $L^{p}{}_q{}^0$ according to
\cite{deWit:2001dj,deWit:2006gn} 
\begin{equation}
  \label{eq:KK-scalar-decomp}
  L^{p}{}_q{}^0 (x,\upsilon,\bar\upsilon)=  \begin{pmatrix}
    -\tfrac12\,\mathrm{i}\,x^0   &   {\upsilon}^0  \\[4mm]
    -\bar\upsilon^0   &    \tfrac12\,\mathrm{i}\,x^0
\end{pmatrix}  \,,
\end{equation}
A priori there is no restriction on the sign of $x^0$ as the phase
factor $\Phi$ can also change $x^0$ to $-x^0$. Under the new
$\mathrm{SU}(2)$ R-symmetry $L^p{}_q{}^0$ transforms as specified in
\eqref{eq:scalar-KKphoton}. For infinitesimal
transformations defined as
\begin{equation}
  \label{eq:infinitesimal-su2}
  V\approx \oneone+\tfrac12 \begin{pmatrix}\mathrm{i} \Sigma^0&
    \Sigma^+\\[3mm] 
      -\Sigma^- &-\mathrm{i}\Sigma^0\end{pmatrix} \,, 
\end{equation}
where $\Sigma^0$ is real and $\Sigma^-= (\Sigma^+)^\ast$, the
components $x^0$, $\upsilon^0$ and $\bar\upsilon^0$ thus transform as
a triplet, 
\begin{equation}
  \label{eq:vector-su2}
  \delta_{\mathrm{SU}(2)} \upsilon^0 = \mathrm{i}(\Sigma^0\,\upsilon^0
  + \tfrac12 \Sigma^+ x^0) \,,
  \qquad \delta_{\mathrm{SU}(2)} x^0= \mathrm{i}( \Sigma^-\upsilon^0
  -  \Sigma^+\bar\upsilon^0)\,.
\end{equation}
The $\mathrm{SU}(2)$ covariant derivative then equals
\begin{equation}
  \label{eq:D-su2-L}
  \mathcal{D}_\mu L^0=  \partial_\mu L^0 +\tfrac12 \big[
  \mathcal{A}_\mu\,,L^0\big] \,.
\end{equation} 

Let us now return to \eqref{eq:susy-e-B-phi} and consider the
supersymmetry transformation of $L^{p}{}_q{\!}^0$. We first note that
$\delta\phi$ in terms of the spinors \eqref{eq:4D-Weyl-fermion-def}
takes the following form,
\begin{equation}
  \label{eq:delta-phi-new}
  \delta\phi= \mathrm{i} \bar\epsilon_{i+} \psi^{i+} -\mathrm{i}
  \bar\epsilon_{i-} \psi^{i-} \,. 
\end{equation}
The supersymmetry transformation of $L^{p}{}_q{}^0$ then follows from
combining \eqref{eq:susy-Phi} with \eqref{eq:delta-phi-new},
\begin{equation}
  \label{eq:delta-KK-scalar}
   \delta L^{p}{}_q{}^0 = 2\,\bar{\epsilon}_{i\, q }\, \psi^{i\,p}
   -\delta^p{}_q\,\bar{\epsilon}_{i\, r}\, \psi^{i\,r}\,,
\end{equation}
where the spinors have been modified by including the phase factor
$\Phi$ by $\psi^{i\,p}\to\Phi^p{}_q \,\psi^{i\,q}$, so that they will
transform consistently under all the {\it local} R-symmetry
transformations.  Hence the Kaluza-Klein vector multiplet consists of
the three fields contained in $L^{p}{}_q{\!}^0$, together with the
modified spinor, $\psi^{i\,p}$, the gauge field $B_\mu$ and the
triplet of auxiliary fields $Y^{i}{}_{j}{\!}^0$ defined in
\eqref{eq:def-two-fields}.  We can now continue and consider the
variation of the spinor $\psi^{i\,p}$, the vector field $B_\mu$, and
the auxiliary fields $Y^{i}{}_j{\!}^0$ using the same conventions. In that
way one finds,
\begin{align}
  \label{eq:KK-mult-3D}
    \delta \psi^{i\,p}
    =&\, \Slash{D} L^{p}{}_q{}^0 \epsilon^{i\, q}
    -\ft12 \, \hat F(B)_{ab} \gamma^{ab}\epsilon^{i\,p} +C\,
    L^{p}{}_q{}^0\,\epsilon^{i\,q}     + Y^{i}{}_j{}^0\,\epsilon^{j\,p}
    + L^{p}{}_q{}^0\,\eta^{i\,q}   \,,   \nonumber\\
    \delta B_\mu=&\,
    \bar\epsilon_{i\,p}\gamma_\mu \psi^{i\,p}
    + L^{p}{}_q{}^0 \,\bar\epsilon_{i\,p} \psi_\mu{}^{i\,q}  \,,  \nonumber\\
    \delta Y^{i}{}_j{}^0 =&\,  2\,\bar{\epsilon}_{j\, p}\Slash{D} \psi^{i\,p}
    - L^{p}{}_q{}^0\, \bar{\epsilon}_{j\, p} \chi^{i\,q}
    -2\,C\, \bar{\epsilon}_{j\,p} \psi^{i\,p}
    - \bar\eta_{j\, p} \psi^{i\,p} - (\mbox{trace})  \,,
\end{align}
where the $3D$ gamma matrices are defined in appendix
\ref{sec:conv-spin-basis}. Here we employ a supercovariant and
$\mathrm{SU}(2)$ covariant derivative, defined by
\begin{equation}
  \label{eq:superco-u1-der}
   D_\mu L^{p}{}_q{}^0=
  (\partial_\mu-b_\mu) L^{p}{}_q{}^0
 +\tfrac12\mathcal{A}_\mu{}^p{}_r\, L^{r}{}_q{}^0 -\tfrac12
 \mathcal{A}_\mu{}^r{}_q\, L^{p}{}_r{}^0
 - \big(\bar\psi_{\mu\, i q}\,\psi^{i p} -\mbox{ trace}\big)  \,.
\end{equation}
Note that, because of the fermion redefinitions that involve the phase
factor $\Phi$, the terms in the supersymmetry transformation rules
proportional to $\tilde\Sigma^{\pm}$ have disappeared. Furthermore,
none of the fields transform under the $\mathrm{U}(1)$ local symmetry
of the $4D$ theory. This completes the derivation of the $3D$
Kaluza-Klein vector supermultiplet. 

Apart from ensuring that the $4D$ and $3D$ fields can transform
consistently under their respective R-symmetry groups, the role of the
phase factor $\Phi$ is also to sweep out the vector defined
by \eqref{eq:scalar-KKphoton} over a sphere $S^2$ such that it will
take the form \eqref{eq:KK-scalar-decomp}. This requirement fixes the
phase factor in terms of the tensor $L^{p}{}_q{\!}^0$ up to a single
phase which is related to the $\mathrm{U}(1)_\mathrm{R}$ local
symmetry of the $4D$ theory. The result is as follows,
\begin{align}
  \label{eq:phase-factor-Phi}
   \Phi^{p}{}_q(x^0,\upsilon^0,\bar\upsilon^0) =&\,
  \frac1{\sqrt{2\,L^0\,\big(L^0 +\tfrac12 x^0)}} \,\begin{pmatrix}
    \mathrm{e}^{-\mathrm{i}\Lambda_\mathrm{A}/2}\,\big(L^0 +\tfrac12
    x^0\big) &
    - \mathrm{e}^{\mathrm{i}\Lambda_\mathrm{A}/2}\,\mathrm{i}\upsilon^0  \\[8mm]
    -\mathrm{e}^{-\mathrm{i}\Lambda_\mathrm{A}/2}\,\mathrm{i} 
    \bar\upsilon^0 & \mathrm{e}^{\mathrm{i}\Lambda_\mathrm{A}/2}\,
    \big(L^0 +\tfrac12 x^0\big)
   \end{pmatrix} \,, \nonumber\\[1ex]
%%%%%%
     \phi =&\, L^0\equiv \sqrt{\det[L^{0\,p}{}_q]} = \sqrt{-\tfrac12
    L^{0\,p}{}_q\,L^{0\,q}{}_p}=\sqrt{\vert\upsilon^0\vert^2
    +\tfrac14 (x^0)^2} \,.
\end{align}
Note that there is no obvious singularity in the limit $\upsilon^0\to
0$ when $x^0>0$ . In that case $\Phi$ becomes equal to the identity
matrix. Considering the same limit when $x^0<0$, there are obviously
some factors that become singular in \eqref{eq:phase-factor-Phi}, but
the final result for the phase factor remains well defined and tends
to a different finite matrix, whose effect is to interchange the two
two eigenvalues of the matrix $L^{p}{}_q{}^0$ in
\eqref{eq:KK-scalar-decomp}.

Since the phase factor $\Phi$ is only defined up to the phase
$\Lambda_\mathrm{A}$, which is related to the exact
$\mathrm{U}(1)_\mathrm{R}$ symmetry of the $4D$ theory, it is an
element of the $\mathrm{SU}(2)/\mathrm{U}(1)$ coset space, which is
indeed isomorphic to the sphere $S^2$. This aspect gives rise to some
subtle features. For instance, we have already derived that the
$\mathrm{SU}(2)$ acts on $(x^0, \upsilon^0, \bar\upsilon^0)$ according
to \eqref{eq:vector-su2}, while on the other hand we have defined the
$\mathrm{SU}(2)$ transformation on $\Phi$ in
\eqref{eq:local-Phi}. However, it turns out that the change of $\Phi$
induced by the transformations of $(x^0, \upsilon^0, \bar\upsilon^0)$
will only be consistent with \eqref{eq:local-Phi}, if one introduces
at the same time a change of the phase $\Lambda_\mathrm{A}$. To see
this we explicitly perform the transformations \eqref{eq:vector-su2}
on $\Phi$ and note that they are subject to the following equation,
\begin{equation}
  \label{eq:su2-vs-su2}
  \delta_{\mathrm{SU}(2)} \Phi = \frac12 \begin{pmatrix}
    \mathrm{i}\Sigma^0& \Sigma^+\\[2mm]
    -\Sigma^- & -\mathrm{i}\Sigma^0  \end{pmatrix} \Phi 
  + \delta\Lambda_\mathrm{A} \,\frac{\partial\Phi}{\partial
    \Lambda_\mathrm{A}} \,,
\end{equation}
where 
\begin{equation}
  \label{eq:def-delta-Lambda-A}
  \delta\Lambda_\mathrm{A} = \Sigma^0
  -\frac{\Sigma^+\,\bar\upsilon^0+\Sigma^-\,\upsilon^0}
  {2(L^0+ \tfrac12 x^0)} \,.   
\end{equation}
The first term on the right-hand side of \eqref{eq:su2-vs-su2}
corresponds to \eqref{eq:local-Phi}. 

The same structure is repeated for all scalar triplets, since
the phase factor $\Phi$ is used to consistently translate the 
4D fields to 3D fields that are covariant with respect to the
emergent $SU(2)$. In order to illustrate this, let us now
consider the following convenient formula for a general triplet,
$(x, \upsilon, \bar\upsilon)$, that is repeatedly used later on,
\begin{align}
    \label{eq:FLF}
    &\Phi^{-1}(x^0,\upsilon^0,\bar\upsilon^0)
    \,L(x,\upsilon,\bar\upsilon) \,\Phi(x^0,\upsilon^0,\bar\upsilon^0) \nonumber\\[2mm]
    &= \frac1{2\,L^0} \begin{pmatrix} 
      -\tfrac12\mathrm{i}\big(x\,x^0+2\,\upsilon\,\bar\upsilon^0
      +2\,\bar\upsilon\,\upsilon^0\big) \
      &    -x\,\upsilon^0 +\upsilon\,x^0 
      - \displaystyle{\frac{\bar\upsilon\,\upsilon^0
        -\upsilon\,\bar\upsilon^0} {L^0+\tfrac12 x^0} }\,
      \upsilon^0 \\[6mm]
      x\,\bar\upsilon^0 -\bar\upsilon\,x^0 - 
      \displaystyle{
        \frac{\bar\upsilon\,\upsilon^0-\upsilon\,\bar\upsilon^0} 
        {L^0+\tfrac12 x^0} } \, \bar\upsilon^0
      &\tfrac12\mathrm{i}\big(x\,x^0+2\,\upsilon\,\bar\upsilon^0
      +2\,\bar\upsilon\,\upsilon^0\big) 
    \end{pmatrix}\,.
\end{align}
Here we have suppressed the phase factor parametrized by
$\Lambda_\mathrm{A}$, which is subject to the exact R-symmetry of the
$4D$ theory.  The result \eqref{eq:FLF} indeed reduces to
\eqref{eq:KK-scalar-decomp} when $x=x^0$ and $\upsilon=\upsilon^0$,
upon identifying $\phi$ with $L^0$, thus confirming the correctness of
\eqref{eq:phase-factor-Phi}.  Under the $\mathrm{SU}(2)$
transformations \eqref{eq:vector-su2} the expression in
\eqref{eq:FLF} is, however, not invariant. As follows from
\eqref{eq:su2-vs-su2} the phase $\Lambda_\mathrm{A}$ is again switched on
under the $\mathrm{SU}(2)$ transformations \eqref{eq:vector-su2} by an amount
$\delta\Lambda_\mathrm{A}$ specified by \eqref{eq:def-delta-Lambda-A}, thus
leading to
\begin{align}
  \label{eq:su(2)-Phi-L-Phi}
  &\delta_{\mathrm{SU}(2)} \Big(\Phi^{-1}(x^0,\upsilon^0,\bar\upsilon^0)
    \,L(x,\upsilon,\bar\upsilon) \,\Phi(x^0,\upsilon^0,\bar\upsilon^0)\Big)\nonumber\\
    &= \tfrac1{2} \mathrm{i} \delta\Lambda_\mathrm{A}\, \Big[\begin{pmatrix}
      1&0\\0&-1\end{pmatrix} \,, \Big(\Phi^{-1}(x^0,\upsilon^0,\bar\upsilon^0)
    \,L(x,\upsilon,\bar\upsilon)
    \,\Phi(x^0,\upsilon^0,\bar\upsilon^0)\Big) \,\Big] \,.
\end{align}
This is the expected result, because it indicates that the $3D$
$\mathrm{SU}(2)$ transformation of a $4D$ field takes the result of a
field-dependent $\mathrm{U}(1)$ transformation associated with the
$4D$ R-symmetry.

Substituting \eqref{eq:phase-factor-Phi} into \eqref{eq:new-R-conn}
one obtains the explicit expressions for the $4D$ quantities $\mathcal{A}_\mu{}^0$ and
$\mathcal{A}_\mu{}^-$, defined in
\eqref{eq:def-weyl-connections}. In principle this result must be 
expressed in terms of the $\mathrm{SU}(2$ covariant derivatives of the
components of $L^p{}_q(x^0,\upsilon^0,\bar\upsilon^0)$, but in view of
the above, this will only be the case up to a field-dependent
$\mathrm{U}(1)$ transformation of $\mathcal{A}_\mu{}^0$ and
$\mathcal{A}_\mu{}^-$. To exhibit the complexities let us first
calculate the explicit expressions  for $\mathcal{A}_\mu{}^0$ and
$\mathcal{A}_\mu{}^-$, following the definition \eqref{eq:new-R-conn}, 
\begin{align}
  \label{eq:T-A} 
  \mathcal{A}_\mu{}^- =&\,   \frac{-\mathrm{i}\bar\upsilon^0 (\upsilon^0
    \stackrel{\leftrightarrow}{\mathcal{D}}_\mu \bar\upsilon^0) + \mathrm{i}
    (L^0+\tfrac12x^0) (\bar\upsilon^0
    \stackrel{\leftrightarrow}{\mathcal{D}}_\mu x^0 )} {2\,(L^0)^2
    (L^0+\tfrac12 x^0)} \,, \nonumber\\[2mm]
  \mathcal{A}_\mu{}^0=&\, \frac{1} {2\,L^0(L^0+\tfrac12 x^0)}
  \left\{\mathrm{i}  \upsilon^0
    \stackrel{\leftrightarrow}{\mathcal{D}}_\mu \bar\upsilon^0 +
    \mathrm{Tr}\left[
    \mathcal{A}_\mu\;\begin{pmatrix} \mathrm{i}(L^0+\tfrac12 x^0)
      &-\upsilon^0 \\
      \bar\upsilon^0&  -\mathrm{i}(L^0+\tfrac12 x^0) 
    \end{pmatrix} \right] \right\} \,,
\end{align}
where covariant derivatives are defined according to
\eqref{eq:D-su2-L} and we again suppressed the phase
$\Lambda_\mathrm{A}$. As is indicated by the structure of
\eqref{eq:su(2)-Phi-L-Phi} the $4D$ quantity $\mathcal{A}_\mu{}^-$
transforms covariantly under $\mathrm{SU}(2)$. However, this is not
the case for $\mathcal{A}_\mu{}^0$ in view of the fact that the
definition \eqref{eq:new-R-conn} contains a space-time derivative
which has not been made explicit in the master formula
\eqref{eq:FLF}. Indeed explicit calculation reveals that
$\mathcal{A}_\mu{}^-$ and $\mathcal{A}_\mu{}^0$ transform as follows
under $\mathrm{SU}(2)$,
\begin{equation}
  \label{eq:u1-on-A}
  \delta_{\mathrm{SU}(2)} \mathcal{A}_\mu{}^0 = \partial_\mu\delta
  \Lambda_\mathrm{A}  \,,\qquad  
   \delta_{\mathrm{SU}(2)} \mathcal{A}_\mu^\pm =
   \pm\mathrm{i}\,\delta\Lambda_\mathrm{A} \,
   \mathcal{A}_\mu^\pm\,,
\end{equation}
which takes precisely the form of the $4D$ infinitesimal
$\mathrm{U}(1)$ transformation. We should emphasize that the
matrix-valued $\mathrm{SU}(2)$ connection $\mathcal{A}_\mu{}^p{}_q$ is
implicitly contained in the covariant derivatives on the right-hand
side of \eqref{eq:T-A} and only appears explicitly in the last term
for $\mathcal{A}_\mu{}^0$. 

%%%%%%%%%%%%%%%%%%%%%%%%%%%%%%%%%%%%%%%%%%%%%%%%%%%%%%%%%
\section{\texorpdfstring{$\boldsymbol{N=4}$}{N=4} Conformal
  supergravity in three dimensions}
\label{sec:3d-weyl-multiplet}
\setcounter{equation}{0}
%%%%%%%%%%%%%%%%%%%%%%%%%%%%%%%%%%%%%%%%%%%%%%%%%%%%%%%%%
In the previous section we have already identified all the fields
belonging the the $3D$ Weyl multiplet. For the composite gauge fields
associated with S-supersymmetry and conformal boosts,
$\phi_\mu{}^{i\,p}$ and $f_\mu{}^a$, we did not yet present explicit
expressions. The proper $3D$ spinor field $\phi_\mu{}^{i\,p}$ follows
from the same redefinition that led to $\eta^{i\,p}$ given in
\eqref{eq:4D-Weyl-fermion-def}, followed by a multiplication with
the matrix $\Phi$. The additional fermion field $\chi$ was defined in
\eqref{eq:def-chi}, and its $3D$ definition was already given in 
\eqref{eq:4D-Weyl-fermion-def} (again, up to the uniform
multiplication with the matrix $\Phi$).

Rather than to recast the $4D$ fields into the $3D$ fields we present
the resulting $3D$ superconformal theory directly in $3D$. We
emphasize that the results we are about to describe are consistent
with results previously reported in the literature, such as in
\cite{Howe:1995zm,Bergshoeff:2010ui,Kuzenko:2011xg, Gran:2012mg}. In
particular they fully reproduce the non-linear results that were
obtained in \cite{Butter:2013goa,Butter:2013rba}.

We start with the Q- and S-supersymmetry transformations and the
conformal boosts, acting on the independent fields,
\begin{align}
\label{eq:Weyl-3D}
   \delta e_\mu{}^a  =&\,
   \bar\epsilon_{i\, p}\gamma^a\psi_\mu{}^{i\,p} \, , \nonumber\\
   \delta\psi_\mu{}^{i\,p}=&\, 2\,\mathcal{D}_\mu \epsilon^{i\,p}
   - \gamma_\mu\,\eta^{i\,p} \,, \nonumber\\
   \delta b_\mu =&\,
   \tfrac12\,\bar\epsilon_{i\, p} \phi_\mu{}^{i\, p}
   -\tfrac12\,\bar\eta_{i\, p}\psi_\mu{}^{i\, p}
   +\Lambda_\mathrm{K}{}^a\,e_{\mu a} \, ,
   \nonumber\\
   \delta \mathcal{V}_\mu{}^{i}{}_j =&\,
   \bar\epsilon_{j\, p} \phi_\mu{}^{i\, p}
   -2\,C\, \bar\epsilon_{j\, p} \psi_\mu{}^{i\, p}
   -  \bar\epsilon_{j\, p} \,\gamma_\mu\, \chi^{i\, p}
   +  \bar\eta_{j\, p}\psi_\mu{}^{i\, p}
   - (\mbox{trace}) \, ,
   \nonumber\\
   \delta \mathcal{A}_\mu{}^{p}{}_q =&\,
   \bar\epsilon_{i\, q} \phi_\mu{}^{i\, p}
   +2\,C\, \bar\epsilon_{i\, p} \psi_\mu{}^{i\, q}
   + \bar\epsilon_{i\, p} \,\gamma_\mu\, \chi^{i\, q}
   + \bar\eta_{i\, p} \psi_\mu{}^{i\, q}
   - (\mbox{trace}) \nonumber\\
   \delta C =&\, \tfrac12\, \bar\epsilon_{i\, p} \, \chi^{i\, p} \,, \nonumber \\
   \delta \chi^{i\, p} =&\, 2\,\Slash{D}C \,\epsilon^{i\, p}
   + D \, \epsilon^{i\, p}
   +\tfrac12\,R(\mathcal{A})_{ab}{}^p{}_q\gamma^{ab} \epsilon^{i\, q}
   -\tfrac12\,R(\mathcal{V})_{ab}{}^i{}_j\gamma^{ab} \epsilon^{j\, p}
   +2 \,C\, \eta^{i\, p} \, , \nonumber \\
   \delta D =&\, \bar\epsilon_{i\, p} \Slash{D} \chi^{i\, p}
   - \bar\eta_{i\, p} \, \chi^{i\, p} \, ,
\end{align}
where $\mathcal{V}_\mu{}^{i}{}_j$ and $\mathcal{A}_\mu{}^{p}{}_q$ are
the $\mathrm{SU}(2)\times \mathrm{SU}(2)$ R-symmetry gauge fields with
corresponding field strengths $R(\mathcal{V})_{\mu\nu}{}^i{}_j$ and
$R(\mathcal{A})_{\mu\nu}{}^p{}_q$, respectively.  Furthermore we will use
covariant derivatives with respect to Lorentz, dilatation, and
R-symmetry transformations, such as 
\begin{equation}
  \label{eq:cov-der-epsilon}
  \mathcal{D}_\mu \epsilon^{i\,p}
  =\big(\partial_\mu - \tfrac{1}{4} \omega_\mu{}^{ab} \, \gamma_{ab}
  + \tfrac1{2} \, b_\mu \big)\epsilon^{i\,p}
  +\tfrac12 \mathcal{V}_{\mu}{}^i{}_j \, \epsilon^{j\,p}
  + \tfrac12 \mathcal{A}_{\mu}{}^p{}_q \, \epsilon^{i\,q}\,,
\end{equation}
while $D_\mu$ denotes the covariant derivative with respect to all
superconformal symmetries. We stress that the Q-supersymmetry
transformation of $b_\mu$ does not coincide with the result that one
obtains from the $4D$ variation given in
\eqref{eq:weyl-multiplet}. This is not an issue because the difference
can be viewed as a field-dependent shift that can be absorbed into the
conformal boosts. Since $b_\mu$ is the only independent field that
transforms under conformal boosts, this has no effect somewhere else,
other than that it changes the field-dependent terms in the
commutation relations (which we are not making use of
explicitly). Finally the scalar fields $C$ and $D$ were identified in
the $4D$ theory in \eqref{eq:def-two-fields} and \eqref{eq:def-D},
respectively. We have summarized the field content of the $3D$ Weyl
multiplet in table \ref{table:weyl3}.

%%%%%%%%%%%%%%%%%%%%%%%%%%%%%%%%%%%%%%%%%%%%%%%%%%%%%%%%%%
%
\begin{table}[t]
\begin{tabular*}{\textwidth}{@{\extracolsep{\fill}} |c||cccccccc|ccc||ccc| }
\hline
  $3D$& &\multicolumn{9}{c}{Weyl multiplet} & &
 \multicolumn{2}{c}{parameters} & \\[1mm]  \hline \hline
 field & $e_\mu{}^{a}$ & $\psi_\mu{\!}^{i\,p}$ & $b_\mu$ &
 $\mathcal{V}_\mu{\!}^i{}_j $ &
 $\mathcal{A}_\mu{}^p{}_q$ & $C$ &
 $ \chi^{i\,p} $ & $D$ & $\omega_M^{AB}$ & $f_M{}^A$ & $\phi_M{}^i$ &
 $\epsilon^i$ & $\eta^i$
 & \\[.5mm] \hline
$w$  & $-1$ & $-\tfrac12 $ & 0 &  0 & 0 & 1 & $\tfrac{3}{2}$ & 2 & 0 &
1 & $\tfrac12 $ & $ -\tfrac12 $  & $ \tfrac12  $ & \\ \hline
\end{tabular*}
\vskip 2mm
\renewcommand{\baselinestretch}{1}
\parbox[c]{\textwidth}{\caption{\label{table:weyl3}{\footnotesize
      Weyl multiplet component fields and supersymmetry parameters
      with their corresponding Weyl weights in three space-time
      dimensions.}}}  
\end{table}
%%%%%%%%%%%%%%%%%%%%%%%%%%%%%%%%%%%%%%%%%%%%%%%%%%%%%%%%%%

Note that the two $\mathrm{SU}(2)_\mathrm{R}$ gauge fields do not
appear symmetrically in \eqref{eq:Weyl-3D}, as the terms proportional
to the auxiliary fields are odd under the exchange of indices. We
therefore find that the Weyl multiplet is symmetric under the
interchange
\begin{equation}
\label{eq:su2-interchange}
 \mathcal{V}_\mu{}^{i}{}_j \longleftrightarrow
 \mathcal{A}_\mu{}^{i}{}_j\,,
\quad
C \rightarrow -C\,,
\quad
\chi^{i\,p} \rightarrow -\chi^{i\,p}\,,
\quad
D \rightarrow -D\,,
\end{equation}
while the vielbein and the gravitini are invariant. Similar properties
have been observed before in three-dimensional extended supergravity
\cite{Bergshoeff:2010ui}, but actually this property could also have
been inferred from \cite{Howe:1995zm} (see also
\cite{Kuzenko:2011xg,Gran:2012mg}). Altogether this Weyl multiplet has
$16+16$ degrees of freedom. The gravitational field $e_\mu{}^a$, the
R-symmetry connections and the two scalar fields comprise 2, 6, 6, 2
bosonic degrees of freedom, respectively. The gravitini, and the
spinor $\chi$ comprise 4 and 4 fermionic degrees of freedom,
respectively. The covariant fields are described by a real scalar
superfield of Weyl weight $w=1$, subject to the constraint that the
second-order superspace derivative $D_\alpha{}^{i\,p} \,D^{\alpha
  j\,q}$ is proportional to the trace with respect to the
$\mathrm{SO}(4)$ indices
\cite{Gran:2012mg,Butter:2013rba}. Furthermore the transformation
\eqref{eq:su2-interchange} can be understood as a parity
transformation in the internal euclidean four-space parametrized by
coordinates $X^I\sim X^{i\, p}$. The fields $C$ and $D$ transform both
as pseudoscalars under this parity operation whereas the R-symmetry
gauge fields consist of linear combinations of three vectors and three
pseudovectors. Note that the $3D$ results that have appeared in the
literature are usually in the context of the $\mathrm{SO}(N)$
R-symmetry group; the case $N=4$ is special because the R-symmetry
group factorizes. 

%They have mainly been obtained using the superspace
%framework and concern both conformal and non-conformal supergravities
%\cite{Kuzenko:2011xg,Gran:2012mg,Butter:2013goa,Butter:2013rba}. Needless
%to say, also our subsequent $3D$ results will be consistent with these
%results.

As in $4D$, the gauge fields associated with local Lorentz transformations,
S-supersymmetry and special conformal boosts, $\omega_{\mu}{}^{ab}$,
$\phi_\mu{}^{ip}$ and $f_{\mu}{}^a$, respectively, are composite and
determined by conventional constraints. These constraints
are S-supersymmetry invariant and they take the following form,
\begin{align}\label{eq:3D-conv-constr}
 R(P)_{\mu\nu}{}^a = R(Q)_{\mu\nu}{}^{i\,p}= R(M)_{\mu\nu}{}^{a b}= 0\,,
\end{align}
where the relevant curvatures appearing in \eqref{eq:3D-conv-constr}
are given by
\begin{align}
  \label{eq:curvatures-3}
  R(P)_{\mu\nu}{}^a = & \, 2 \, \partial_{[\mu} \, e_{\nu]}{}^a + 2 \,
  b_{[\mu} \, e_{\nu]}{}^a -2 \, \omega_{[\mu}{}^{ab} \, e_{\nu]b} -
  \tfrac1{2} \, \bar\psi_{[\mu i\,p} \gamma^a \psi_{\nu]}{}^{ip}
  \, , \nonumber\\[.2ex]
  R(Q)_{\mu \nu}{}^{i\,p} = & \, 2 \, \mathcal{D}_{[\mu} \psi_{\nu]}{}^{i\,p} -
  \gamma_{[\mu}   \phi_{\nu]}{}^{i\,p}  \, , \nonumber\\[.2ex]
  R(M)_{\mu \nu}{}^{ab} = & \,
  2 \,\partial_{[\mu} \omega_{\nu]}{}^{ab} - 2\, \omega_{[\mu}{}^{ac}
  \omega_{\nu]c}{}^b
  - 4 f_{[\mu}{}^{[a} e_{\nu]}{}^{b]}
  + \tfrac12 \,\bar{\psi}_{[\mu i\,p} \, \gamma^{ab} \, \phi_{\nu]}{}^{i\,p}\,.
\end{align}
%%%%%%%
The constraints \eqref{eq:3D-conv-constr} can be solved directly,
\begin{align}
  \label{eq:dependent}
  \omega_\mu^{ab} =&\, -2e^{\nu[a}\partial_{[\mu}e_{\nu]}{}^{b]}
     -e^{\nu[a}e^{b]\sigma}e_{\mu c}\partial_\sigma e_\nu{}^c
     -2e_\mu{}^{[a}e^{b]\nu}\,b_\nu  % \nonumber\\
      -\ft{1}{4}\,\left(2\bar{\psi}_{\mu\,i\,p}\gamma^{[a}\psi^{b] i\,p}
     +\bar{\psi}^{a}_{i\,p}\gamma_\mu\psi^{b\,i\,p} \right) \,,\nonumber\\
     %%%
     \phi_\mu{}^{i\,p}  =& \, \tfrac12 \gamma^{\rho \sigma} \gamma_\mu
      \mathcal{D}_\rho \psi_\sigma{}^{i\,p}
    \,,  \nonumber\\
    %%%%%%%%
    f_\mu{}^{a}  =& \, R(\omega,e)_\mu{}^a
         -\tfrac14\,e_\mu{}^a\,R(\omega,e)
    +\tfrac12 \,e_b^{\nu} \, \bar{\psi}_{[\mu i\,p} \,
                      \gamma^{ab} \, \phi_{\nu]}{}^{i\,p}
    -\tfrac18 \,e_{\mu}{}^{a}\, \bar{\psi}_{\rho i\,p} \,
                      \gamma^{\rho \sigma} \, \phi_{\sigma}{}^{i\,p} \, ,
\end{align}
where $R(\omega,e)_\mu{}^a= R(\omega)_{\mu\nu}{}^{ab} e_b{}^\nu$ is
the non-symmetric Ricci tensor, and $R(\omega,e)$ the corresponding
Ricci scalar. The curvature $R(\omega)_{\mu\nu}{}^{ab}$ is associated
with the spin connection field $\omega_\mu{}^{ab}$.

The transformations of $\omega_{\mu}{}^{ab}$, $\phi_\mu{}^{i\,p}$ and
$f_{\mu}{}^a$ are induced by the constraints
\eqref{eq:3D-conv-constr}. We present their Q- and S-supersymmetry
variations up to terms cubic in fermions, as well as the
transformations under conformal boosts, as
\begin{align}
  \label{eq:dep-variations}
  \delta\omega_\mu{}^{ab} =&\,-\tfrac12 \bar \epsilon_{i\,p}\gamma^{ab}
  \phi_{\mu}{}^{i\,p}
  - \tfrac12\,\bar \eta_{i\,p}\gamma^{ab} \psi_{\mu}{}^{i\,p}
  + 2\,\Lambda_\mathrm{K}{}^{[a} e_\mu{}^{b]}\,, \nonumber \\
   %%%%%%
  \delta\phi_\mu{}^{i\,p} =&\, - 2\,f_\mu{}^a\gamma_a\epsilon^{i\,p}
  + \ft12\, \gamma^{ab}\gamma_\mu \left(
   R({\cal V})_{ab}{}^{\, i}{}_{\!j} \,\epsilon^{j\,p}
 + R({\cal A})_{ab}{}^{\, p}{}_{\!q} \,\epsilon^{i\,q} \right)
 \nonumber\\
  &\, + 2\,\mathcal {D}_\mu\eta^{i\,p} +
  \Lambda_\mathrm{K}{}^a\gamma_a\psi_\mu{}^{i\,p} \,,\nonumber\\
  %%%%%%
  \delta f_\mu{}^a =&\,
  -\tfrac12\, \bar\epsilon_{i\,p} \gamma^{ab} R(S)_{\mu b}{}^{i\,p}
  +\tfrac14\,\mathrm{i}\, \varepsilon^{abc}
  \bar\epsilon_{i\,p}
  \left( R({\cal V})_{bc}{}^{\, i}{}_{\!j}\psi_\mu{}^{j\,p}
    + R({\cal A})_{bc}{}^{\, p}{}_{\!q} \psi_\mu{}^{i\,q}\right)
\nonumber\\
  &\,  +\tfrac12\, \bar\eta_{i\,p} \gamma^a\phi_{\mu}{}^{i\,p}
  +\mathcal{D}_\mu \Lambda_\mathrm{K}{}^a\,.
\end{align}
Note that the curvature, $R(S)_{\mu b}{}^{i\,p}$, of the S-supersymmetry
gauge field appears explicitly, as one cannot solve for it in terms of
other fields. This is familiar from the curvature $R(M)_{\mu\nu}{}^{ab}$
in higher dimensional theories, but in the three dimensional theory this
property extends to the field strengths of all composite gauge fields.

In order to exhibit this difference with higher dimensions, we consider
the Bianchi identities for $R(P)_{\mu\nu}{}^{a}$, $R(Q)_{\mu\nu}{}^{ip}$
and $R(M)_{\mu\nu}{}^{ab}$, which lead to
\begin{align}
 D_{[a}R(P)_{b c]}{}^{d} + R(M)_{[bc\,a]}{}^{d} = \delta^d_{[a} R(D)_{b c]}
\quad \Rightarrow &\, \quad
 R(D)_{ab} =0\,,
\nonumber\\
 D_{[a}R(Q)_{b c]}{}^{i\,p} = \gamma_{[a} R(S)_{b c]}{}^{i\,p}
\quad \Rightarrow &\, \quad
\gamma^{ab} R(S)_{ab}{}^{i\,p}=0\,,
\nonumber\\
\varepsilon^{abc} D_a R(M)_{bc}{}^{de} = 2\,
  \varepsilon^{bc[d} \, R(K)_{bc}{}^{e]}
\quad \Rightarrow &\, \quad
 \varepsilon^{bc[d} \, R(K)_{bc}{}^{e]}=0\,,
\end{align}
where in the second step in each line we used the conventional
constraints \eqref{eq:3D-conv-constr}, to obtain simple constraints on
the curvatures $R(D)_{\mu\nu}$, $R(S)_{\mu\nu}{}^{ip}$ and
$R(K)_{\mu\nu}{}^{a}$. In particular, the constraints for
$R(S)_{\mu\nu}{}^{ip}$ and $R(K)_{\mu\nu}{}^{a}$ are identically
satisfied for the composite gauge fields in
\eqref{eq:dependent}. These results are generally in agreement with
earlier off-shell results
\cite{Bergshoeff:2010ui,Howe:1995zm,Gran:2012mg,Butter:2013goa,Butter:2013rba}
as well as with on-shell results \cite{Rocek:1985bk,Lindstrom:1989eg}.

%%%%%%%%%%%%%%%%%%%%%%%%%%%%%%%%%%%%%%%%%%%%%%%%%%%%%%%%%%%%%%
\section{Off-shell dimensional reduction: matter multiplets}
\label{sec:summary-4d-matter}
\setcounter{equation}{0}
%%%%%%%%%%%%%%%%%%%%%%%%%%%%%%%%%%%%%%%%%%%%%%%%%%%%%%%%%%%%%%
In this section we consider the off-shell reduction of three $4D$
supermultiplets in a background of conformal supergravity: the vector
multiplet, the tensor multiplet and the hypermultiplet. The strategy
is the same as previously followed for the Weyl and the Kaluza-Klein
vector multiplet and the results turn out to be mutually
consistent. At the end we note that there exists a second $3D$
hypermultiplet that arises upon applying
\eqref{eq:su2-interchange}.

\paragraph{\it The vector multiplet:\\} 
In four space-time dimensions the vector supermultiplet consists of a
complex scalar $X$, a chiral spinor doublet $\Omega_i$, a gauge field
$W_M$ and a triplet of auxiliary field $Y_{ij}$ which transform
under Q- and S-supersymmetry transformations as follows,
\begin{align}
  \label{eq:variations-vect-mult}
  \delta X =&\, \bar{\epsilon}^i\Omega_i \,,\nonumber\\
  \delta\Omega_i =&\, 2 {\Slash{D}} X\epsilon_i
     +\ft12 \varepsilon_{ij}  \big( F_{AB}^- -\tfrac14 \bar X
     T_{AB}{}^{kl}\varepsilon_{kl} \big) \gamma^{AB}\epsilon^j +Y_{ij}
     \epsilon^j  +2X\eta_i\,,\nonumber\\
  \delta W_{M} = &\, \varepsilon^{ij} \bar{\epsilon}_i
  (\gamma_{M} \Omega_j+2\,\psi_{M j} X)
  + \varepsilon_{ij}
  \bar{\epsilon}^i (\gamma_{M} \Omega^{j} +2\,\psi_M{}^j
  \bar X)\,,\nonumber\\
\delta Y_{ij}  = &\, 2\, \bar{\epsilon}_{(i}
  {\Slash{D}}\Omega_{j)} + 2\, \varepsilon_{ik}
  \varepsilon_{jl}\, \bar{\epsilon}^{(k} {\Slash{D}}\Omega^{l)
  } \,,
\end{align}
where $(Y^{ij})^\ast\equiv Y_{ij}= \varepsilon_{ik}\varepsilon_{jl}
Y^{kl}$, and $F_{MN}^-$ denotes the anti-selfdual supersymmetrized
component of the field strength $F_{MN} =\partial_MW_N-\partial_NW_M$.
Under local scale and chiral transformations the fields transform
according to the weights shown in table
\ref{table:w-weights-matter-4D}.

Upon reduction to three dimensions, the vector field decomposes into a
$3D$ vector field $W_\mu$ and a scalar field $W= W_{\hat 4}$ according
to the standard Kaluza-Klein decomposition
\begin{equation}
  \label{eq:vec-KK}
    W_{M }=
    \begin{pmatrix}W_{\mu}+ B_\mu W \\[4mm]
      W \end{pmatrix}\,.
\end{equation}
As it will turn out, it is convenient to define the following
linear combinations for the fermions and the auxiliary triplet,
\begin{align}
  \label{eq:redef-vector}
%   \breve W_\mu=&\, W_\mu - B_\mu \,W\,, \nonumber \\ 
  \breve\Omega_i=&\,\Omega_i+ \phi^2\,W\,\varepsilon_{ij}\psi^j\,,
  \nonumber\\
  \breve Y_{ij} =&\, Y_{ij} + \phi^2\,W\,\varepsilon_{ik} {\cal V}^k{}_j
  +\, \phi^3\,W\,\bar\psi^k\gamma^4\psi_{(j}\varepsilon_{i)k}
  -\phi^2\,X\,\bar\psi_{(i}\psi_{j)} -\phi^2\,\bar
  X\,\varepsilon_{ik}\varepsilon_{jl} \bar\psi^{(k}\psi^{l)}
  \nonumber\\
  &\, - \phi\,\bar\psi_{(i}\gamma^4\breve\Omega_{j)}
  -\varepsilon_{ik}\varepsilon_{jl}\, 
  \phi\,\bar\psi^{(k}\gamma^4\breve\Omega^{l)}
  \,.
\end{align}
The supersymmetry variations then take the following form,
\begin{align}
  \label{eq:vect-mult-3D}
  \delta X =&\, \bar{\epsilon}^i\breve \Omega_i
  - \tfrac12\tilde\Sigma^-  \, \phi W \,,\nonumber\\
  \delta (\phi W) = &\,
  \varepsilon^{ij} \bar{\epsilon}_i\gamma_{4} \breve\Omega_j
 +\varepsilon_{ij} \bar{\epsilon}^i\gamma_{4} \breve\Omega^j
  + \tilde\Sigma^- \bar X +  \tilde\Sigma^+  X \,, \nonumber\\
  \delta\breve\Omega_i
  =&\, 2\, \Slash{D}(\mathcal{A}^0) X\,\epsilon_i
 + \phi\,W\,\Slash{\mathcal{A}}^{-} \epsilon_i
 \nonumber\\
  &\,
  +\varepsilon_{ij} \Slash{D} (\phi\,W) \gamma_4 \epsilon^j
  -\big( X\, \Slash{\mathcal{A}}^+ + \bar{X}\, \Slash{\mathcal{A}}^{-} \big)
      \gamma_4 \varepsilon_{ij} \epsilon^j
 \nonumber\\
  &\,
  +\ft12 \varepsilon_{ij} \hat F(W)_{ab} \gamma^{ab}\epsilon^j
  +\breve Y_{ij}\epsilon^j
 +\mathrm{i} \,C \,\big( 2X \gamma_4\epsilon_i
   +\phi\,W\,\varepsilon_{ij}\epsilon^j\big) \nonumber\\
&\,
     +2X\,\eta_i  -\phi\,W\,\gamma_4\varepsilon_{ij}
   \eta^j  +\tfrac12\,\tilde\Sigma^+ \,\varepsilon_{ij}\gamma_4\,
    \breve\Omega^j  \,,  \nonumber\\
    \delta W_{\mu} = &\, \varepsilon^{ij} \bar{\epsilon}_i
    \big(\gamma_{\mu} \breve\Omega_j  +2\,\psi_{\mu j} X
      + \phi\,W\,\varepsilon_{jk}\gamma_4\psi_\mu{}^k\big)
    +\mathrm{h.c.} \,,   \nonumber\\
    \delta \breve Y_{ij}  = &\, 2\, \bar{\epsilon}_{(i}
    \Slash{D}(\mathcal{A}^0)\breve\Omega_{j)}
    -\bar{\epsilon}^{k}\Slash{\mathcal{A}}^+\,\gamma_4
    \breve\Omega_{(i} \varepsilon_{j)k}
   \nonumber\\
   &\,
    +2\mathrm{i}\,C\,\bar{\epsilon}_{(i} \gamma_4 \breve\Omega_{j)}
   - \bar\epsilon_{(i}\big( 2\,X\, \breve\chi_{j)} 
     + \phi\,W\,\varepsilon_{j)k}\breve\chi^k \big)
    -\eta_{(i} \breve\Omega_{j)}    + \mbox{h.c.}   \,,
\end{align}
where we have again suppressed the field-dependent S-supersymmetry
and $\mathrm{SU}(2)$ transformations with parameters $\tilde \eta_i$
and $\tilde\Lambda_i{}^j$ as in \eqref{eq:field-dep-S-R}. In deriving
the above result all higher-order terms in the fermions were taken
into account, with the exception of those appearing in the variation of
the auxiliary fields. The covariant derivative $D_\mu(\mathcal{A}^0)$
is $3D$ Lorentz covariant and contains the modified $\mathrm{U}(1)$
gauge field defined in \eqref{eq:def-weyl-connections}. The latter
appears always in combination with terms proportional to
$\mathcal{A}^\pm_\mu$ that will eventually provide the full
$\mathrm{SU}(2)$ covariantization. Furthermore, we have used various
expressions defined in \eqref{eq:def-two-fields},
\eqref{eq:field-dep-S-R} and \eqref{eq:def-chi}. Finally, note that
the spinor $\breve\chi^i$ was defined in \eqref{eq:def-chi} and that
in all the expressions we use the $3D$ Q- and S-supersymmetry gauge
fields identified in section \ref{sec:off-shell-dim-red-Weyl}.

The expressions found above can be compared to the expressions given
for the Kaluza-Klein multiplet, derived in the previous section. As it
turns out, the $3D$ supersymmetry transformations for the latter
coincide with those given above upon introducing the appropriate
redefinitions of fields in three dimensions, 
\begin{equation}
  \label{eq:3d-vec-scal}
  L^{p}{}_q= \Phi
  \begin{pmatrix}
    \mathrm{i}\,\phi\, W\, & 2\,\mathrm{i}\,\bar X   \\[4mm]
    2\,\mathrm{i}\,X & -\mathrm{i}\,\phi\, W
  \end{pmatrix}
  \Phi^{-1}\,,
  \quad
  Y^i{}_j=\varepsilon^{ik}\,\breve Y_{kj}\,, \quad
 \Omega^{i\,p}= \Phi
 \begin{pmatrix}  -\gamma_4\,\breve\Omega^j  \\[4mm]
   \varepsilon^{ij}\,\breve\Omega_j 
    \end{pmatrix} \,.
\end{equation}
We can then write the supersymmetry variations \eqref{eq:vect-mult-3D}
as
\begin{align}
  \label{eq:susy-vector-3D}
  \delta L^{p}{}_q =&\,
  2\, \bar{\epsilon}_{i\, q }\, \Omega^{i\,p}
  -\delta^p{}_q\,\bar{\epsilon}_{i\, r}\, \Omega^{i\,r}
  \,,\nonumber\\
  \delta \Omega^{i\,p}
  =&\,
  \Slash{D} L^{p}{}_q \epsilon^{i\, q}
  -\ft12 \, F(W)_{ab} \gamma^{ab}\epsilon^{i\,p}  %sign changed%%%
  +\hat Y^i{}_j\epsilon^{j\,p}
  + C\, L^{p}{}_q\,\epsilon^{i\,q}
  + \,L^{p}{}_q\, \eta^{i\,q}
  \,,\nonumber\\
  \delta W_\mu=&\,
  \bar\epsilon_{i\,p}\gamma_\mu \Omega^{i\,p}
  + L^p{}_q \,\bar\epsilon_{i\,p} \psi_\mu{}^{i\,q}
  \,,\nonumber\\
  \delta  Y^i{}_j
  =&\,
  2\, \bar{\epsilon}_{j\, p}\Slash{D} \Omega^{i\,p}
  - L^{p}{}_q\, \bar{\epsilon}_{j\, p} \chi^{i\,q}
  -2\,C\, \bar{\epsilon}_{j\,p} \Omega^{i\,p}
  - \bar\eta_{j\, p} \Omega^{i\,p} - (\mbox{trace})
  \,.
\end{align}
In these relations, we employ the $3D$ gamma matrices defined in
appendix \ref{sec:conv-spin-basis} and the derivatives, ${D}_\mu$ are
covariant with respect to all $3D$ superconformal transformations
including the emergent $\mathrm{SU}(2)_\mathrm{R}$, as in
\eqref{eq:superco-u1-der} with corresponding gauge fields defined
according to \eqref{eq:new-R-conn}. Note that the terms
$\tilde{\Sigma}^\pm$ have been cancelled by the variation of the phase
factor $\Phi$, just as before. The Weyl weights for the $3D$ fields
are given in table \ref{table:w-weights-matter-3D}. 

%%%%%%%%%%%%%%%%%%%%%%%%%%%%%%%%%%%%%%%%%%%%%%%%%%%%%%%%%%%%%%%%
%%%%%%%%%%%%%%%%%%%%%%%%%%%%%%%%%%%%%%%%%%%%%%%%%%%%%%%%%%%%%%%
%
\begin{table}[t]
\begin{center}
\begin{tabular*}{13.5cm}{@{\extracolsep{\fill}}|c|cccc|cccc|cc| }
\hline
 $4D$ & \multicolumn{4}{c|}{vector multiplet} &
 \multicolumn{4}{c|}{tensor multiplet} &
 \multicolumn{2}{c|}{hypermultiplet} \\
 \hline \hline
 field & $X$ & $W_\mu$  & $\Omega_i$ & $Y^{ij}$& $L^{ij}$ &
 $E_{\mu\nu}$ & $\varphi_i$ & $G$ & $A_i{}^\alpha$ & $\zeta^\alpha$ \\[.5mm] \hline
$w$  & $1$ & $0$ & $\tfrac32$ & $2$ & $2$& $0$ &$\tfrac52$& $3$ & $1$
&$\tfrac32$
\\[.5mm] \hline
$c$  & $-1$ & $0$ & $-\tfrac12$ & $0$ &$0$&$0$&$-\tfrac12$ &$1$ & $0$ &$-\tfrac12$
\\[.5mm] \hline
$\gamma_5$   & && $+$  &   &&&$-$&&  &$-$ \\ \hline
\end{tabular*}
\vskip 2mm
\renewcommand{\baselinestretch}{1}
\parbox[c]{13.5cm}{\caption{\footnotesize Weyl and chiral weights ($w$
    and $c$) and fermion chirality $(\gamma_5)$ of the vector
    multiplet, the tensor multiplet and the hypermultiplet component
    fields in four space-time
    dimensions. \label{table:w-weights-matter-4D} } } 
\end{center}
\end{table}
%%%%%%%%%%%%%%%%%%%%%%%%%%%%%%%%%%%%%%%%%%%%%%%%%%%%%%%%%%%%%%

\paragraph{\it The tensor multiplet:\\} 
The tensor multiplet in four dimensions comprises an $\mathrm{SU}(2)$
triplet of scalars $L_{ij}$, a chiral spinor doublet $\varphi_i$, a
two-form gauge field $E_{MN}$ and an auxiliary complex scalar field,
$G$. The Weyl and chiral weights of these fields are summarized in
table \ref{table:w-weights-matter-4D}. In a superconformal background
the $Q$- and $S$-supersymmetry transformations of the $4D$ tensor
supermultiplet fields take the following form,
\begin{align}
  \label{eq:tensor-tr}
  \delta L_{ij} =& \,2\,\bar\epsilon_{(i}\varphi_{j)} +2
  \,\varepsilon_{ik}\varepsilon_{jl}\,
  \bar\epsilon^{(k}\varphi^{l)}  \,,\nonumber \\
  \delta\varphi^{i} =& \,\Slash{D} L^{ij} \,\epsilon_j +
  \varepsilon^{ij}\,\Slash{\hat E}\,\epsilon_j - G \,\epsilon^i
  + 2 L^{ij}\, \eta_j \,,\nonumber \\
  \delta G =& \,-2 \,  \bar\epsilon_i \Slash{D} \, \varphi^{i} \,
  - \bar\epsilon_i  ( 6 \, L^{ij} \, \chi_j + \tfrac1{4} \,
    \gamma^{AB}  T_{AB jk} \, \varphi_l \,
    \varepsilon^{ij} \varepsilon^{kl}) + 2 \, \bar{\eta}_i\varphi^{i}
    \, ,\nonumber\\
  \delta E_{MN} =& \, \mathrm{i}\bar\epsilon^i\gamma_{MN}
  \varphi^{j} \,\varepsilon_{ij} - \mathrm{i}\bar\epsilon_i\gamma_{MN}
  \varphi_{j} \,\varepsilon^{ij} \,  + \,2 \mathrm{i} \, L_{ij} \,
  \varepsilon^{jk} \, \bar{\epsilon}^i \gamma_{[M} \psi_{N ]k}
  - 2 \mathrm{i}\,  L^{ij} \, \varepsilon_{jk} \, \bar{\epsilon}_i
    \gamma_{[M} \psi_{N]}{}^k \,,\nonumber\\
    \delta\hat E^A=&\, \varepsilon_{ij} \,\bar\epsilon^i\gamma^{AB}
    D_B\varphi^j + \tfrac14
    \bar\epsilon^i\gamma^A\big(6\,\varepsilon_{ij} \chi_k\,L^{jk}
    -\tfrac14 T_{BC\,ij} \gamma^{BC} \varepsilon^{jk}\varphi_k \big)
    %\nonumber\\    &\,
    + \tfrac32 \bar\eta^i \gamma^A \varphi^j\varepsilon_{ij} +
    \mbox{h.c.} \,.
\end{align}
Here, the derivatives $\mathcal{D}_M$ are covariant with respect to
Lorentz transformations, dilatations and R-symmetry
transformations. The vector, $\hat E^M$, denotes the superconformally
covariant field strength
\begin{align}
  \label{eq:cov-qu}
  \hat E^M =&\,\tfrac{1}{2}\mathrm{i}\, e^{-1} \, \varepsilon^{MNPQ}
  \nonumber\\
  &\, \times \Big[\partial_N E_{PQ} - \tfrac{1}{2} \mathrm{i}
  \bar{\psi}{}^i_N \gamma_{PQ} \varphi^j \varepsilon_{ij} +
  \tfrac{1}{2}\mathrm{i} \bar{\psi}_{N\, i} \gamma_{PQ} \varphi_{j}
  \varepsilon^{ij} -\mathrm{i} \, L_{ij} \varepsilon^{jk}
  \bar{\psi}_N{}^i \gamma_P \psi_{Q\, k}\Big] \,,
\end{align}
associated with the tensor field $E_{MN}$; the latter is subject to
tensor gauge transformations parametrized by a vector $\lambda_M$,
\begin{equation}
  \label{eq:4D-gauge-tr}
  \delta E_{MN} = 2\, \partial_{[M}\lambda_{N]} \,.
\end{equation}

The Kaluza-Klein decomposition of the tensor gauge field reads 
\begin{equation}
  \label{eq:KK-tensor}
  E_{MN} =
 \begin{pmatrix} E_{\mu\nu}+  2\,B_{[\mu}\, E_{\nu]}&- E_\mu\\[4mm]
    E_{\nu}& 0 \end{pmatrix}\;,
\end{equation}
where $E_\mu\equiv  E_{\hat 4\mu}$. The tensor gauge transformation
parameter decomposes accordingly into $\lambda_M= (\lambda_\mu,
\lambda)$, where $\lambda\equiv \lambda_{\hat 4}$, so that the $3D$ tensor
and vector fields, $E_{\mu\nu{}}$ and $E_\mu$ transform as
\begin{equation}
  \label{eq:3D-gauge-tr}
  \delta E_{\mu\nu} = 2\, \partial_{[\mu}\lambda_{\nu]} - 2\,B_{[\mu}
  \,\partial_{\nu]} \lambda  \,, \qquad 
  \delta E_\mu= \partial_\mu\lambda \,. 
\end{equation}
Let us now proceed and determine the reduction of the $4D$ tensor
field strength. First we note that $E_{ABC}= E_A{}^M E_B{}^N
E_C{}^P \, \partial_{[M} E_{NP]}$ decomposes as follows,
\begin{align}
  \label{eq:cov-E}
   E_{abc}  =&\,e_a{}^\mu e_b{}^\nu e_c{}^\rho \big( 
   \partial_{[\mu}E_{\nu\rho]}  + {F}(B)_{[\mu\nu} \,E_{\nu]} \big) \,,
 \nonumber \\
  E_{ab 4}=&\, -\tfrac13 \phi\,F(E)_{ab}\,.
\end{align}
where $F(E)_{\mu\nu} = \partial_\mu E_\nu - \partial_\nu E_\mu$. Note
that $E_{abc}$ and $E_{ab4}$ have Weyl weight $w=3$. , and they
correspond to the bosonic part of the field strength \eqref{eq:cov-qu}
written with tangent-space indices. We can now write $E_{abc}$ as a
$3D$ real scalar of weight $w=3$ by defining $E^4\equiv \tfrac12\mathrm{i}
\varepsilon^{abc} \,E_{abc}$. Indeed, a comparison with a two-rank
tensor field in three dimensions shows that it represents only one
degree of freedom (the $3D$ tensor has three degrees of freedom from
which one must subtract two gauge degrees of freedom). Hence, we can
base ourselves exclusively on the scalar $E^4$ and ignore the underlying
tensor field, without loss of generality. Therefore the field
strength \eqref{eq:cov-qu} written with tangent-space indices is
precisely equal to 
\begin{equation}
  \label{eq:scalar-tensor}
  \hat E^4 = \tfrac12 \mathrm{i} \varepsilon^{abc} \,E_{abc} + \cdots \,,
  \qquad 
  \hat E^a=  \tfrac12 \mathrm{i} \phi\,\varepsilon ^{abc} \,\hat F(E)_{bc} +\cdots \,, 
\end{equation}
where the dots denote the fermionic bilinears. Henceforth $\hat E^4$
will be regarded as a supercovariant scalar, as was explained above,
whereas $\hat F(E)_{ab}$ denotes the supercovariant field strength
associated with the $3D$ gauge field $E_\mu$. 

We are now ready to present the result of the reduction to three
dimensions for the supersymmetry transformation rules. To this end,
we find it useful to rescale the scalar triplet by $\phi$ to obtain
a triplet of unit Weyl weight, and introduce the quantities
\begin{align}
\breve\varphi_{i} =&\, \phi^{-1}\,\varphi_{i} + L_{ij}\gamma_4\psi^j\,,
\nonumber\\
 \breve E =&\, \phi^{-1} \hat E^4
 + \tfrac12\, \varepsilon_{ij}{\cal V}^{i}{}_k\,L^{jk}
 - \varepsilon_{ij}\bar\psi^i\varphi^j
 - \varepsilon^{ij}\bar\psi_i\varphi_j
 -\tfrac12\, \phi\,L^{ij}\,\bar\psi_{i}\gamma^4\psi^{k}\varepsilon_{jk}\,,
\nonumber\\
\breve G=&\, \phi^{-1} G - \bar\psi_{i}\gamma_4\varphi^i
         - \tfrac12\, \phi\,L^{ij}\,\bar\psi_{i}\psi_{j}\,.
\end{align}
With these definitions, we can write the reduced result as follows,
\begin{align}
\label{eq:tensor-tr-3D}
  \delta (\phi^{-1}L_{ij}) =&\,
  2\,\bar\epsilon_{(i}\hat\varphi_{j)}
  +2  \,\varepsilon_{ik}\varepsilon_{jl}\,
  \bar\epsilon^{(k}\hat\varphi^{l)} \,, \nonumber\\
  \delta\breve\varphi^{i}=&\,\Slash{D} ( \phi^{-1} L^{ij} ) \,\epsilon_j
  +\tfrac12 \mathrm{i}\,e^{-1}\varepsilon^{\mu\nu\rho}F(E)_{\mu\nu}\,
  \gamma_\rho\, \varepsilon^{ij}\epsilon_j
  + \breve E\, \varepsilon^{ij}\gamma_4 \,\epsilon_j
  - \breve G \,\epsilon^i \,, \nonumber  \\
  &\,
 -\mathrm{i}\,\phi^{-1} L^{ij}\,C\,\gamma_4\epsilon_j
 + \phi^{-1} L^{ij}\, \eta_j
 +\tfrac12\,\tilde\Sigma^-\varepsilon^{ij}\gamma_4\,\breve\varphi_{j}\,,
 \nonumber \\
  \delta E_{\mu} =&\,
  \mathrm{i}\,\varepsilon_{ij}
  \bar\epsilon^i\gamma_{\mu}\gamma_4 \breve\varphi^{j}
 - \,\mathrm{i} \, \phi^{-1} L_{ij} \,
  \varepsilon^{jk} \, \bar{\epsilon}^i \gamma_{4} \psi_{\mu k}
  +\text{h.c.} \,, \nonumber \\
  \delta \breve E =&\, \varepsilon_{ij}\bar{\epsilon}^i\gamma_{4}
  \left(\Slash{D}(\mathcal{A}^0) \hat \varphi^{j}
    - \tfrac12\, \Slash{A}^+\gamma_4\,\varepsilon^{jk}\breve \varphi_{k}
  \right)
  - \mathrm{i}\, C\,\varepsilon_{ij}
  \bar{\epsilon}^i \breve \varphi^{j}
  + \ft12\,\varepsilon_{ij} \phi^{-1}L^{jk}
  \,\bar\epsilon^i\gamma_4 \breve\chi_k \nonumber\\ 
  &\,
  + \ft12 \,\varepsilon_{ij}\,\bar \varphi^i\gamma_4 \eta^j
  -\tfrac12\,\tilde\Sigma^+\,\breve G
  +  {\rm h.c.}\,, \nonumber \\
  \delta \breve G =& \,-2 \,  \bar\epsilon_i\,
  \left(\Slash{D}(\mathcal{A}^0)\breve\varphi^{i}
    - \tfrac12\, \Slash{A}^+\gamma_4\,\varepsilon^{ij}\breve \varphi_{j}
  \right)
  +2\,\mathrm{i} \, C\,\bar\epsilon_i\gamma_4\,\breve \varphi^{i}
  - \phi^{-1}L^{ij} \,\bar\epsilon_i \breve\chi_j \nonumber\\
  & \,+ \bar\eta_i\breve\varphi^{i}\,    + \tilde\Sigma^-\,\breve E \, ,
\end{align}
where we retained all fermionic terms in the variations of $L_{ij}$,
$\breve\varphi$ and $E_\mu$, but considered only the variations linear
in the fermions for $\breve{G}$ and $\breve{E}$.  In these expressions
we used once again the covariant derivative $D_\mu(\mathcal{A}^0)$
that contains the modified $U(1)$ gauge field in
\eqref{eq:def-weyl-connections}.  We have again suppressed the
field-dependent S-supersymmetry and $\mathrm{SU}(2)$ transformations
in the above result.

To write the supersymmetry variations \eqref{eq:tensor-tr-3D}
in three-dimensional form, we employ a definition of fields that
transform covariantly under the the local R-symmetry,
\begin{equation}
\label{eq:tensor-scals}
L^i{}_{j}=\phi^{-1}\, \varepsilon^{ik} L_{kj}\,,
\qquad
 Y^{p}{}_q = \Phi 
 \begin{pmatrix}
   \mathrm{i}\,\breve E\, & -\mathrm{i}\,\breve G   \\[4mm]
   -\mathrm{i}\,\breve{\bar G} & -\mathrm{i}\,\breve E
\end{pmatrix} \Phi^{-1}\,,\qquad 
  \varphi^{i\,p} = \Phi
 \begin{pmatrix} \mathrm{i}\,\breve\varphi^{i}\\[4mm]
  \mathrm{i}\gamma_4\,\varepsilon^{ij}\,\breve\varphi_{j}
       \end{pmatrix}\,.
\end{equation}
The supersymmetry variations resulting upon use of these definitions
are as follows
\begin{align}
\label{eq:susy-tensor-3D}
 \delta L^i{}_{j} =&\,
 2\, \bar{\epsilon}_{j\, p }\, \varphi^{i\,p}
 -\delta^{i}{}_{j}\, \bar{\epsilon}_{k\, p }\, \varphi^{k\,p}
 \,,\nonumber\\
  \delta\varphi^{i\,p} =& \,
  \Slash{D} L^i{}_{j} \,\epsilon^{j\,p}
  - \ft12 \, F(E)_{ab} \gamma^{ab}\epsilon^{i\,p}  %sign changed%
 + Y^p{}_q\, \epsilon^{i\,q}
  -C\, L^i{}_{j}\,\epsilon^{j\,p}
  + L^i{}_{j}\, \eta^{j\,p}
 \,,\nonumber\\
  \delta E_\mu=&\,
\bar\epsilon_{i\,p}\gamma_\mu \varphi^{i\,p}
+ L^i{}_j \,\bar\epsilon_{i\,p} \psi_\mu{}^{j\,p}
 \,,\nonumber\\
 \delta  Y^p{}_q
=&\,
 2\, \bar{\epsilon}_{i\, q}\Slash{D} \varphi^{i\,p}
 + L^{i}{}_j\, \bar{\epsilon}_{j\, q} \chi^{i\,p}
+2\,C\, \bar{\epsilon}_{i\,q} \varphi^{i\,p}
   - \bar\eta_{i\, q} \varphi^{i\,p} - (\mbox{trace})\,.
\end{align}
Once again, the variations proportional to $\tilde{\Sigma}^\pm$
cancel. All the fields and the gamma matrices refer to $3D$; the
covariant derivative, ${D}_\mu$ also includes the gauge fields
associated with the extra local $\mathrm{SU}(2)_\mathrm{R}$ symmetry,
just as in \eqref{eq:superco-u1-der}. The Weyl weights of the component
fields are summarized in table \ref{table:w-weights-matter-3D}. 

Note that the supersymmetry transformations for the tensor multiplet
\eqref{eq:susy-tensor-3D} are very similar to those of the vector
multiplet given in \eqref{eq:susy-vector-3D}. In fact they are related
precisely by the exchange symmetry noted for the Weyl multiplet
in \eqref{eq:su2-interchange}. We will return to this issue later in
this section. 

%%%%%%%%%%%%%%%%%%%%%%%%%%%%%%%%%%%%%%%%%%%%%%%%%%%%%%%%%%%%%%%
%
\begin{table}[t]
\begin{center}
\begin{tabular*}{13.5cm}{@{\extracolsep{\fill}}|c|cccc|cccc|cc| }
\hline
 $3D$ & \multicolumn{4}{c|}{vector multiplet} &
 \multicolumn{4}{c|}{tensor multiplet} &
 \multicolumn{2}{c|}{hypermultiplet} \\
 \hline \hline
 field & $L^p{\!}_q$ & $W_\mu$  & $\Omega^{i\,p}$ & $Y^i{\!}_j$& $L^i{\!}_j$ &
 $E_{\mu}$ & $\varphi^{i\,p}$ & $Y^p{\!}_q$ & $A_i{}^\alpha$ &
 $\zeta^\alpha$ \\[.5mm] \hline 
$w$  & $1$ & $0$ & $\tfrac32$ & $2$ & $1$& $0$ &$\tfrac32$& $2$ &
$\tfrac12$ &$1$
\\ [.5mm] \hline
\end{tabular*}
\vskip 2mm
\renewcommand{\baselinestretch}{1}
\parbox[c]{13.5cm}{\caption{\footnotesize Matter multiplet fields with
    corresponding Weyl weights of the vector multiplet, the tensor
    multiplet and the hypermultiplet fields in three space-time
    dimensions. \label{table:w-weights-matter-3D} } }
\end{center}
\end{table}
%%%%%%%%%%%%%%%%%%%%%%%%%%%%%%%%%%%%%%%%%%%%%%%%%%%%%%%%%%%%%%

\paragraph{\it The hypermultiplet:\\} 
The hypermultiplets are not realized off-shell, but they can be
coupled to conformal supergravity provided the target-space geometry
is restricted to a hyperk\"ahler cone. For rigid supersymmetry it is
sufficient that the target space is hyperk\"ahler, but in order for
the action to be superconformally invariant the target space must also
admit a homothetic conformal Killing vector. This in turn implies
that the homothetic Killing vector can locally be expressed in terms
of the so-called hyperk\"ahler potential which also defines the
target-space metric \cite{deWit:1999fp,deWit:2001dj} . Assuming that
these conditions are met, let us now introduce the local Q- and
S-supersymmetry transformations of the hypermultiplet fields, which
only close modulo the equations of motion of the fermion fields,
\begin{align}
  \label{eq:4dsusy}
  \delta A_i{}^\alpha+ \delta\phi^B
  \Gamma_B{}^\alpha{}_\beta A_i{}^\beta = &\,
  2\,\bar\epsilon_i\zeta^\alpha +2\,\varepsilon_{ij}
  G^{\alpha\bar\beta}\Omega_{\bar\beta\bar\gamma}\,\bar\epsilon^j
  \zeta^{\bar\gamma}
  \,,\nonumber\\
  \delta\zeta^\alpha +\delta\phi^A\,
  \Gamma_{A}{}^{\!\alpha}{}_{\!\beta}\, \zeta^\beta =&\,
  \Slash{D} A_i{}^\alpha\,\epsilon^i   +A_i{}^\alpha \,\eta^i\,,
  \nonumber\\
  \delta\zeta^{\bar \alpha}+\delta\phi^A\,
  \bar\Gamma_{A}{}^{\!\bar\alpha}{}_{\!\bar \beta} \,\zeta^{\bar\beta}
  =&\, \Slash{D}A^{i\bar \alpha}\, \epsilon_i  +
    A^{i\bar\alpha} \,\eta_i \,.
\end{align}
where we employ the local sections of an
$\mathrm{Sp}(r)\times\mathrm{Sp}(1)$ bundle, denoted by
$A_i{}^\alpha$, for $\alpha= 1,2,\ldots,2r$. The Weyl and chiral
weights of these quantities are shown in table
\ref{table:w-weights-matter-4D}. We also note the existence of a
covariantly constant hermitian tensor $G^{\alpha\bar\beta}$ (which is
used in raising and lowering indices) and of a covariantly constant
skew-symmetric tensor $\Omega_{\alpha\beta}$ (and its complex
conjugate $\bar\Omega^{\bar\alpha\bar\beta}$ satisfying
$\Omega_{\bar\alpha\bar\gamma}\bar\Omega^{\bar\beta\bar\gamma}=
-\delta_{\bar\alpha}{}^{\bar\beta}$).  Covariant derivatives contain
the $\mathrm{Sp}(r)$ connection $\Gamma_A{}^\alpha{}_\beta$,
associated with rotations of the fermions. The sections $A_i{}^\alpha$
are pseudo-real, i.e. they are subject to the constraint,
$\varepsilon^{ij} \bar\Omega^{\bar\alpha\bar\beta}
G_{\bar\beta\gamma}A_i{}^\gamma = A^j{}^{\bar\beta} \equiv
(A_j{}^\beta)^\ast$. For our purpose the geometry of the hyperk\"ahler
cone is not relevant and we assume for simplicity that the cone is
flat, so that the target-space connections and curvatures will
vanish. The sections can then be identified with the fields, and the
tensors $G^{\alpha\bar\beta}$ and $\Omega_{\alpha\beta}$ are constant
\cite{deWit:1980lyi,deWit:1999fp}. The extension to non-trivial
hyperk\"ahler cone geometries is straightforward.

The Weyl and chiral weights of the sections and the fermion fields
are listed in table \ref{table:w-weights-matter-4D}.  The $3D$ 
hypermultiplet fields will be rescaled, however, so that they have the
appropriate canonical dimension in $3D$. This motivates the following
field redefinitions,
\begin{align}
 \breve{A}_i{}^\alpha = &\, \phi^{-1/2} A_i{}^\alpha\,,
\nonumber\\
 \breve{\zeta}^\alpha = &\, \phi^{-1/2} \zeta^\alpha
  +\tfrac{1}2 \, \phi^{1/2} A_j{}^\alpha \gamma^4 \psi^j \,,
\nonumber\\
 \breve{\zeta}^{\bar\alpha}=&\, \phi^{-1/2}\, \,\zeta^{\bar\alpha}
   +\tfrac{1}2 \, \phi^{1/2} A^i{}^{\bar\alpha} \gamma^4 \psi_i\,.
\end{align}
Upon dimensional reduction the supersymmetry transformation rules
\eqref{eq:4dsusy} can be cast in the form, 
\begin{align}
  \delta \breve{A}_i{}^\alpha =&\,
  2\,\bar\epsilon_i \breve{\zeta}^\alpha
  +2\,\varepsilon_{ij}
  G^{\alpha\bar\beta}\Omega_{\bar\beta\bar\gamma}\,
  \bar\epsilon^j \breve{\zeta}^{\bar\gamma} \,,   \nonumber\\
%%%%%%%%%
  \delta \breve{\zeta}^\alpha =&\,
  \Slash{D}\breve{A}_i{}^\alpha\,\epsilon^i
  +\tfrac12\,\mathrm{i}\,C\,\breve{A}_i{}^\alpha
   \gamma_4 \epsilon^i
 +\tfrac12\,\tilde\Sigma^-
 G^{\alpha\bar\beta}\Omega_{\bar\beta\bar\gamma}\,
 \gamma_4\, \breve{\zeta}^{\bar\gamma}
 +\tfrac12\,\breve{A}_i{}^\alpha \eta^i \,,
\end{align}
where all fermionic terms were taken into account and where $D_\mu$
denotes the supercovariant derivative in three dimensions. Again we
suppressed the field-dependent $\mathrm{SU}(2)$ and S-supersymmetry
transformations. Subsequently we further redefine the fermion such as
to incorporate their consistent transformation behaviour under local R-symmetry,
\begin{equation}
  \zeta^{\alpha\,p} = \Phi  \begin{pmatrix}
    -\mathrm{i}\,G^{\alpha\bar\beta}\Omega_{\bar\beta\bar\gamma}\,
    \breve{\zeta}^{\bar\gamma}
    \\[4mm]
    \mathrm{i}\gamma_4\,\breve{\zeta}^\alpha
       \end{pmatrix}\,.
\end{equation}
The Weyl weights of the $3D$ quantities $A_i{}^\alpha$ and
$\zeta^\alpha$ have been shown in table
\ref{table:w-weights-matter-3D}. 
With these redefinitions we obtain the following $3D$ supersymmetry
transformations of the $3D$ fields, 
\begin{align}
  \label{eq:hyper-3D}
  \delta {A}_i{}^\alpha =&\,
  2\,\bar\epsilon_{i\,p} \zeta^{\alpha\,p}\,,   \nonumber\\
%%%%%%%%%
  \delta \zeta^{\alpha\,p} =&\,
  \Slash{D} {A}_i{}^\alpha\,\epsilon^{i\,p}
  -\tfrac12\,C \,{A}_i{}^\alpha \epsilon^{i\,p}
  +\tfrac12\,{A}_i{}^\alpha \eta^{i\,p} \,,
\end{align}
expressed in terms of $3D$ gamma matrices and supercovariant
derivatives. The terms proportional to $\tilde\Sigma^\pm$ have again
disappeared as they should. 

As we alluded to at the beginning of the section, there is an
alternative hypermultiplet that transforms under the other
$\mathrm{SU}(2)$ factor of the R-symmetry, which cannot emerge under
dimensional reduction. For future reference we give its transformation
rules below, relying on the reflection \eqref{eq:su2-interchange}, 
\begin{align}
  \label{eq:hyper-3D-new}
  \delta \tilde{A}_p{}^\alpha =&\,
  2\,\bar\epsilon_{i\,p} \tilde\zeta^{\alpha\,i}\,,   \nonumber\\
%%%%%%%%%
  \delta \tilde\zeta^{\alpha\,i} =&\,
  \Slash{D} \tilde{A}_p{}^\alpha\,\epsilon^{i\,p}
  +\tfrac12\,C \,\tilde{A}_p{}^\alpha \epsilon^{i\,p}
  +\tfrac12\,\tilde{A}_p{}^\alpha \eta^{i\,p} \,.
\end{align}

%%%%%%%%%%%%%%%%%%%%%%%%%%%%%%%%%%%%%%%%%%%%%%%%%%%%%%%%%%%%%%%%

%%%%%%%%%%%%%%%%%%%%%%%%%%%%%%%%%%%%%%%%%%%%%%%%%%%%%%%%%%%%%%
\section{Four and three-dimensional fields and invariant Lagrangians}
\label{sec:five-four-dimens-fields-lagr}
\setcounter{equation}{0}
%%%%%%%%%%%%%%%%%%%%%%%%%%%%%%%%%%%%%%%%%%%%%%%%%%%%%%%%%%%%%%%%
In this section we express the $4D$ fields in terms of the $3D$ ones;
subsequently we convert the known supersymmetric $4D$ Lagrangians by
direct substitution in terms of the supersymmetric $3D$
Lagrangians. The section is divided into three subsections. In the
first one we express the $4D$ bosonic fields in terms of the $3D$ ones
that were identified in the previous sections based on the off-shell
supersymmetry transformations.  The fermionic fields are ignored, as
the supersymmetry transformations are fully known in both $3D$ and
$4D$. In the second subsection we consider three $4D$ supersymmetric
Lagrangians quadratic in derivatives and derive the corresponding
expressions for the bosonic terms belonging to the reduced $3D$
Lagrangians. Because the reduction procedure is fully off-shell (with
the exception of the hypermultiplets that are not defined as genuine
off-shell multiplets) there is no need for additional adjustments. In
the third subsection we discuss some features of the c-map and compare
to results in the literature.

%%%%%%%%%%%%%%%%%%%%%%%%%%%%%%%%%%%%%%%%%%%%%%%%%%%%%%%
\subsection{The supercovariant dictionary: expressing
  \texorpdfstring{$\boldsymbol{4D}$}{4D} fields 
  in terms of \texorpdfstring{$\boldsymbol{3D}$}{3D} fields}
\label{sec:4d-fields-terms}
%%%%%%%%%%%%%%%%%%%%%%%%%%%%%%%%%%%%%%%%%%%%%%%%%%%%%%%
Some of the $4D$ and $3D$ fields are identical, except that the latter
will no longer depend on the fourth coordinate. For instance, the
vierbein fields, which contain the three-dimensional fields
$e_\mu{}^a$ as a submatrix, belong to this class. For other fields the
relation is more involved. In particular for the Kaluza-Klein scalar
$\phi$, matters are more subtle, as this field is contained in a
$2\times2$ anti-hermitian traceless matrix $L^{p}{}_q{\!}^0$, defined in
\eqref{eq:scalar-KKphoton} by absorbing a phase factor $\Phi$,
introduced in subsection \ref{sec:g-compensator-KKvector} to realize
the new local $\mathrm{SU}(2)$ factor of the $3D$ R-symmetry group.
As a result the expressions for the $4D$ fields are invariant under
the new local $\mathrm{SU}(2)$ R-symmetry that emerges in the
reduction, up to a term that takes the form of a $4D$ $\mathrm{U}(1)$
R-symmetry.

The $3D$ vector and tensor matter supermultiplets contain scalar fields that were
conveniently written in terms of anti-hermitian traceless $2\times2$
matrices. For instance, the Kaluza-Klein multiplet contains the scalar
$L^{p}{}_q{\!}^0$ with Weyl weight $w=1$, as well as a similar field
$Y^{i}{}_j{\!}^0$ of weight $w=2$. The vector multiplets corresponding to
the matter $4D$ vector supermultiplets also contain these fields,
$L^{p}{}_q$ and $Y^{i}{}_j$, which depend linearly on the components
of the $4D$ vector multiplet. The same situation arises for the tensor
multiplet, but with the indices $p,q$ and $i,j$ interchanged. Hence
these multiplets contain fields $L^{i}{}_j$ and $Y^{p}{}_q$,
with Weyl weights $w=1$ and $w=2$, respectively. Obviously, in the
context of Lagrangians that are at most quadratic in derivatives, the
fields $L$ correspond to the physical scalars and the fields $Y$ to
the auxiliary fields.

Although the matrix form of the scalar is convenient when considering
the supersymmetry transformations, it is not always easy to write the
results in the form of matrix products and traces thereof. Therefore
we will use a uniform decomposition in terms of the three independent
components transforming as a vector under the appropriate
$\mathrm{SU}(2)$ R-symmetry group. Hence for the vector multiplet we
have
\begin{equation}
  \label{eq:L-decomp}
  L^{p}{}_q(x,\upsilon,\bar\upsilon) =
  \begin{pmatrix}
    -\tfrac12\,\mathrm{i}\,x    &   {\upsilon}  \\[.6ex]
    -\bar\upsilon   &    \tfrac12\,\mathrm{i}\,x
  \end{pmatrix}
  \,,\qquad
%%%%%%%%%%%%%%%%%
    Y^{i}{}_j (y,w,\bar w) =  
   \begin{pmatrix}
     -\tfrac12\,\mathrm{i}\,y  &   {w}   \\[.6ex]
     -\bar{w}   &    \tfrac12\,\mathrm{i}\,y 
   \end{pmatrix}
 \,, 
\end{equation}
and for the tensor multiplet we have corresponding definitions for
$L^{i}{}_j(x,\upsilon,\bar\upsilon)$ and $Y^{p}{}_q (y,w,\bar
w)$. Obviously, in Lagrangians with both vector and tensor multiplets,
these multiplets should in principle be labelled by different indices.
We recall the components of $\mathcal{D}_\mu L^p{}_q$ and
$\mathcal{D}_\mu L^i{}_j$ for convenience,
\begin{align}
  \label{eq:DL-pq}
    \mathcal{D}_\mu L^p{}_q=&\,  (\partial_\mu -b_\mu) L^p{}_q +\tfrac12 \big[
  \mathcal{A}_\mu\,,L\big] ^p{}_q \,, \nonumber\\
  \mathcal{D}_\mu L^i{}_j=&\,  (\partial_\mu -b_\mu) L^i{}_j +\tfrac12 \big[
  \mathcal{V}_\mu\,,L\big]^i{}_j \,,
\end{align}
which is in agreement with \eqref{eq:D-su2-L}.

Let us now consider the $4D$ Weyl multiplet, whose fields can be
expressed in terms of the fields of the $3D$ Weyl multiplet and the
Kaluza-Klein vector multiplet (restricting ourselves to the bosonic
fields and ignoring that the redefined bosonic fields may also contain
fermionic bilinears),
\begin{align}
  \label{eq:D4to3Weyl}
  e_M{}^A=&\, \left\{ \begin{array}{rl} 
                      e_\mu{}^a=&\!\! e_\mu{}^a\,,\\
                      e_\mu{}^4=& \!\! W_\mu{}^0 \, (L^{0})^{-1}  \,,\\
                      e_{\hat 4}{}^a=&\!\! 0\,,\\
                      e_{\hat 4}{}^4=&\!\! (L^0)^{-1} \,, 
                    \end{array} \right. \nonumber\\[.8ex]
                  L^0 =&\,\sqrt{\vert\upsilon^0\vert^2
                    +\tfrac14(x^0)^2} = \sqrt{-\tfrac12
                    L^p{}_q{\!}^0\,L^q{}_p{\!}^0}  \,, \nonumber\\[.6ex]
                  B_\mu =&\, W_\mu{}^0 \,,\nonumber \\[.6ex] 
                  \mathcal{V}_{M}{}^i{}_j\big\vert_{4D}  =&\, \left\{ \begin{array}{rl}
                      \mathcal{V}_{\mu}{}^i{}_j =&\!\! \mathcal{V}_{\mu}{}^i{}_j +
                      W_\mu{}^0 \,Y^{i}{}_{j}{\!}^0\, (L^0)^{-2}  \,, \\[6mm]
                      \mathcal{V}_{\hat 4}{}^i{}_j  =&\!\!
                      Y^{i}{}_{j}{\!}^0 \, (L^0)^{-2}    \,,
         \end{array} \right.   \nonumber\\[.6ex]
       A_M =&\, \left\{ \begin{array}{rl}
          A_\mu  =&\!\! \mathcal{A}_\mu{}^0 
          +\displaystyle{\frac1{ L^0}}\,\big[\tfrac14 F(W^0)_\mu
          +W_\mu{}^0 \,C\big]  \,,\\[6mm]
           A_{\hat 4} =&\!\! (L^0)^{-1}\, C\,, 
           \end{array}\right.
         \nonumber\\[.6ex] 
          T_{AB}{}^{ij} =&\, \left\{ \begin{array}{rl}
           T_{ab}{}^{ij} =&\!\! \tfrac12\mathrm{i} \,(L^0)^{-2}
           \,\varepsilon^{ij} \,\varepsilon_{abc} \, \bigg[  (\bar\upsilon^0
    \stackrel{\leftrightarrow}{\mathcal{D}}{\!}^c  x^0 ) -
           \displaystyle{ \frac{\bar\upsilon^0 } { L^0+\tfrac12 x^0 }}
           \,(\upsilon^0 \stackrel{\leftrightarrow}{\mathcal{D}}{\!}^c
           \bar\upsilon^0) \bigg] \,,  \\[4mm] 
           T_{a4}{}^{ij}  =&\!\! \tfrac12\mathrm{i}\,(L^0)^{-2} \, 
           \varepsilon^{ij}\, \bigg[  (\bar\upsilon^0
           \stackrel{\leftrightarrow}{\mathcal{D}}{\!}_a  x^0 ) -
           \displaystyle{ \frac{\bar\upsilon^0  } {L^0+\tfrac12
               x^0}}\,
           (\upsilon^0 \stackrel{\leftrightarrow}{\mathcal{D}}{\!}_a
           \bar\upsilon^0)   \bigg] \,, 
           \end{array}\right.
         \nonumber\\[.6ex]
         D\big\vert_{4D}=&\, \tfrac12 D -\tfrac1{12} R -\tfrac13
         (L^0)^{-1} \mathcal{D}^a \mathcal{D}_a L^0 + \tfrac16
         (L^0)^{-2} \big(\mathcal{D}_a L^0\big)^2 \nonumber\\
         &\, -\tfrac1{12} (L^0)^{-2} \,F(W^0)_{ab}{}^2 -\tfrac12
         C^2 \, -\tfrac14 (L^0)^{-2}\, Y^{i}{}_j{\!}^0 \,Y^{j}{}_i{\!}^0 \,,
\end{align}
where the covariant derivatives contain the connections
$\mathcal{A}_\mu{}^p{}_q$ associated with the second $3D$
$\mathrm{SU}(2)$ R-symmetry group and $R$ denotes the
three-dimensional Ricci scalar. As explained in section
\ref{sec:off-shell-dim-red-Weyl} the fields
$L^{p}{}_q{}^0(x,\upsilon,\bar\upsilon)$, $W_\mu{}^{0}$ and
$Y^{i}{}_j{}^0$ denote the bosonic fields of the Kaluza-Klein
supermultiplet. The remaining fields belong to the $3D$ Weyl multiplet
and were discussed in section \ref{sec:3d-weyl-multiplet}. The
connection $\mathcal{A}_\mu{}^0$ was defined in the second equation of
\eqref{eq:T-A}. Its explicit form is not relevant, but it is important
to realize that under local $3D$ $\mathrm{SU}(2)$ it transforms as the
connection of the $4D$ $\mathrm{U}(1)$ R-symmetry.
%%%%%%%%%%%%%
\vskip 3mm
%%%%%%%%%%%%%
Subsequently we consider the $4D$ vector multiplet, which upon
reduction leads to a $3D$ vector multiplet. The bosonic fields of the
latter are denoted by $L^p{}_q(x,\upsilon,\bar\upsilon)$, $W_\mu$ and
$Y^i{}_j$, 
\begin{align}
  \label{eq:D4to3vector}
         X =&\, -\tfrac14 \mathrm{i}  \Big[\frac{x\,\bar\upsilon^0
             -\bar\upsilon\,x^0 }{L^0} -
             \frac{\bar\upsilon\, \upsilon^0 
        -\upsilon\,\bar\upsilon^0} {L^0( L^0+\tfrac12 x^0)} \,
      \bar\upsilon^0\Big] \,,\nonumber \\[2mm] 
      W_M =&\, \left\{ \begin{array}{rl} W_\mu =&\!\! W_\mu +  
          W_\mu{}^0 \; \displaystyle \frac{L^p{}_q\, L^q{}_p{\!}^0} { 2\, (L^0)^2}
            \,,  \\[4mm] 
          W_{\hat 4} =&\!\! \displaystyle \frac{L^p{}_q\, L^q{}_p{\!}^0} { 2\, (L^0)^2}
          \,,
           \end{array}\right.
         \nonumber\\[2mm]
         Y_{ij} =&\, - \varepsilon_{ik} \Big[Y^k{}_j +
          Y^{k}{}_j{\!}^0    \; \displaystyle \frac{L^p{}_q\, L^q{}_p{\!}^0} { 2\, (L^0)^2}
         \Big]\,,
\end{align}
where we note that 
\begin{equation}
  \label{eq:innder-prod-LL}
  L^p{}_q\, L^q{}_p{\!}^0 = -\tfrac12 x\,x^0 - \upsilon\,\bar\upsilon^0
  -\bar\upsilon \,\upsilon^0\,. 
\end{equation}
For the $3D$ vector multiplet the following transformations will define
an invariance,
\begin{equation}
  \label{eq:L-L)-invariance}
  \begin{array}{rcl}
   \delta L^p{}_q &\!\!= \!\!&  \alpha\, L^p{}_q{}^0\,,\\
  \delta W_\mu &\!\!=  \!\!& \alpha\, W_\mu{\!}^0\,,\\
   \delta Y^i{}_j &\!\!=  \!\!& \alpha\, Y^i{}_j{}^0\,,
\end{array} 
\qquad\quad 
  \begin{array}{rcl}
    \delta L^p{}_q{}^0 &\!\!=\!\!&   0 \,,\\
    \delta W_\mu{}^0 &\!\!=  \!\!&   0\,,\\
    \delta Y^i{}_j{}^0  &\!\!= \!\!& 0\,,
\end{array} 
\end{equation}
where $\alpha$ is constant parameter, because the $4D$
vector components remain invariant under \eqref{eq:L-L)-invariance},
with the exception of $W_{\hat4}$. The latter is shifted by a constant
which represents a remnant of the full $4D$ gauge
transformations. This shows that \eqref{eq:L-L)-invariance} defines an
invariance for {\it any} $4D$ locally supersymmetric Lagrangian that
involves vector multiplets upon dimensional reduction. In principle
there are additional invariances, as discussed in \cite{deWit:1992wf}, but
those are not immediately relevant for what follows. The tensor multiplet and the
hypermultiplet do not give rise to symmetries such as
\eqref{eq:L-L)-invariance}.
%%%%%%%%%%%
\vskip 3mm
%%%%%%%%%%%
The $4D$ tensor multiplet reduces to the $3D$ tensor multiplet. The
bosonic fields of the latter are $L^i{}_j$, $E_\mu$ and $Y^p{}_q(y,w,\bar{w})$,
\begin{align}
  \label{eq:D4to3tensor}
     L_{ij} =&\, -\varepsilon_{ik} \,L^k{}_j \, L^0\,,  \nonumber\\[.4ex]
     E^A =&\, \left\{ \begin{array}{rl} E^a
         =&\!\! \tfrac12\mathrm{i}\,L^0\, \varepsilon^{abc} \,F(E)_{bc}  \,,\\[4mm]
         E^{4} =&\!\!  \tfrac12  Y^p{}_q\, L^q{}_p{\!}^0   +\tfrac12 L^i{}_j
          \,Y^{j}{}_i{\!}^0 \,, 
           \end{array}\right.
         \nonumber\\[.4ex]
         E_{\hat 4\mu} =&\, E_\mu\,, \nonumber\\[.4ex] 
         G= &\,  \tfrac12 \mathrm{i} \Big[-y\,\upsilon^0
             +{w}\,x^0 -
             \frac{\bar w\,\upsilon^0 
        -{w}\,\bar\upsilon^0} { L^0+\tfrac12 x^0} \,
      \upsilon^0\Big] \,,
\end{align}
We note an alternative expression for the scalar $X$ of the $4D$
vector multiplet and a corresponding one for the auxiliary scalar
$\bar G$ of the tensor multiplet,
\begin{align}
  \label{eq:X(L)-1}
  X=&\,\tfrac12\mathrm{i}\, \bar\upsilon -\tfrac14\mathrm{i}
  \bigg[x -\frac{L^p{}_q{}\,L^q{}_p{}^0}{L^0} \bigg]
  \frac{\bar\upsilon^0}{L^0+\tfrac12 x^0} \,,\nonumber\\[2mm]
    \bar G=&\, 2\, L^0\left[\tfrac12 \mathrm{i}\, \bar w -\tfrac14\mathrm{i}
  \bigg[y -\frac{Y^p{}_q\,L^q{}_p{}^0}{L^0} \bigg]
  \frac{\bar\upsilon^0}{L^0+\tfrac12 x^0} \right] \,,
\end{align}
which will turn out to be useful later on. 

%%%%%%%%%%%
\vskip 3mm
%%%%%%%%%%%
Finally we consider the $4D$ hypermultiplet which reduces to a $3D$
hypermultiplet, where we have only a single bosonic quantity
represented by the local sections $A_i{}^\alpha$. Upon the reduction
these sections are redefined according to 
\begin{equation}
  \label{eq:sections-hm}
  A_i{}^\alpha \vert_{4D} =  (L^0)^{1/2}  A_i{}^\alpha \,. 
\end{equation}
%%%%%%%%%%%
\vskip 3mm
%%%%%%%%%%%
This completes the dictionary between the $4D$ and $3D$ fields. The
reader may now verify explicitly that under an $\mathrm{SU}(2)$
R-symmetry transformation of the $3D$ fields, the $4D$ fields remain
invariant up to a $4D$ $\mathrm{U}(1)$ R-symmetry transformation. This
is guaranteed by the relations discussed in subsection
\ref{sec:g-compensator-KKvector} and more in particular by the equations
\eqref{eq:su2-vs-su2}-\eqref{eq:u1-on-A}.

%%%%%%%%%%%%%%%%%%%%%%%%%%%%%%%%%%%%%%%%
\subsection{Lagrangians quadratic in derivatives}
\label{sec:lagrangians}
%%%%%%%%%%%%%%%%%%%%%%%%%%%%%%%%%%%%%%%%
We now turn to the $4D$ Lagrangians quadratic in space-time
derivatives and reduce them to three dimensions in terms of the $3D$
fields that we have derived. We will restrict ourselves to the bosonic
expressions, because supersymmetry is ensured in the off-shell
reduction. We start with the hypermultiplet Lagrangian, because that
is the simplest one. Subsequently we will discuss the tensor multiplet
Lagrangian and finally the vector Lagrangian.
%%%%%%%%%%%%%%%%%%%%%%%%%%%%%%%%%%%%%%%%%%%%%%
\subsubsection{The hypermultiplet Lagrangian} 
\label{subsubsection:hyper-multiplet-lagr}
%%%%%%%%%%%%%%%%%%%%%%%%%%%%%%%%%%%%%%%%%%%%%%
The $4D$ bosonic Lagrangian for hypermultiplets reads \cite{deWit:1999fp},
\begin{equation}
  \label{eq:4D-hyper-lagr}
  \mathcal{L}_\mathrm{hyper}\big\vert_{4D} = -\tfrac12 E\,
  \Omega_{\alpha\beta}\, \varepsilon^{ij}\Big[
  {\cal D}_M A_i{}^\alpha\, {\cal D}^M
  A_j{}^{\beta}- A_i{}^\alpha A_j{}^\beta \big[\tfrac{1}{6} R
  + \tfrac12\, D \big] \Big] \,.
\end{equation}
Upon reduction to three dimensions, the first term becomes
\begin{align}
  \label{eq:DADA-terms}
  &-\tfrac12 E\,\Omega_{\alpha\beta}\, \varepsilon^{ij} \, {\cal D}_M
  A_i{}^\alpha\, {\cal D}^M A_j{}^{\beta} =  -\tfrac12
  e\,\Omega_{\alpha\beta}\, \varepsilon^{ij} \nonumber\\
  &\qquad \times \Big[ {\cal D}_\mu
  A_i{}^\alpha\, {\cal D}^\mu A_j{}^{\beta}  
  +  A_i{}^\alpha A_j{}^\beta
  \big[-\tfrac12 \phi^{-1} \mathcal{D}^\mu\mathcal{D}_\mu \phi
  +\tfrac34\phi^{-2} (\mathcal{D}_\mu\phi)^2
  -\tfrac18 \phi^{-2}\, Y^{i}{}_j{\!}^0  \,Y^{j}{}_i{\!}^0 \,\big]\Big] \,,\nonumber
\end{align}
where we suppressed a total derivative. Note that the covariant
derivatives $\mathcal{D}_\mu A_i{}^\alpha$ on the right-hand side
contain the $3D$ $\mathrm{SU}(2)$ gauge fields
$\mathcal{V}_\mu{}^i{}_j$. Next we turn to the second term in
\eqref{eq:4D-hyper-lagr}. Making use of \eqref{eq:contracted-R}, which
relates the $4D$ and $3D$ Ricci scalars, and of \eqref{eq:D4to3Weyl},
which gives the relation between the $4D$ and $3D$ $D$-fields, the two
terms readily combine into the $3D$ Lagrangian,
\begin{equation}
  \label{eq:3D-hyper-L}
    e^{-1}\mathcal{L}_\mathrm{hyper}\big\vert_{3D} =
  -\tfrac12\, \Omega_{\alpha\beta}\, \varepsilon^{ij} \Big[
  {\cal D}_\mu A_i{}^\alpha\, {\cal D}^\mu A_j{}^{\beta}
   - \tfrac14\,A_i{}^\alpha A_j{}^\beta \big( \tfrac12\,R + D - C^2  \big)
   \Big] \,,
\end{equation}
which agrees with the expression given in \cite{Bergshoeff:2010ui}.
Observe that all the components of the Kaluza-Klein vector multiplet
decouple from the hypermultiplet Lagrangian, so that the well-known
property that vector multiplets and hypermultiplets have no direct
interaction in the ungauged case, is preserved under the reduction. We
will return to this feature in due course. 

Of course, there exists a second $3D$ hypermultiplet Lagrangian, which is
obtained from applying the reflection symmetry
\eqref{eq:su2-interchange}. We have already given its transformation
rules in \eqref{eq:hyper-3D-new}. The corresponding Lagrangian takes the form 
\begin{equation}
  \label{eq:3D-hyper-L-reflected}
    e^{-1}\tilde{\mathcal{L}}_\mathrm{hyper}\big\vert_{3D} =
  -\tfrac12\, \tilde\Omega_{\alpha\beta}\, \varepsilon^{pq} \Big[
  {\cal D}_\mu \tilde A_p{}^\alpha\, {\cal D}^\mu \tilde A_q{}^{\beta}
   - \tfrac14\,\tilde A_p{}^\alpha \tilde A_q{}^\beta \big( \tfrac12\,R - D - C^2\big)
   \Big] \,.
\end{equation}
Such a Lagrangian can only be obtained from $4D$ upon reducing a
vector multiplet and applying $3D$ vector-scalar duality. However, as
we shall see in subsection \ref{subsubsection:vector-multiplet-lagr},
these hypermultiplet Lagrangians will belong to a restricted class.
  
%%%%%%%%%%%%%%%%%%%%%%%%%%%%%%%%%%%%%%%%%
\subsubsection{The tensor multiplet Lagrangian}
\label{subsubsection:tensor-multiplet-lagr}
%%%%%%%%%%%%%%%%%%%%%%%%%%%%%%%%%%%%%%%%%
Here we consider the tensor multiplet Lagrangian in four
space-time dimensions, which reads as (we follow the notation of
\cite{deWit:2006gn}),
\begin{align}
  \label{eq:4D-lagrangian-tensor}
  \mathcal{L}_{\mathrm{tensor}}\big\vert_{4D} = &\, - \tfrac{1}{2} E\,
  F (L)_{IJ} \,\mathcal{D}_M L_{ij}{}^{I} \, \mathcal{D}^M L^{ijJ} +
  F(L)_{IJ} \, L_{ij}{}^I \, L^{ijJ} \, \big(\tfrac{1}{3} R + D \big)
  \nonumber\\
  &\,+E\, F(L)_{IJ} \, \big( E_{M}{}^I \, E^{M\, J} -E^{M\,I}\,
  \mathcal{V}_M{}^i{}_j \,L_{ik}{}^I\,\varepsilon^{jk} + G^I \bar{G}^J
  \big)
  \nonumber\\
  &\, + \tfrac12 \mathrm{i} \varepsilon^{M N P Q} \, F (L)_{IJK}{}^{ij}
  \,E_{MN}{}^I \, \, \partial_P L_{ik}{}^J\, \partial_Q
  L_{jl}{}^K\,\varepsilon^{kl} \, ,
\end{align}
where the tensor multiplets have been labelled with indices
$I,J,\ldots$. Here the functions $F_{IJ}(L)$ depend on the tensor
multiplet scalars $L_{ij}{}^I$ and are invariant under the
$\mathrm{SU}(2)$ R-symmetry group and homogeneous of degree $-1$. Furthermore $F_{IJK}{}^{ij}$
denotes the derivative of $F_{IJ}$ with respect to $L_{ij}{}^K$. The
$E^{M\,I}$ are the bosonic field strengths associated with the tensor
fields $E_{MN}{}^I$, which follow from \eqref{eq:cov-qu}. 

For any $4D$ rigidly or locally supersymmetric tensor multiplet
Lagrangian, the functions $F_{IJ}(L)$ must satisfy the following
equations \cite{Lindstrom:1983rt,deWit:2006gn},  
\begin{equation}
  \label{eq:properties-FIJ}
  F_{IJK}{}^{ij} = F_{(IJK)}{}^{ij} \,,\qquad F_{IJKL}{}^{i[jk]l}
  =0\,. 
\end{equation}
These conditions suffice to prove that there must exist a function
$F(x,\upsilon,\bar\upsilon)$ such that 
\begin{equation}
  \label{eq:F-second-derivatives}
  F_{IJ} = \frac{\partial^2 F(x,\upsilon,\bar\upsilon)}{\partial x^I\,\partial x^J} =
  -\frac{\partial^2 F(x,\upsilon,\bar\upsilon)}{\partial\upsilon^I\,\partial
    \bar\upsilon^J}\,, \qquad  \frac{\partial^2 F(x,\upsilon,\bar\upsilon)}{\partial
    x^I\,\partial\upsilon^J} = \frac{\partial^2 F(x,\upsilon,\bar\upsilon)}{\partial
    x^J\,\partial\upsilon^I} \,,
\end{equation}
where we have used $L^{21\,I}= \tfrac12 \mathrm{i}\,x^I$ and
$L^{11\,I}= \upsilon^I$, which is consistent with earlier definitions.
For superconformally invariant Lagrangians (as well as all locally
supersymmetric Lagrangians) the function $F(x,\upsilon,\bar\upsilon)$
can be chosen to be homogeneous of first degree and invariant under
phase transformations of the components $\upsilon^I$ and
$\bar\upsilon^I$ \cite{deWit:2001dj}, so that
\begin{align}
  \label{eq:F-scale-phase-inv}
  x^I\,\frac{\partial F(x,\upsilon,\bar\upsilon)}{\partial x^I}
  +\upsilon^I\,\frac{\partial F(x,\upsilon,\bar\upsilon)}{\partial \upsilon^I}
  +\bar\upsilon^I\,\frac{\partial F(x,\upsilon,\bar\upsilon)}{\partial
    \upsilon^I}  =&\, F(x,\upsilon,\bar\upsilon)\,,\nonumber\\
  \upsilon^I\,\frac{\partial F(x,\upsilon,\bar\upsilon)}{\partial \upsilon^I}
  - \bar\upsilon^I\,\frac{\partial F(x,\upsilon,\bar\upsilon)}{\partial
    \upsilon^I}  =&\, 0\,. 
\end{align}
The $\mathrm{SU}(2)$ invariance and the homogeneity of the functions
$F_{IJ}(L)$ imply the equation
\begin{equation}
  \label{eq:FIJ-conformal}
  F_{IJK}{}^{ik} \, L_{kj}{}^K = -\tfrac12 \delta^i{}_j\, F_{IJ} \,,  
\end{equation}
Unlike the functions $F_{IJ}$, the function
$F(x,\upsilon,\bar\upsilon)$ is {\it not} invariant under the full
$\mathrm{SU}(2)$ R-symmetry group, but only under a $\mathrm{U}(1)$
subgroup.

In the superconformal case it is convenient to introduce also an
$\mathrm{SU}(2)$ invariant quantity whose second derivative generates
the metric of the tensor multiplet scalars,
\begin{equation}
  \label{eq:tensor-potential}
  \chi_\mathrm{tensor}(L)\equiv 2\, F_{IJ}\, L^{ij I}\,L_{ij}{}^J\,,
\end{equation}
which is a homogeneous function of first degree invariant under
$\mathrm{SU}(2)$. This tensor-potential
$\chi_\mathrm{tensor}$ satisfies the following equations,
\begin{equation}
  \label{eq:derivative-potential}
  \frac{\partial \chi_\mathrm{tensor}(L)}{\partial L_{ij}{}^I} = 2\,
  F_{IJ}(L)\, L^{ij J}\,, \qquad \varepsilon_{kl} \,\frac{\partial^2
    \chi_\mathrm{tensor}(L)}{\partial L_{ik}{}^I\,\partial L_{jl}{}^J
  } = 2\,F_{IJ}(L)\, \varepsilon^{ij} \,.    
\end{equation}
Not surprisingly there exists a relation between $\chi_\mathrm{tensor}(L)$
and the function $F(x,\upsilon,\bar\upsilon)$,
\begin{equation}
  \label{eq:tensor-potential-2}
  \chi_{\mathrm{tensor}}(L) = F_{IJ} \,(x^I x^J + 4
  \,\upsilon^I\bar\upsilon^J) =  -
  F(\upsilon,\bar \upsilon, x) + x^I \,\frac{\partial
    F(x,\upsilon,\bar \upsilon)}{\partial x^I} \,, 
\end{equation}
which can be established by making use of the equations
\eqref{eq:F-second-derivatives} and \eqref{eq:F-scale-phase-inv}. The
right-hand side of \eqref{eq:tensor-potential-2} coincides with the
expression for the hyperk\"ahler potential
$\chi_\mathrm{hyper}(\upsilon,\bar \upsilon, w,\bar{w})$ given in
\cite{deWit:2001dj} for the hyperk\"ahler cones that one obtains upon
dualizing the tensor fields to scalars. In that case $\partial
F(x,\upsilon,\bar \upsilon)/\partial x^I $ is identified with $w^I +
\bar w^I$, so that one is performing a Legendre transformation
\cite{deWit:2006gn}. The reason that only the real part of $w^I$
appears is that the hypermultiplet Lagrangian will have an abelian
isometry for every tensor multiplet.

Observe that the last term in \eqref{eq:4D-lagrangian-tensor}
specifies a coupling of the tensor gauge fields to $F_{IJK}{}^{ij}$,
the derivative of $F_{IJ}$ with respect to $L_{ij}{}^K$. This coupling
does not involve the tensor field strengths, but is nevertheless
invariant under tensor gauge transformations. The reason is that the
term $F_{IJK}{}^{ij} \,\partial_P L_{ik}{}^J \,\partial_Q L_{jl}{}^K
\,\varepsilon^{kl}$ satisfies the equation,
\begin{equation}
  \label{eq:bianchi-partial-L2}
  \partial_{[M}\big( F_{IJK}{}^{ij} \,\partial_P L_{ik}{}^J
  \,\partial_{Q]} L_{jl}{}^K \,\varepsilon^{kl} \big)=0\,,
\end{equation}
by virtue of the properties satisfied by $F_{IJ}$. This result implies
that one can write (locally)
\begin{equation}
  \label{eq:expl-2-form}
  F_{IJK}{}^{ij} \,\partial_P L_{ik}{}^J
  \,\partial_Q L_{jl}{}^K \,\varepsilon^{kl} = \partial_{[P}
  A(x,\upsilon,\bar\upsilon)_{Q] I} \,.
\end{equation}
One particular solution for the space-time vectors
$A(x,\upsilon,\bar\upsilon)_{M\,I}$ is defined in terms of second
derivatives of the function $F(x,\upsilon,\bar\upsilon)$ introduced in
\eqref{eq:F-second-derivatives}, and reads
\begin{equation} 
  \label{eq:special-A_M-sol}
   A(x,\upsilon,\bar\upsilon)_{M\,I} =
    \frac{\partial^2 F(x,\upsilon,\bar\upsilon)}{\partial x^I\,\partial\bar\upsilon^J}
    \,\partial_M \bar\upsilon^J - \frac{\partial^2F(x,\upsilon,\bar\upsilon)}{\partial
      x^I\,\partial \upsilon^J} \,\partial_M \upsilon^J \,, 
\end{equation}
Clearly these space-time vectors are only determined up to a gauge
transformation $A_{M\,I} \to A_{M\,I}+ \partial_M \Lambda_I$. They are,
however, only manifestly invariant under a (rigid) $\mathrm{U}(1)$
subgroup of the full $\mathrm{SU}(2)$ R-symmetry
transformations. Nevertheless, when applying an infinitesimal
$\mathrm{SU}(2)$ transformation with (local) parameter
$\Lambda_i{}^j(x)$ on the left-hand side of \eqref{eq:expl-2-form},
one obtains
\begin{equation}
  \label{eq:su2-var-expl-2-form}
  \delta_{\mathrm{SU}(2)} \Big(F_{IJK}{}^{ij} \,\partial_P L_{ik}{}^J
  \,\partial_Q L_{jl}{}^K \,\varepsilon^{kl} \Big) = \partial_{[P} \left[
    \partial_{Q]} \Lambda_i{}^j \, F_{IJ}(L) \,L_{jk}{}^J\, \varepsilon^{ki} \right] \,,
\end{equation}
where we made use of the $\mathrm{SU}(2)$ invariance of
$F_{IJ}(L)$. This result, which is in line with
\eqref{eq:bianchi-partial-L2}, implies that the vectors
$A(x,\upsilon,\bar\upsilon)_{M\,I}$ are invariant under (rigid)
$\mathrm{SU}(2)$ up to an abelian gauge transformation with
field-dependent parameter. For the solution \eqref{eq:special-A_M-sol}
one can calculate this transformation explicitly in terms of multiple
derivatives of the function $F(x,\upsilon,\bar\upsilon)$.

A relevant question is whether it is possible to apply such a gauge
transformation to the $3D$ vector fields
$A(x,\upsilon,\bar\upsilon)_{M\,I}$ such that the results become
exactly $\mathrm{SU}(2)$ invariant. As was observed long ago
\cite{deWit:1982na}, the answer to this question is in general
negative: it is not always possible to satisfy \eqref{eq:expl-2-form}
with $\mathrm{SU}(2)$ invariant `potentials'
$A(x,\upsilon,\bar\upsilon)_{M\,I}$.  However, as we will establish in
the next subsection, there do exist specific models where both the
gauge invariance and the $\mathrm{SU}(2)$ invariance is
manifest. Hence we may distinguish two distinct classes of tensor
interactions characterized by the fact whether or not the vector
fields $A_I(x,\upsilon,\bar\upsilon)_M$ can be globally extended to
full $\mathrm{SU}(2)$ invariants or will be at most be invariant under
a $\mathrm{U}(1)$ subgroup.

Many of the $4D$ features related to tensor gauge invariance and
R-symmetry invariance remain relevant in $3D$. Hence let us turn to
the reduction of the Lagrangian \eqref{eq:4D-lagrangian-tensor} to
three dimensions by using the expressions for the $4D$ fields
summarized in \eqref{eq:D4to3tensor} and \eqref{eq:D4to3Weyl} and the
Ricci scalar in \eqref{eq:contracted-R}. Starting from the first line
in \eqref{eq:4D-lagrangian-tensor}, we find
\begin{align}
  \label{eq:first-term-tensor}
  &  - \tfrac{1}{2} E \, F_{IJ} \,\mathcal{D}_M L_{ij}{}^{I} \,
     \mathcal{D}^M L^{ij\,J} + E\, F_{IJ} \, L_{ij}{}^I \, L^{ij\,J} \,
     \big(\tfrac{1}{3} R +  D \big)  \nonumber \\ 
%%%%%%%%%%%%% 
  &\;=    \tfrac{1}{2} e \, F_{IJ} \,\mathcal{D}_\mu L^i{}_j{}^{I} \,
     \mathcal{D}^\mu L^j{}_i{}^{J}  -\tfrac12\,e\,F_{IJ} \,
     L^i{}_j{}^I \, L^j{}_i{}^{J}\, \big(\tfrac12\,R + D - C^2\big)\nonumber\\
   &\qquad   +\tfrac14 e\,F_{IJ} \,\phi^{-2} \big(L^i{}_j{}^I\,
   Y^{0\,j}{}_i\big)    \,\big(L^k{}_l{}^J\,  Y^{0\,l}{}_k\big)   \,,
\end{align}
where we used the homogeneity of $F_{IJ}$ to express it in terms the
$3D$ scalars $L^i{}_j$ defined in
\eqref{eq:D4to3tensor}.\footnote{ %%%%%%%%%%%%%%%%%%%%%%%%
  We remind the reader that the $L^i{}_j$ are anti-hermitian. } %%%%%%%%
We also suppressed a total derivative term (for this it is convenient
to make use of the first equation
\eqref{eq:derivative-potential}). Subsequently we consider the next
few terms of \eqref{eq:4D-lagrangian-tensor} and reduce them to three
dimensions,
\begin{align}
  \label{eq:reduction-tensor}
    & E\, F (L)_{IJ} \, \big( E_{M}{}^I \, E^{M\, J} -E^{M\,I}\,
  \mathcal{V}_M{}^i{}_j \,L_{ik}{}^J\,\varepsilon^{jk} + G^I \bar{G}^J  \big)
  \nonumber\\
  &= -\tfrac12 e\,F (L)_{IJ} \big[ \tfrac12 \phi^{-2} \big(L^i{}_j{}^I\,
   Y^{j}{}_i{\!}^0\big)    \,\big(L^k{}_l{}^J\,  Y^{l}{}_k{\!}^0\big)
   +F(E)_{\mu\nu}{}^I\,F(E)^{\mu\nu J} + Y^p{}_q{}^I\,Y^q{}_p{}^J  
   \big] \nonumber\\
   &\qquad -\tfrac12 \mathrm{i}\varepsilon^{\mu\nu\rho}\,F (L)_{IJ}
   \,F(E)_{\mu\nu}{}^I \,L^i{}_j{}^J\, \mathcal{V}_\rho{}^j{}_i \,. 
\end{align}
Combining all these contributions with those coming form the last term
in \eqref{eq:4D-lagrangian-tensor}, we obtain the final result, 
\begin{align}
  \label{eq:3D-lagrangian-tensor}
  \mathcal{L}_{\mathrm{tensor}}\big\vert_{3D} = &\,
  \tfrac{1}{2} e \, F(L)_{IJ} \,\mathcal{D}_\mu L^i{}_j{}^{I} \,
     \mathcal{D}^\mu L^j{}_i{}^{J}  -\tfrac12\,e\,F(L)_{IJ} \,
     L^i{}_j{}^I \, L^j{}_i{}^{J}\, \big(\tfrac12\,R + D - C^2\big)\nonumber\\
   &\, -\tfrac12 e\,F(L)_{IJ} \big[F(E)_{\mu\nu}{}^I\,F(E)^{\mu\nu J} +
  Y^p{}_q{}^I\,Y^q{}_p{}^J   
   \big] \nonumber\\
   &\, -\tfrac12 \mathrm{i}\varepsilon^{\mu\nu\rho}\,F(L)_{IJ}
   \,F(E)_{\mu\nu}{}^I \,L^i{}_j{}^J\, \mathcal{V}_\rho{}^j{}_i
   \nonumber\\
   &+\mathrm{i}\varepsilon^{\mu\nu\rho} F(L)_{IJKi}{}^j \, 
   \,\partial_\mu L^i{}_k{}^I\, \partial_\nu L^k{}_j{}^J\, E_\rho{}^K\,  \,, 
\end{align}
where we note that the Kaluza-Klein vector multiplet again manifestly
decouples. This should not come as a surprise as one can dualize the
$3D$ vector field $E_\mu$ to a scalar and then obtain a hypermultiplet
Lagrangian, for which we have noted the same decoupling
phenomenon. Let us stress that all the properties of the $4D$ tensor
Lagrangians related to the tensor potential $\chi_\mathrm{tensor}(L)$
carry over to the three-dimensional context. However, the definition
\eqref{eq:tensor-potential} in terms of the fields $L^i{}_j$ acquires
an explicit minus sign because $L^{ijI} L_{ij}{}^J= -L^i{}_j{}^I
\,L^j{}_i{}^J$. As a result the equations \eqref{eq:tensor-potential}
and \eqref{eq:derivative-potential} take the following form,
\begin{align}
  \label{eq:3D-tensor-pot}
  \chi_\mathrm{tensor}(L) =&\, -2\, F_{IJ}\, L^i{}_j{}^I \,L^j{}_i{}^J 
  \,,\nonumber\\[2mm]
  \frac{\partial \chi_\mathrm{tensor}(L)}{\partial L^i{}_j{}^I} =&\, - 2\,
  F_{IJ}(L)\, L^j{}_i{}^ J\,, \nonumber\\ 
  \frac{\partial^2
    \chi_\mathrm{tensor}(L)}{\partial L^i{}_k{}^I\,\partial
    L^k{}_{j}{}^J} =&\, - 2\,F_{IJ}(L)\, \delta_i{}^j  \,.     
\end{align}

%%%%%%%%%%%%%%%%%%%%%%%%%%%%%%%%%%%%%%%%%%%%%%%%%%%%%%%
\subsubsection{The vector multiplet Lagrangian}
\label{subsubsection:vector-multiplet-lagr}
%%%%%%%%%%%%%%%%%%%%%%%%%%%%%%%%%%%%%%%%%%%%%%%%%%%%%%%
Finally, we turn to the bosonic Lagrangian for vector multiplets,
whose evaluation is considerably more complicated, as the Kaluza-Klein
vector multiplet will not decouple in this case. Therefore the number
of vector multiplets will increase by one under the reduction. To
avoid confusion with the discussion in subsection
\ref{subsubsection:tensor-multiplet-lagr}, we will use indices
$\Lambda,\Sigma,\Xi,\ldots$ to label the $n+1$ off-shell $4D$ vector
multiplets.  With the Kaluza-Klein vector multiplet we will thus
obtain $n+2$ $3D$ vector multiplets. As before we start from the
bosonic terms of the $4D$ Lagrangian, which take the form,
\begin{align}
  \label{eq:4D-lagrangian-vectors}
  \mathcal{L}_\mathrm{vector}\big\vert_{4D} =&\,
  E\, N_{\Lambda\Sigma} \big[X^\Lambda\bar X^\Sigma
  \,\big(\tfrac{1}{6} R -D\big) + 
  \tfrac1{8}Y_{ij}{}^\Lambda Y^{ij \Sigma} -\mathcal{D}_M
  X^\Lambda\,\mathcal{D}^M \bar X^\Sigma 
  \big]  \nonumber\\[.3ex] 
   &\,- \ft18 E\, N_{\Lambda\Sigma} \,F_{MN}{}^{\Lambda} \,\,F^{MN}{}^{\Sigma} -\tfrac1{16}
  \mathrm{i} \varepsilon^{MNPQ}
  \,R_{\Lambda\Sigma} \, F_{MN}{}^\Lambda \, F_{PQ}{}^\Sigma   \nonumber\\[.3ex]
  &\,+\tfrac18 E\,\big[ \bar X^\Lambda  N_{\Lambda\Sigma} F^{AB
    \Sigma} \, T_{AB}{}^{ij}\varepsilon_{ij} 
  -\tfrac1{8} \bar X^\Lambda  N_{\Lambda\Sigma}  \bar X^\Sigma 
  \, \big(T_{AB}{}^{ij}\varepsilon_{ij}\big)^2 +\mbox{h.c.}
  \big]   \,,
\end{align}
and is encoded in a holomorphic function $F(X)$ that is homogeneous of
second degree. Its multiple derivatives are denoted by
$F_{\Lambda\Sigma\Xi\cdots}$ and the second derivatives are decomposed
into two real tensors, $N_{\Lambda\Sigma}= -\mathrm{i}
F_{\Lambda\Sigma} + \mathrm{i} \bar F_{\Lambda\Sigma}$ and
$R_{\Lambda\Sigma}= F_{\Lambda\Sigma} + \bar F_{\Lambda\Sigma}$, which
we have used in the above expression.

As before, we reduce \eqref{eq:4D-lagrangian-vectors} in steps,
starting with the first two terms,
\begin{align}
  \label{eq:first-two-terms-vector}
  & E\, N_{\Lambda\Sigma} \big[X^\Lambda\bar X^\Sigma \,\big(\tfrac{1}{6} R -D\big) +
  \tfrac1{8}Y_{ij}{}^\Lambda Y^{ij \Sigma}   \big]  \nonumber\\
  %%%%%%%%%%%%% 
  &\,=\tfrac1{2} e\,\phi^{-1} N_{\Lambda\Sigma} X^\Lambda \bar{X}^\Sigma\,
  \big[\tfrac12\,R - D + C^2 
  + \tfrac14 \, \phi^{-2}\big(F(W)_{ab}{}^0\big)^2 +
  \phi^{-2} \big(\mathcal{D}_\mu \phi\big)^2  \big]  \nonumber \\
  %%%%%%%%%%%%%%% 
  &\quad -\tfrac1{8} e\, \phi^{-1}N_{\Lambda\Sigma}\, \big[ Y^i{}_{j}{}^\Lambda\,
  Y^j{}_{i}{}^\Sigma + \phi^{-2}  L^p{}_q{}^\Lambda\,L^q{}_p{}^0\; Y^{i}{}_{j}{}^\Sigma\,
  Y^j{}_{i}{}^0 \big] \nonumber\\[.1ex]
  &\qquad +\tfrac14 e \,\phi^{-3} N_{\Lambda\Sigma} \big[ X^\Lambda \bar X^\Sigma
  - \tfrac18\phi^{-2} 
  L^p{}_q{}^\Lambda\,L^q{}_p{}^0  \;  L^r{}_s{}^\Sigma\,L^s{}_r{}^0 \big]\,
  Y^i{}_{j}{}^0 \,Y^j{}_{i}{}^0   \,, 
\end{align}
where we made use of \eqref{eq:D4to3Weyl}, \eqref{eq:D4to3vector} and
\eqref{eq:contracted-R}, and we employed the identifications
$\phi=L^0$ and $L^p{}_q{}^\Lambda\,L^q{}_p{}^0= -\tfrac12 x^\Lambda
x^0 -\upsilon^\Lambda\bar\upsilon^0
-\bar\upsilon^\Lambda\upsilon^0$. The next terms related to the
kinetic terms of $X^\Lambda$ and the various field-strengths reduce to
\begin{align}
  \label{eq:F-2-vector}
       &-E\, N_{\Lambda\Sigma} \,\mathcal{D}_M X^\Lambda\,\mathcal{D}^M \bar X^\Sigma
       - \ft18 E\, N_{\Lambda\Sigma} \,F_{MN}{}^{\Lambda} \,\,F^{MN \Sigma} -\tfrac1{16}
  \mathrm{i} \varepsilon^{MNPQ}
  \,R_{\Lambda\Sigma} \, F_{MN}{}^\Lambda \, F_{PQ}{}^\Sigma   \nonumber\\[.3ex]
  &\,= -e\,
   \phi^{-1} N_{\Lambda\Sigma}\big[  X^\Lambda\bar X^\Sigma \,C^2  + \mathcal{D}_\mu X^\Lambda
   \mathcal{D}_\mu \bar X^\Sigma \big] \nonumber\\
   &\,\quad +\tfrac14 \phi^{-2} N_{\Lambda\Sigma}
   \big(X^\Lambda \stackrel{\leftrightarrow}{\mathcal{D}}{\!}_\mu\bar X^\Sigma \big) \,
   \varepsilon^{\mu\nu\rho}  F(W)_{\nu\rho}{}^0 \nonumber\\
  &\,\quad - \tfrac1{8} e\, \phi^{-1}N_{\Lambda\Sigma}\, \big[
  F(W)_{\mu\nu}{}^\Lambda  F(W)^{\mu\nu \Sigma } 
  + \phi^{-2}  L^p{}_q{}^\Lambda\,L^q{}_p{}^0\; F(W)_{\mu\nu}{}^\Sigma F(W)^{\mu\nu 0}   
 \big] \nonumber\\[.1ex]
  &\,\quad +\tfrac14 e\,\phi^{-3} N_{\Lambda\Sigma} \big[ \tfrac12 X^\Lambda \bar X^\Sigma
  -\tfrac18\phi^{-2} 
  L^p{}_q{}^\Lambda\,L^q{}_p{}^0  \;  L^r{}_s{}^\Sigma\,L^s{}_r{}^0 \big]\,
  F(W)_{\mu\nu}{}^0 F(W)^{\mu\nu 0}  \nonumber\\
  &\,\quad + \tfrac18\mathrm{i} \varepsilon^{\mu\nu\rho}\,R_{\Lambda\Sigma} 
  \big[F(W)_{\mu\nu}{}^\Lambda +\tfrac12 \phi^{-2}  L^p{}_q{}^\Lambda\,L^q{}_p{}^0\, 
   F(W)_{\mu\nu}{}^0 \big] \,\partial_\rho \big(\phi^{-2}
   L^r{}_s{}^\Sigma\,L^s{}_r{}^0\big)\nonumber\\
  &\, \quad -\tfrac1{16}  e\,\phi \, N_{\Lambda\Sigma} \, \partial_\mu \big(\phi^{-2}
   L^p{}_q{}^\Lambda\,L^q{}_p{}^0\big)\,\partial^\mu \big(\phi^{-2}
   L^r{}_s{}^\Sigma\,L^s{}_r{}^0\big)\,.
\end{align}
In the left-hand side of this relation, $\mathcal{D}_\mu X^\Lambda$
denotes a $\mathrm{U}(1)$ covariant derivative with the connection
$\mathcal{A}_\mu{\!}^0$ that was defined in \eqref{eq:T-A}, 
\begin{equation}
  \label{eq:D-cov-X}
  \mathcal{D}_\mu X^\Lambda= \left(\partial_\mu -b_\mu +\mathrm{i}
    \mathcal{A}_\mu{\!}^0   \right) X^\Lambda\,. 
\end{equation}
Finally we reduce the remaining terms
\begin{align}
  \label{eq:T-terms}
  &\tfrac18 E\,\big[ \bar X^\Lambda N_{\Lambda\Sigma} F^{AB \Sigma} \,
  T_{AB}{}^{ij}\varepsilon_{ij} -\tfrac1{8} \bar X^\Lambda
  N_{\Lambda\Sigma} \bar X^\Sigma
  \, \big(T_{AB}{}^{ij}\varepsilon_{ij}\big)^2  \big]  +\mathrm{h.c.}  \nonumber\\
  &\,= \tfrac18\mathrm{i} e\, \phi^{-3} N_{\Lambda\Sigma} \bar
  X^\Lambda\, \bigg[ (\bar\upsilon^0
  \stackrel{\leftrightarrow}{\mathcal{D}}{\!}_\mu x^0 ) -
  \displaystyle{ \frac{\bar\upsilon^0 } {L^0+\tfrac12 x^0}}\,
  (\upsilon^0 \stackrel{\leftrightarrow}{\mathcal{D}}{\!}_\mu
  \bar\upsilon^0)   \bigg] \nonumber\\
  &\qquad\times \Big[ \phi\,\partial^\mu \big(\phi^{-2}
  L^p{}_q{}^\Sigma\,L^q{}_p{}^0\big) +e^{-1} \varepsilon^{\mu\nu\rho}
  \big(F(W)_{\nu\rho}{}^\Sigma + \tfrac12\phi^{-2}
  L^p{}_q{}^\Sigma\,L^q{}_p{}^0
  F(W)_{\nu\rho}{}^0\big) \Big]\nonumber\\
  &\,\quad +\tfrac1{16} e\,\phi^{-5}\,N_{\Lambda\Sigma}\,\bar X^\Lambda
  \bar X^\Sigma \, \bigg[ (\bar\upsilon^0
  \stackrel{\leftrightarrow}{\mathcal{D}}{\!}_a x^0 ) - \displaystyle{
    \frac{\bar\upsilon^0 } {L^0+\tfrac12 x^0}}\, (\upsilon^0
  \stackrel{\leftrightarrow}{\mathcal{D}}{\!}_a \bar\upsilon^0)
  \bigg]^2 +\mathrm{h.c.} \,.
\end{align}

Because the number of vector multiplets is increased by the presence
of the Kaluza-Klein vector multiplet, we extend the range of the
indices $\{\Lambda\}$ to $\{A\}=\{0,\Lambda\}$, where the index $A=0$
refers to the Kaluza-Klein vector multiplet. Up to terms that involve
derivatives of the scalar fields and the epsilon tensor, the $3D$
Lagrangian can then be written as
\begin{align}
  \label{eq:Lvector-3d-partial}
  \mathcal{L}_\mathrm{vector}\big\vert_{3D}=&\, -\tfrac12   e\, 
     \mathcal{F}_{AB}(L) \,L^p{}_q{}^A\,L^q{}_p{}^B\big[ \tfrac12 R -D - C^2\big]
     \nonumber\\
       &\, -\tfrac12 e\, \mathcal{F}_{AB}(L)  \big[
       F(W)_{\mu\nu}{}^A\,F(W)^{\mu\nu B}  +   Y^i{}_j{}^A\,Y^j{}_i{}^B  \big]\,.
\end{align}
Here we have simply collected all the corresponding terms of
\eqref{eq:first-two-terms-vector} and \eqref{eq:F-2-vector}, which
lead to the following expressions for the tensor $\mathcal{F}_{AB}(L)$,
\begin{align}
  \label{eq:F-AB}
  \mathcal{F}_{\Lambda\Sigma} =&\, \frac1{4\,L^0} \,N_{\Lambda\Sigma} \,, \nonumber\\[2mm] 
  \mathcal{F}_{\Lambda 0} =&\,\mathcal{F}_{0\Lambda}=  \frac1{8\,(L^0)^3} \,N_{\Lambda\Sigma} \, 
  L^p{}_q{}^\Sigma\,L^q{}_p{}^0 \,,\nonumber\\ 
  \mathcal{F}_{00} =&\, \frac1{16\,(L^0)^3} \,N_{\Lambda\Sigma} \bigg[
  L^p{}_q{}^\Lambda\,L^q{}_p{}^\Sigma +  \frac{3\,  L^p{}_q{}^\Lambda\,L^q{}_p{}^0\;
    L^r{}_s{}^\Sigma\,L^s{}_r{}^0}{2\,(L^0)^2} \bigg]\,. 
\end{align}
Furthermore, one easily verifies that the direct analogue of the
tensor potential that was introduced earlier in
\eqref{eq:tensor-potential}, {\footnote{ %%%%%%%%%%%%%%%%%%
    Note that a minus sign has to be introduced in the definition
    below because the quadratic form $L^p{\!}_q\,L^p{\!}_q$ that we are
    using here is non-positive!}  %%%%%%%%%%%%%%%%%%%%%%%%
\begin{align}
  \label{eq:FLL-NXX}
  \chi_\mathrm{vector}\equiv&\, - 2\, \mathcal{F}_{AB} \,L^p{}_q{}^A\,L^q{}_p{}^B\nonumber\\
  =&\,-\frac{N_{\Lambda\Sigma}} {4\, L^0} \left[ 
    L^p{}_q{}^\Lambda\,L^q{}_p{}^\Sigma + \frac{L^p{}_q{}^\Lambda\,L^q{}_p{}^0\;
      L^r{}_s{}^\Sigma\,L^s{}_r{}^0}{2\,(L^0)^2} \right]\nonumber\\
  =&\,   \frac{2\,N_{\Lambda\Sigma} \,X^\Lambda\bar X^\Sigma }{L^0} \,,
\end{align}
is a homogeneous function of first degree  and is manifestly invariant under the symmetries
\eqref{eq:L-L)-invariance}. To prove the identity
between the second and third line one may use the following convenient
expression for $X^\Lambda$ (cf. \eqref{eq:X(L)-1}), 
\begin{equation}
  \label{eq:X(L)}
  X^\Lambda=\tfrac12\mathrm{i}\, \bar\upsilon^\Lambda -\tfrac14\mathrm{i}
  \bigg[x^\Lambda -\frac{L^p{}_q{}^\Lambda\,L^q{}_p{}^0}{L^0} \bigg]
  \frac{\bar\upsilon^0}{L^0+\tfrac12 x^0} \,.
\end{equation}
The reader may verify that the application of a $3D$ $\mathrm{SU}(2)$
transformation on the right-hand side of \eqref{eq:X(L)} takes the
form of an $\mathrm{U}(1)$ transformation on the left-hand side with
parameter $\Lambda_A$ defined in
\eqref{eq:def-delta-Lambda-A}. Therefore $\mathrm{U}(1)$ invariant
products such as $X^\Lambda\,\bar X^\Sigma$ should take an
$\mathrm{SU}(2)$ invariant form. In particular we find
\begin{equation}
  \label{eq:X-barX-LL}
  -8\,  X^{(\Lambda} \,\bar X^{\Sigma)}=
    L^p{}_q{}^\Lambda\,L^q{}_p{}^\Sigma +
    \frac{L^p{}_q{}^\Lambda\,L^q{}_p{}^0\; 
      L^r{}_s{}^\Sigma\,L^s{}_r{}^0}{2\,(L^0)^2} \,,
\end{equation}
which indeed confirms the last identity in \eqref{eq:FLL-NXX}. 
Likewise $F_{\Lambda\Sigma}(X)$ is also $\mathrm{U}(1)$ invariant, and
must therefore be $\mathrm{SU}(2)$ invariant as well. From this
observation it follows directly that $N_{\Lambda\Sigma}= -\mathrm{i}
(F_{\Lambda\Sigma}-\bar F_{\Lambda\Sigma})$, the functions
$\mathcal{F}_{AB}$ and $\chi_\mathrm{vector}$ are all $\mathrm{SU}(2)$
invariant as well. 

We should point out that all properties derived above are so far
consistent with the fact that there exists a reflection associated
with \eqref{eq:su2-interchange} that correspondingly interchanges the
vector and the tensor multiplets. This explains the different sign of
the terms proportional to the field $D$ in the two Lagrangians
\eqref{eq:3D-lagrangian-tensor} and \eqref{eq:Lvector-3d-partial}. We
will see that this relation between tensor and vector supermultiplets
is also valid for the remaining terms in the full
Lagrangians. Therefore, the remainder of this section will be
devoted to a detailed derivation of the bosonic terms of the vector
Lagrangian in order to isolate the intricate features that are crucial
for establishing the relationship with the tensor Lagrangian.

Before specifying the remaining terms in the Lagrangian
\eqref{eq:Lvector-3d-partial}, we present a convenient expression
based on the derivatives of $X^\Lambda$ with respect to the components
of $L^p{}_q{}^A$,
\begin{align}
  \label{eq:variation-X-generic}
  \delta X^\Lambda =&\, \frac{\upsilon^0\,\delta\bar\upsilon^0 -
    \bar\upsilon^0\,\delta\upsilon^0} {2\,L^0(L^0+\tfrac12 x^0)}
  \,X^\Lambda  \nonumber\\
  &\,-\frac{\mathrm{i}\,\bar\upsilon^0}{4\,L^0} \bigg[ \Big(\delta
  x^\Lambda + \delta x^0
  \,\frac{L^p{}_q{}^0\,L^q{}_p{}^\Lambda}{2(L^0)^2}\Big)
  +\frac{\bar\upsilon^0}{L^0+\frac12 x^0} \Big(\delta
  \upsilon^\Lambda + \delta \upsilon^0
  \,\frac{L^p{}_q{}^0\,L^q{}_p{}^\Lambda}{2(L^0)^2}\Big) \nonumber\\
  &\,\qquad\qquad\qquad \qquad\qquad \qquad\qquad 
  - \frac{\upsilon^0}{L^0-\frac12 x^0} \Big(\delta
  \bar\upsilon^\Lambda + \delta \bar\upsilon^0
  \,\frac{L^p{}_q{}^0\,L^q{}_p{}^\Lambda}{2(L^0)^2}\Big) \bigg]\,.
\end{align}
Using that $F_{\Lambda\Sigma}(X)$ is the second derivative of a
holomorphic homogeneous function of degree two, we derive the
following two identities,
\begin{align}
  \label{eq:0-Xi-derivative}
  \frac{\partial F_{\Lambda\Sigma}(X)}{\partial L^p{}_q{}^\Xi}  = 
 \frac{\partial F_{\Lambda\Xi}(X)}{\partial L^p{}_q{}^\Sigma} \,, \qquad
  \frac{\partial F_{\Lambda\Sigma}(X)}{\partial L^p{}_q{}^0} = 
  \frac{\partial F_{\Lambda\Sigma}(X)}{\partial L^p{}_q{}^\Xi}
  \,\frac{L^r{}_s{}^0\,L^s{}_r{}^\Xi}{2(L^0)^2}  \,,
\end{align}
where the second equation follows directly from
\eqref{eq:variation-X-generic}.  Furthermore we note the identities
\begin{align}
  \label{eq:F-der-times-L0}
  \frac{\partial F_{\Lambda\Sigma}(X)}{\partial L^r{}_p{}^\Xi}
  \,L^r{}_q{\!}^0  =&\,
  - L^p{}_r{\!}^0\, \frac{\partial F_{\Lambda\Sigma}(X)}{\partial
    L^q{}_r{}^\Xi}  = \mathrm{i}L^0 \,
  \frac{\partial F_{\Lambda\Sigma}(X)}{\partial L^q{}_p{}^\Xi}  \,,
  \nonumber\\[2mm] 
  %%%
  \frac{\partial \bar{F}_{\Lambda\Sigma}(\bar{X})}{\partial L^r{}_p{}^\Xi}
  \,L^r{}_q{\!}^0  =&\,
  - L^p{}_r{\!}^0\, \frac{\partial
    \bar{F}_{\Lambda\Sigma}(\bar{X})}{\partial L^q{}_r{}^\Xi}  = -
  \mathrm{i} L^0\,
  \frac{\partial\bar{F}_{\Lambda\Sigma}(\bar{X})}{\partial L^q{}_p{}^\Xi}  \,.
\end{align}

Subsequently we make use of the fact that $F_{\Lambda\Sigma}(X)$ is a
homogeneous function of zeroth degree, so that it is invariant under
complex scale transformations of the $4D$ fields $X^\Xi$. When
regarding $F_{\Lambda\Sigma}(X)$ as function of the $3D$ fields, it
must be an $\mathrm{SU}(2)$ invariant homogeneous function of zeroth
degree. Moreover, close inspection based on \eqref{eq:X(L)} shows that
it must be a homogeneous function of the $L^p{}_q{}^\Xi$ and
$L^p{}_q{}^0$ separately. Exploiting the second equation
\eqref{eq:0-Xi-derivative}, we thus derive the following results based
on homogeneity and $\mathrm{SU}(2)$ invariance,
\begin{align}
  \label{eq:FX-holo-su}
  &\frac{\partial F_{\Lambda\Sigma}} {\partial L^p{}_q{}^\Xi} \,
  L^{p}{}_q{}^0   =0 \,, \quad 
  \frac{\partial F_{\Lambda\Sigma}} {\partial L^{p}{}_q{}^\Xi } \,
  L^{p}{}_q{}^\Xi = 0\,,\quad 
   \frac{\partial F_{\Lambda\Sigma}} {\partial L^{p}{}_q{}^0 } \,
  L^{p}{}_q{}^0 = 0\,,
  \nonumber\\[2mm]
  &\frac{\partial F_{\Lambda\Sigma}} {\partial L^{q}{}_r{}^\Xi} \Big(
  L^{p}{}_r{}^\Xi + L^{p}{}_r{}^0 \, \frac{L^{s}{}_t{}^0\, L^{t}{}_s{}^\Xi}{2(L^0)^2} \Big) =0 \,,
\end{align}
where the first equation, while consistent with homogeneity, is
actually derived from \eqref{eq:variation-X-generic}. Furthermore the
homogeneity of $F(X)$ implies that $\delta F_{\Lambda\Sigma}$ under
any variations $\delta X^\Xi$ must satisfy $\delta F_{\Lambda\Sigma}
\,X^\Lambda=0$, so that
\begin{equation}
  \label{eq:partial-F-LL}
  \frac{\partial F_{\Lambda\Sigma}} {\partial L^{A\,t}{}_u } \, \left[ 
    L^p{}_q{}^\Lambda\,L^q{}_p{}^\Sigma + \frac{L^p{}_q{}^\Lambda\,L^q{}_p{}^0\;
      L^r{}_s{}^\Sigma\,L^s{}_r{}^0}{2(L^0)^2} \right]= 0\,,
\end{equation}
where we again made use of \eqref{eq:X-barX-LL}.

The above results can straightforwardly be used to derive a number of
specific results that confirm the relation with the tensor multiplet
Lagrangians. First of all, we may verify by using \eqref{eq:0-Xi-derivative}
and \eqref{eq:partial-F-LL} that the derivative\footnote{ %%%%%%%%%%%%
  Observe that with the definitions of this paper we have
  \begin{equation}
    \label{eq:dL-dL}
    \frac{\partial L^p{\!}_q}{\partial L^r{\!}_t}  = \delta^p{\!}_r\,
    \delta^t{\!}_q \,, \qquad
    \frac{\partial L^0}{\partial L^p{\!}_q{}^0} = - \frac{L^q{}_p{}^0}
    {2\,L^0} \,. \nonumber 
  \end{equation}
  } %%%%%%%%%%%%%%%%%%%
of $\mathcal{F}_{AB}$ in \eqref{eq:F-AB} with respect to $L^p{}_q{}^C$,
denoted by $\mathcal{F}_{ABC}{}^q{}_p$, satisfies
\begin{equation}
  \label{eq:partial-F-AB}
  \mathcal{F}_{ABC}{}^p{}_q = \mathcal{F}_{(ABC)}{}^p{}_q \,,
\end{equation}
which corresponds to the first equation given in
\eqref{eq:properties-FIJ} in the context of the tensor
multiplets. Then we have already argued that the $F_{AB}(L)$ must be
$\mathrm{SU}(2)$ invariant; moreover they are manifestly homogeneous
functions of degree $-1$ in terms of the $3D$ fields
$F^p{}_q{}^A$. Therefore we derive the identity
\begin{equation}
  \label{eq:SU2-FAB}
  \mathcal{F}_{ABC}{}^q{}_r \, L^r{}_p{\!}^C  = -\tfrac12 \delta^p{}_q \, \mathcal{F}_{AB}\,, 
\end{equation}
which is precisely analogous to \eqref{eq:FIJ-conformal}, considered
in the context of the tensor Lagrangian.  Furthermore, from
\eqref{eq:partial-F-LL} one can verify the following relations,
\begin{equation}
  \label{eq:der-chi}
  \frac{\partial \chi_\mathrm{vector}} {\partial L^p{}_q{}^A} = - 2\,
  \mathcal{F}_{AB}\,  L^q{}_p{}^B\,, \qquad   
    \frac{\partial^2
    \chi_\mathrm{vector}(L)}{\partial L^p{}_r{}^A\,\partial
    L^r{}_{q}{}^B} = - 2\,\mathcal{F}_{AB}(L)\, \delta_p{}^q  \,.   
\end{equation}
which are analogous to the equations \eqref{eq:3D-tensor-pot} derived
for tensor multiplets. For future use we give the explicit expressions for the
independent components of $\mathcal{F}_{ABC}{}^p{}_q$,
\begin{align}
  \label{eq:calF-ABC-comp}
  \mathcal{F}_{\Lambda\Sigma\Xi}{}^p{}_q =&\,   \frac1{4\,L^0}
  \,\frac{\partial N_{\Lambda\Sigma}}{\partial L^q{}_p{}^\Xi}  \,,
  \nonumber\\[2mm]
  %%%%%
  \mathcal{F}_{\Lambda\Sigma 0}{}^p{}_q = &\,  \frac1{8\,(L^0)^3} \,
  N_{\Lambda\Sigma} \, L^p{}_q{\!}^0 + \frac1{8\,(L^0)^3} \,\frac{\partial
    N_{\Lambda\Sigma}}{\partial L^q{}_p{}^\Xi}  \,L^r{}_s{\!}^\Xi\,
  L^s{}_r{\!}^0 \,, \nonumber\\[2mm] 
  %%%%%%
   \mathcal{F}_{\Lambda 00}{}^p{}_q = &\, \frac1{8\,(L^0)^3}
   \,N_{\Lambda\Sigma} \,\bigg[ L^p{}_q{\!}^\Sigma + \frac{3\,
   L^r{}_s{\!}^\Sigma \,L^s{}_r{\!}^0 }{2\, (L^0)^2}\,
   L^p{}_q{\!}^0     \bigg]  \nonumber\\
   &\, -\frac1{8\,(L^0)^3} \,\frac{\partial
    N_{\Sigma\Xi}}{\partial L^q{}_p{}^\Lambda}  \; L^r{}_s{\!}^\Sigma\,
  L^s{}_r{\!}^\Xi     \,, \nonumber\\[2mm]
 %%%%%%%
  \mathcal{F}_{000}{}^p{}_q = &\, \frac3{32\,(L^0)^5}
  \,N_{\Lambda\Sigma} \,\bigg[ L^r{}_s{\!}^\Lambda
  \,L^s{}_r{\!}^\Sigma \, L^p{}_q{\!}^0
  +2\,L^r{}_s{\!}^\Lambda\,L^s{}_r{\!}^0\, L^p{}_q{\!}^\Sigma + \frac{5\,
    L^r{}_s{\!}^\Lambda\,L^s{}_r{\!}^0 \,L^t{}_u{\!}^\Sigma
    \,L^u{}_t{\!}^0} {2\,
    (L^0)^2} \,L^p{}_q{\!}^0   \bigg]  \nonumber\\
  &\, - \frac1{16\,(L^0)^5} \,\frac{\partial
    N_{\Lambda\Sigma}}{\partial L^q{}_p{}^\Xi} \;
  L^r{}_s{\!}^\Lambda\, L^s{}_r{\!}^\Sigma \, L^t{}_u{\!}^\Xi \,
  L^u{}_t{\!}^0 \,,
\end{align}
where we made use of the relations \eqref{eq:0-Xi-derivative} and
\eqref{eq:partial-F-LL}.  To verify their correctness one can,
for instance, verify the validity of \eqref{eq:SU2-FAB}.

To continue we will also need the following result for the covariant
derivatives $\mathcal{D}_\mu X^\Lambda$ in terms of the
three-dimensional fields,
\begin{align}
  \label{eq:DX}
  \mathcal{D}_\mu  X^\Lambda = &\, 
  \tfrac12\mathrm{i} \,\mathcal{D}_\mu\bar\upsilon^\Lambda -\tfrac14\mathrm{i}
  \bigg[\mathcal{D}_\mu x^\Lambda - \frac{\mathcal{D}_\mu
    L^p{}_q{}^\Lambda\,L^q{}_p{}^0}{L^0} \bigg] 
  \frac{\bar\upsilon^0}{L^0+\tfrac12 x^0} \nonumber\\
  &\, -\mathrm{i}\,
  \frac{ L^p{}_q{}^\Lambda\,L^q{}_p{}^0} {8\,(L^0)^3}  \,
  \bigg[ (\bar\upsilon^0
  \stackrel{\leftrightarrow}{\mathcal{D}}{\!}_\mu x^0 ) -
  \displaystyle{ \frac{\bar\upsilon^0 } {L^0+\tfrac12 x^0}}\,
  (\upsilon^0 \stackrel{\leftrightarrow}{\mathcal{D}}{\!}_\mu
  \bar\upsilon^0)   \bigg]   \,, 
\end{align}
which has been derived by making use again of
\eqref{eq:variation-X-generic}. One then proceeds to evaluate the
remaining terms of the action which all involve derivatives of the
scalar fields. First let us collect all the terms quadratic in these
derivatives from \eqref{eq:first-two-terms-vector},
\eqref{eq:F-2-vector} and \eqref{eq:T-terms},
\begin{align}
  \label{eq:Ds-Ds}
  e\, N_{\Lambda\Sigma} &\bigg[ \tfrac12 \phi^{-3}\, X^\Lambda\bar
  X^\Sigma\,(\mathcal{D}_\mu \phi)^2 
   -  \phi^{-1} \,  \mathcal{D}_\mu X^\Lambda\,
   \mathcal{D}^\mu \bar X^\Sigma  \nonumber\\
  &\;  +\tfrac1{16} \phi^{-5}\, \bar X^\Lambda
  \bar X^\Sigma \, \Big[ (\bar\upsilon^0
  \stackrel{\leftrightarrow}{\mathcal{D}}{\!}_\mu x^0 ) - \displaystyle{
    \frac{\bar\upsilon^0 } {L^0+\tfrac12 x^0}}\, (\upsilon^0
  \stackrel{\leftrightarrow}{\mathcal{D}}{\!}_\mu \bar\upsilon^0)
  \Big]^2 +\mathrm{h.c.} \nonumber\\
  &\;
  + \tfrac18\mathrm{i}  \phi^{-2}  \bar
  X^\Lambda \Big[ (\bar\upsilon^0
  \stackrel{\leftrightarrow}{\mathcal{D}}{\!}_\mu x^0 ) -
  \displaystyle{ \frac{\bar\upsilon^0 } {L^0+\tfrac12 x^0}}\,
  (\upsilon^0 \stackrel{\leftrightarrow}{\mathcal{D}}{\!}_\mu
  \bar\upsilon^0)   \Big] 
  \partial^\mu \big(\phi^{-2}
  L^p{}_q{}^\Sigma\,L^q{}_p{}^0\big) +\mathrm{h.c.} \nonumber\\
  &\;-\tfrac1{16}  \phi \, \partial_\mu \big(\phi^{-2}
   L^p{}_q{}^\Lambda\,L^q{}_p{}^0\big)\,\partial^\mu \big(\phi^{-2}
   L^r{}_s{}^\Sigma\,L^s{}_r{}^0\big) \bigg]\,.
\end{align}
To write this expression in terms of the $3D$ fields we first derive
the following three identities, 
\begin{align}
  \label{eq:X-A+and more}
  \mathrm{i} 
  & X^\Lambda\, \Big[ (\upsilon^0
  \stackrel{\leftrightarrow}{\mathcal{D}}{\!}_\mu x^0 ) -
  \displaystyle{ \frac{\upsilon^0 } {L^0+\tfrac12 x^0}}\,
  (\bar\upsilon^0 \stackrel{\leftrightarrow}{\mathcal{D}}{\!}_\mu
  \upsilon^0)   \Big] +\mathrm{h.c.} \nonumber\\
  &\qquad = - L^0 \,L^p{}_q{}^\Lambda   \,\mathcal{D}_\mu L^q{}_p{}^0  
  - \frac{L^r{}_s{}^\Lambda L^s{}_r{}^0}{2\,L^0} \,
  L^p{}_q{}^0\,\mathcal{D}_\mu L^q{}_p{}^0 \,, \nonumber\\[2mm]
%%%%%
  & \mathcal{D}_\mu X^{(\Lambda}\,  \mathcal{D}^\mu \bar X^{\Sigma)}\nonumber\\
   &\qquad = -\frac18  \mathcal{D}_\mu
  L^p{}_q{}^\Lambda\,\mathcal{D}^\mu L^q{}_p{}^\Sigma
  -\frac1{16\,(L^0)^2} \, L^p{}_q{}^0\, \mathcal{D}_\mu L^q{}_p{}^\Lambda \,
  L^r{}_s{}^0\,   \mathcal{D}^\mu L^s{}_r{}^\Sigma
  \nonumber\\
  &\qquad\qquad  -\frac{L^r{}_s{}^0\, L^s{}_r{}^{(\Lambda}} {8\,(L^0)^2} \,
    \mathcal{D}_\mu L^p{}_q{}^{\Sigma)} \,\mathcal{D}^\mu L^q{}_p{}^0 - 
    \frac{L^r{}_s{}^0 \,L^s{}_r{}^{(\Lambda}} {16\,(L^0)^4} \,
    \mathcal{D}_\mu L^p{}_q{}^{\Sigma)}\, L^q{}_p{}^0
    \,\mathcal{D}^\mu \,L^t{}_u{}^0\,L^u{}_t{}^0   
 \nonumber\\
 &\qquad \qquad + \frac1{64\,(L^0)^6} \, L^p{}_q{}^\Lambda\,L^q{}_p{}^0
 \,  L^r{}_s{}^\Sigma\,L^s{}_r{}^0  \;
  \Big\vert (\bar\upsilon^0
  \stackrel{\leftrightarrow}{\mathcal{D}}{\!}_\mu x^0 ) -
  \displaystyle{ \frac{\bar\upsilon^0 } {L^0+\tfrac12 x^0}}\,
  (\upsilon^0 \stackrel{\leftrightarrow}{\mathcal{D}}{\!}_\mu
  \bar\upsilon^0)  \Big\vert^2  \,,\nonumber\\[2mm]
%%%%%%%%
  &\Big\vert (\bar\upsilon^0
  \stackrel{\leftrightarrow}{\mathcal{D}}{\!}_\mu x^0 ) - \displaystyle{
    \frac{\bar\upsilon^0 } {L^0+\tfrac12 x^0}}\, (\upsilon^0
  \stackrel{\leftrightarrow}{\mathcal{D}}{\!}_\mu \bar\upsilon^0)
  \Big\vert^2  \nonumber\\
  &\qquad    =-2\, (L^0)^2 \big(\mathcal{D}_\mu
  L^p{}_q{}^0\,\mathcal{D}^\mu L^q{}_p{}^0\big) 
  -\big(  L^p{}_q{}^0\,\mathcal{D}_\mu L^q{}_p{}^0\big)^2 \,.
\end{align}
The right-hand side of these expressions is manifestly invariant under
the emergent $3D$ $\mathrm{SU}(2)$ R-symmetry, as is to be expected
because the expressions on the left-hand side are invariant under the
the $4D$ $\mathrm{U}(1)$ R-symmetry. Collecting the various terms one
can verify that all the terms quadratic in the derivatives of the scalar
fields combine into the following form,
\begin{equation}
  \label{eq:DL-DL}
  \mathcal{L}_\mathrm{vector}\big\vert_{3D} =  \tfrac12 e \,\mathcal{F}_{AB}(L)
  \;\mathcal{D}_\mu L^p{}_q{}^A \,\mathcal{D}^\mu  L^q{}_p{}^B \,. 
\end{equation}

What remains to evaluate are the terms linear in the field
strengths. Collecting those terms gives rise to
\begin{align}
  \label{eq:F-Ds}
  \mathrm{i} \varepsilon^{\mu\nu\rho} \,N_{\Lambda\Sigma} \bigg[&
  -\tfrac14\mathrm{i}\, \phi^{-2}  
   \big(X^\Lambda \stackrel{\leftrightarrow}{\mathcal{D}}{\!}_\mu\bar X^\Sigma \big) \,
   F(W)_{\nu\rho}{}^0 \nonumber\\
  &\; + \tfrac18  \phi^{-3}  \Big(\bar
  X^\Lambda\, \Big[ (\bar\upsilon^0
  \stackrel{\leftrightarrow}{\mathcal{D}}{\!}_\mu x^0 ) -
  \displaystyle{ \frac{\bar\upsilon^0 } {L^0+\tfrac12 x^0}}\,
  (\upsilon^0 \stackrel{\leftrightarrow}{\mathcal{D}}{\!}_\mu
  \bar\upsilon^0)   \Big] +  \mathrm{h.c} \Big)\nonumber\\
  &\qquad\qquad \times 
  \big(F(W)_{\nu\rho}{}^\Sigma + \tfrac12\phi^{-2}
  L^p{}_q{}^\Sigma\,L^q{}_p{}^0
  F(W)_{\nu\rho}{}^0\big) \bigg] \nonumber\\[2mm]
  + \tfrac18\mathrm{i} \varepsilon^{\mu\nu\rho}\,R_{\Lambda\Sigma} 
  &\big[F(W)_{\mu\nu}{}^\Lambda +\tfrac12 \phi^{-2}  L^p{}_q{}^\Lambda\,L^q{}_p{}^0\, 
   F(W)_{\mu\nu}{}^0 \big] \,\partial_\rho \big(\phi^{-2}
   L^r{}_s{}^\Sigma\,L^s{}_r{}^0\big) \,.
\end{align}
These terms can be rewritten by using identities similar to the ones
given in \eqref{eq:X-A+and more}, which lead to the following expression,
\begin{align}
  \label{eq:F-Ds-2}
  \mathrm{i} \varepsilon^{\mu\nu\rho} \,N_{\Lambda\Sigma} \bigg[&  
  \frac1{16\, (L^0)^3}   \Big(L^p{}_q{}^0\,
  L^q{}_r{}^\Lambda\, \mathcal{D}_\mu  L^r{}_p{}^\Sigma  -
  \frac{3\,L^s{}_t{}^\Lambda L^t{}_s{}^0}{2\, (L^0)^2} \,L^p{}_q{}^\Sigma \, 
  L^q{}_r{}^0\, \mathcal{D}_\mu  L^r{}_p{}^0  \Big) 
   F(W)_{\nu\rho}{}^0  \nonumber\\
  &\;   - \frac1{8\, (L^0)^3} \,L^p{}_q{}^\Lambda \,  L^q{}_r{}^0\,
  \mathcal{D}_\mu  L^r{}_p{}^0 \,
  F(W)_{\nu\rho}{}^\Sigma  \bigg]  \nonumber\\[2mm]
  + \tfrac18\mathrm{i} \varepsilon^{\mu\nu\rho}\,R_{\Lambda\Sigma}\,  
  &\,\partial_\mu \Big(\frac{ L^p{}_q{}^\Lambda\,L^q{}_p{}^0}{(L^0)^2}
  \Big) \, \Big[F(W)_{\nu\rho}{}^\Sigma
   +\frac{L^s{}_t{}^\Sigma\,L^t{}_s{}^0}{2\,(L^0)^2}   \,  
   F(W)_{\nu\rho}{}^0 \Big] \,. 
\end{align}
This expression is {\it manifestly} invariant under {\it local}
$\mathrm{SU}(2)$ transformations as well as under gauge
transformations of the fields $W_\mu{\!}^A$. This is in contrast with
the situation encountered for generic tensor multiplets discussed in
subsection \ref{subsubsection:tensor-multiplet-lagr}, where we argued
that this is not the case in general (see, in particular the
discussion related to the equations
\eqref{eq:bianchi-partial-L2}-\eqref{eq:su2-var-expl-2-form}).  Hence
we conclude that the models obtained by dimensional reduction from $4D$
vector multiplets belong to a restricted class.  As we shall discuss
in the next subsection \ref{sec:c-map}, this implies that certain
tensor multiplet models are not in the image of the c-map. This
does not come as a surprise as such a phenomenon has been
noted earlier for hypermultiplets \cite{Cecotti:1988qn}. 

It remains to verify explicitly that \eqref{eq:F-Ds-2} has the same
structure as the two last terms in
\eqref{eq:3D-lagrangian-tensor}. Let us therefore first extract the
terms proportional to the $\mathrm{SU}(2)$ connections
$\mathcal{A}_\mu{}^p{}_q$. We note that the covariant derivatives in
\eqref{eq:F-Ds-2} appear in the form $\mathrm{tr}\big[ L_1
\,L_2\,\mathcal{D}_\mu L_3 \big]$, so that the terms proportional to
the gauge connection $\mathcal{A}_\mu$ take the form
\begin{equation}
  \label{eq:covariantization F-W-A}
  \tfrac12 \mathrm{tr}\big[ L_1 \,L_2\,[\mathcal{A}, L_3] \big]=\tfrac12
  \mathrm{tr}\big[L_1\,L_3\big]\,
  \mathrm{tr}\big[L_2\,\mathcal{A}_\mu\big]
  -\tfrac12 \mathrm{tr}\big[L_2\,L_3\big] \,
  \mathrm{tr}\big[L_1\,\mathcal{A}_\mu \big] \,, 
\end{equation}
where we have used that $L_1$, $L_2$, $L_3$ and $\mathcal{A}_\mu$ are traceless,
anti-hermitian two-by-two matrices. Collecting the various terms from
\eqref{eq:F-Ds-2} linear in the connection is now straightforward and
leads to 
\begin{equation}
  \label{eq:linear-A-term}
  \mathcal{L}\big\vert_{3D} = -\tfrac12\mathrm{i}\,
  \varepsilon^{\mu\nu\rho} \,\mathcal{F}_{AB}(L) \,F(W)_{\mu\nu}{\!}^A
  \,L^p{}_q{\!}^B\, \mathcal{A}_\rho{}^q{}_p\,  \,.
\end{equation}
This term takes exactly the same form as the corresponding term in the
Lagrangian \eqref{eq:3D-lagrangian-tensor}. 

Finally we have to show that the terms in \eqref{eq:F-Ds-2} with an
ordinary derivative are equal to 
\begin{equation}
  \label{eq:two-form-vector}
  \mathcal{L}\big\vert_{3D} = \mathrm{i}\,
  \varepsilon^{\mu\nu\rho} \, \mathcal{F}_{ABC}{}^p{}_q \,\partial_\mu L^q{}_r{\!}^A 
  \,\partial_\nu L^r{}_p{\!}^B \, W_\rho{\!}^C   \,,
\end{equation}
upon adding a total derivative. In this way the terms in
\eqref{eq:F-Ds-2} that involve $R_{\Lambda\Sigma}$ can be written
such that they become proportional to $\partial_\mu
R_{\Lambda\Sigma}$ times a bare gauge field. Making use of
\eqref{eq:F-der-times-L0} one then derives the following identity,
\begin{align}
  \label{eq:closed-R2}
   &\partial_{[\mu} R_{\Lambda\Sigma} \;\partial_{\nu]} 
  \bigg(\frac{ L^p{}_q{}^\Lambda\,L^q{}_p{}^0}{(L^0)^2}
  \bigg)  \nonumber\\[1mm]
  &=\frac1{(L^0)^2}\,\frac{\partial R_{\Lambda\Sigma}}{\partial L^p{}_q{\!}^\Xi}
  \bigg[ \partial_{[\mu} L^p{}_q{\!}^\Lambda
  \, \partial_{\nu]}L^r{}_s{\!}^\Xi \; L^s{}_r{\!}^0 + 
  \partial_{[\mu} L^p{}_q{\!}^\Lambda \, \partial _{\nu]}
  L^r{}_s{\!}^0 \; L^s{}_r{\!}^\Xi \nonumber \\[1mm]
  &\,\qquad + \frac{ L^t{}_u{}^\Lambda\,L^u{}_t{}^0}{2(L^0)^2} 
  \bigg(\partial_{[\mu} L^p{}_q{\!}^0
  \, \partial_{\nu]}L^r{}_s{\!}^\Xi \; L^s{}_r{\!}^0 
  +\partial_{[\mu} L^p{}_q{\!}^0
  \, \partial_{\nu]}L^r{}_s{\!}^0 \; L^s{}_r{\!}^\Xi
  +2\, \partial_{[\mu} L^p{}_q{\!}^\Xi 
  \, \partial_{\nu]}L^r{}_s{\!}^0\; L^s{}_r{\!}^0 \bigg)
  \nonumber\\[1mm]
   &\,\qquad +\frac{ L^t{}_u{}^\Lambda\,L^u{}_t{}^0}{2(L^0)^2} \, \frac{
     L^v{}_w{}^\Xi\,L^w{}_v{}^0}{(L^0)^2}  \; \partial_{[\mu} L^p{}_q{\!}^0 
  \, \partial_{\nu]}L^r{}_s{\!}^0\; L^s{}_r{\!}^0  \bigg]\,. 
\end{align}
Since $\mathcal{F}_{ABC}{}^p{}_q$ is defined in terms of
$N_{\Lambda\Sigma}$ and its derivatives, we have to convert these
terms so that the result is either proportional to $N_{\Lambda\Sigma}$
or to its derivative. This can be achieved by making use of 
\eqref{eq:F-der-times-L0}, from which one derives 
\begin{equation}
  \label{eq:der-R-to der-N}
  \frac{\partial R_{\Lambda\Sigma}}{\partial L^q{}_p{}^\Xi}   =
  \frac1{L^0} \, \frac{\partial N_{\Lambda\Sigma}(X)}{\partial L^r{}_p{}^\Xi}
  \,L^r{}_q{\!}^0  = - \frac1{L^0} \, L^p{}_r{\!}^0\, \frac{\partial N_{\Lambda\Sigma}}{\partial
    L^q{}_r{}^\Xi}  \,. 
\end{equation}
With the above results, upon using \eqref{eq:0-Xi-derivative}, 
\eqref{eq:FX-holo-su}, and 
\begin{align}
  \label{eq:procuct-index-rearr}
  L^{p}{}_q{}^{[A} \, L^{r}{}_s{}^{B]}   = -\tfrac12 \delta^p{}_s \,
  L^r{}_t{}^{[A}\, L^t{}_q{}^{B]}  
  +\tfrac12 \delta^r{}_q \,  L^{p}{}_t{}^{[A} \, L^t{}_s{}^{B]}  \,, 
\end{align}
to rearrange the various contractions of $\mathrm{SU}(2)$ indices,
one can verify that all terms lead indeed to
\eqref{eq:two-form-vector}.

Combining the various results derived in this subsection, the
resulting $3D$ vector multiplet Lagrangian reads as follows,
\begin{align}
  \label{eq:3D-lagrangian-vector}
  \mathcal{L}_{\mathrm{vector}}\big\vert_{3D} = &\,
  \tfrac12 e \,\mathcal{F}(L)_{AB} \;\mathcal{D}_\mu L^p{}_q{}^A \,
  \mathcal{D}^\mu  L^q{}_p{}^B  -\tfrac12   e\,\mathcal{F}(L)_{AB} \,
     L^p{}_q{}^A\,L^q{}_p{}^B\, \big(\tfrac12 R -D - C^2\big)\nonumber\\
 &\, -\tfrac12 e\, \mathcal{F}(L)_{AB}  \big[
  F(W)_{\mu\nu}{}^A\,F(W)^{\mu\nu B}  +   Y^i{}_j{}^A\,Y^j{}_i{}^B  \big]
 \nonumber\\
   &\, -\tfrac12\mathrm{i}\,
  \varepsilon^{\mu\nu\rho} \,\mathcal{F}(L)_{AB} \,F(W)_{\mu\nu}{\!}^A
  \,L^p{}_q{\!}^B\, \mathcal{A}_\rho{}^q{}_p
   \nonumber\\
   &+\mathrm{i}\,
  \varepsilon^{\mu\nu\rho} \, \mathcal{F}(L)_{ABC}{}^p{}_q \,
  \partial_\mu L^q{}_r{\!}^A \,\partial_\nu L^r{}_p{\!}^B \, W_\rho{\!}^C  \,,
\end{align}
which coincides with that of the tensor Lagrangian
\eqref{eq:3D-lagrangian-tensor}, except that the $\mathrm{SU}(2)$
indices $i,j,\ldots$ have been interchanged by $p,q,\ldots$, and the
term proportional to the field $D$ has changed sign. Note that the
above Lagrangian does not represent the most general Lagrangian of
this type. First of all \eqref{eq:3D-lagrangian-vector} can be written
in a form that is manifestly invariant under both the gauge
transformations associated with the gauge fields $W_\mu{\!}^A$ and the
local $\mathrm{SU}(2)$ transformations, as follows from
\eqref{eq:F-Ds-2}. Secondly, this Lagrangian is invariant under
the $n+1$ rigid abelian transformations noted in
\eqref{eq:L-L)-invariance}.\footnote{ %%%%%%%%%%%%%%
  From the previous results in this subsection, the reader can verify
  that this is indeed the case.  In fact the Lagrangian is expected to
  have more rigid symmetries but those are ignored here.
} %%%%%%%%%%%%
Both these properties are characteristic for dimensionally reduced
$4D$ vector multiplet Lagrangians and are not generic for these $3D$
couplings.

Just as for the tensor multiplets (c.f. \eqref{eq:F-second-derivatives}) a function
$\mathcal{F}(x, \upsilon, \bar{\upsilon})$ should exist such that  
\begin{equation}
  \label{eq:F-second-derivatives-vector}
  \frac{\partial^2 \mathcal{F}(x,\upsilon,\bar\upsilon)}{\partial
    x^A\,\partial x^B} = 
  -\frac{\partial^2 \mathcal{F}(x,\upsilon,\bar\upsilon)}{\partial\upsilon^A\,\partial
    \bar\upsilon^B} = \mathcal{F}_{AB} \,, 
  \qquad  \frac{\partial^2 F(x,\upsilon,\bar\upsilon)}{\partial
    x^A\,\partial\upsilon^B} = \frac{\partial^2 F(x,\upsilon,\bar\upsilon)}{\partial
    x^B\,\partial\upsilon^A} \,.
\end{equation}
The function $\mathcal{F}$ can be expressed in terms of the function $F(X)$ that
encodes the $4D$ vector multiplet Lagrangian and it takes the
following form,
\begin{equation}
  \label{eq:non-symm-function}
  \mathcal{F}(x, \upsilon, \bar{\upsilon})= -8\,L^0 \;\mathrm{Im} \bigg[
  \frac{F\big(X(L)\big)}{(\bar{\upsilon}^0)^2}\bigg] \,,
\end{equation}
where $X^\Lambda(L)$ is defined by \eqref{eq:X(L)}. Clearly this
function is homogeneous of degree $+1$ and it is also manifestly
invariant under the shift transformations
\eqref{eq:L-L)-invariance}. Note, however, that it is not invariant
under the full $\mathrm{SU}(2)$ R-symmetry group, but only under its
$\mathrm{U}(1)$ subgroup. One can explicitly show that this function
indeed satisfies the differential equations
\eqref{eq:F-second-derivatives-vector}. Alternatively one can show
that \eqref{eq:non-symm-function} satisfies the relation
\begin{equation}
  \label{eq:vector-potential-2}
  \chi_{\mathrm{vector}}(L) =  -
  \mathcal{F}(\upsilon,\bar \upsilon, x) + x^A \,\frac{\partial
    \mathcal{F}(x,\upsilon,\bar \upsilon)}{\partial x^A} \,, 
\end{equation} 
which is the exact analogue of \eqref{eq:tensor-potential-2}. To prove
this result we note the following useful equations,
\begin{equation}
  \label{eq:x-partial x}
  x^A \,\frac{\partial}{\partial x^A} \,\frac{F(X(L))}{(\bar\upsilon^0)^2}
  = - \frac1{2\,(L^0)^2}\bigg[ \bar X^\Lambda F_\Lambda- 
  \frac{\vert\upsilon^0\vert^2}{(\bar\upsilon^0)^2} \,X^\Lambda
  F_\Lambda\bigg] \,,
\end{equation}
which follows upon using \eqref{eq:variation-X-generic},
\eqref{eq:X(L)} and \eqref{eq:FLL-NXX}. 

Obviously the function \eqref{eq:non-symm-function} is singular when
imposing the $\mathrm{SU}(2)$ gauge condition $\upsilon^0=0$. In that
case we have $X^\Lambda=\tfrac12\mathrm{i}\bar\upsilon^\Lambda$, and
the role of the function \eqref{eq:non-symm-function} is taken over by
a different function,
\begin{equation}
  \label{eq:fixed-no-symm-function}
  \hat{\mathcal{F}}(x,\upsilon,\bar\upsilon)= 
 \frac{ N_{\Lambda\Sigma} \big[ x^\Lambda x^\Sigma -2\,
  \upsilon^\Lambda \bar\upsilon^\Sigma\big]} {4\,x^0}  \,, 
\end{equation}
which satisfies the same equations
\eqref{eq:F-second-derivatives-vector} for $A,B= \Lambda,\Sigma$,  as
well as
\begin{equation}
  \label{eq:F-gauge-second}
    \frac{\partial^2 \hat{\mathcal{F}}(x,\upsilon,\bar\upsilon)}{\partial
    x^0\,\partial x^0} = \mathcal{F}_{00} \,.
\end{equation}
In \cite{deWit:2001dj} the result \eqref{eq:fixed-no-symm-function}
was noted in the case of rigid supersymmetry (where $x^0$
equals a constant) for the c-map between vector and tensor
multiplets. It was later extended to local
supersymmetry in \cite{Rocek:2005ij,Rocek:2006xb}. Note, however, that
the general context in \cite{deWit:2001dj,Rocek:2005ij,Rocek:2006xb}
is somewhat different than in this paper as it is primarily directed
towards the study of hypermultiplets.

Here we should add that different functions
$\mathcal{F}(x,\upsilon,\bar\upsilon)$ (as well as
$\hat{\mathcal{F}}(x,\upsilon,\bar\upsilon)$) will correspond to
different Lagrangians that can, however, still describe the same theory, as we
can deduce from the existence of electric-magnetic duality of the $4D$ vector multiplet
Lagrangians. An analogous situation exists for the $4D$ tensor
Lagrangians because of 'tensor-tensor' duality \cite{deWit:2001dj}
(the existence of such tensor dualities is now also implied by the c-map).

%%%%%%%%%%%%%%%%%%%%%%%%%%%%%%%%%%%%%%%%%%%%%%%%%%%%%%%%%
\subsection{The c-map}
\label{sec:c-map}
%%%%%%%%%%%%%%%%%%%%%%%%%%%%%%%%%%%%%%%%%%%%%%%%%%%%%%%%%
We have now determined the $3D$ Lagrangian for systems of
hypermultiplets, tensor multiplets and vector multiplets quadratic
in space-time derivatives. As noted in subsection
\ref{subsubsection:hyper-multiplet-lagr}, there
exist two different hypermultiplets, distinguished by the fact that
their scalar sections, $A_i{}^\alpha$ and $\tilde A_p{}^\alpha$,
transform under different $\mathrm{SU}(2)$ factors of the R-symmetry
group. Their corresponding Lagrangians are given in
\eqref{eq:3D-hyper-L} and \eqref{eq:3D-hyper-L-reflected}. Let us then
summarize the terms in the combined Lagrangian that contain the Ricci
scalar, the two auxiliary fields of the superconformal multiplet, $C$
and $D$, as well as the kinetic terms of the scalars of the various
supermultiplets,
\begin{align}
  e^{-1}{\cal L} =&\, 
  \tfrac14\,( \chi_\mathrm{hyper} + \chi_\mathrm{tensor} +
  \tilde\chi_\mathrm{hyper} +\chi_\mathrm{vector})\,
  (\tfrac12 R - C^2) \nonumber\\[1mm]
  &\, +\tfrac14\,( \chi_\mathrm{hyper} + \chi_\mathrm{tensor}
  -\tilde\chi_\mathrm{hyper} -  \chi_\mathrm{vector})\,D
  \nonumber\\[1mm]
  &\, 
   -\tfrac12\, \Omega_{\alpha\beta}\, \varepsilon^{ij} \,
  {\cal D}_\mu A_i{}^\alpha\, {\cal D}^\mu A_j{}^{\beta}  
   -\tfrac12\, \tilde\Omega_{\alpha\beta}\, \varepsilon^{ij} \,
  {\cal D}_\mu \tilde A_i{}^\alpha\, {\cal D}^\mu \tilde A_j{}^{\beta}
  \nonumber\\[1mm] 
     &\,
    +\tfrac{1}{2}  F_{IJ} \;\mathcal{D}_\mu L^i{}_j{}^{I} \,
     \mathcal{D}^\mu L^j{}_i{}^{J} + \tfrac12 \mathcal{F}_{AB}
  \;\mathcal{D}_\mu L^p{}_q{}^A \,\mathcal{D}^\mu  L^q{}_p{}^B \,. 
\end{align}
Here we made use of the vector and tensor potentials as well as the
hyperk\"ahler potentials, which are homogeneous in the scalar fields and invariant
under R-symmetry,
\begin{align}
  \chi_\mathrm{hyper} = &\, \tfrac12\, \Omega_{\alpha\beta}\,
  \varepsilon^{ij}\, A_i{}^\alpha\, A_j{}^{\beta}\,,
  \nonumber\\
  \tilde\chi_\mathrm{hyper} = &\, \tfrac12\, \tilde\Omega_{\alpha\beta}\,
  \varepsilon^{pq}\, \tilde A_p{}^\alpha\, \tilde A_q{}^{\beta}\,,
  \nonumber\\
  \chi_\mathrm{tensor} = &\, -2 \, F_{IJ} \, L^i{\!}_j{}^I \,
  L^j{\!}_i{}^J\,,
  \nonumber\\
  \chi_\mathrm{vector} = &\, -2\, \mathcal{F}_{AB} \, L^p{\!}_q{}^A\,
  L^q{\!}_p{}^B\,.
\end{align} 

The above equations represent the generic situation in three
dimensions.\footnote{ %%%%%%%%%%
  For simplicity we are ignoring the option of partially performing
  vector-scalar dualities in which case one obtains an (on-shell)
  Lagrangian that consists of vector multiplets and hypermultiplets
  with mutual interactions beyond the ones induced by the coupling to
  the fields of the superconformal theory.  } %%%%%%%%%%%%%%%%%%%
It seems that there is a symmetric situation between the two sectors
corresponding to $(\chi_\mathrm{hyper}+ \chi_\mathrm{tensor})$ and
$(\tilde\chi_\mathrm{hyper}+ \chi_\mathrm{vector})$, which involves
also the reflection \eqref{eq:su2-interchange} noted for the $3D$
superconformal multiplet.  While we have obtained these results by
dimensional reduction from four dimensions, starting with vector and
tensor multiplets and only one type of hypermultiplets, one may now
consider the inverse procedure and ask which of these $3D$ theories
can be uplifted to four dimensions. A special subclass then consists
of those theories that can be uplifted to $4D$ in two different ways,
meaning that the $3D$ Lagrangian and its dual one with respect to the
reflection \eqref{eq:su2-interchange} can both be uplifted. In that
case there will exist two {\it inequivalent} $4d$ Lagrangians that
yield the same $3D$ theory upon dimensional reduction. Henceforth we
will concentrate on this subclass.

To have two possible uplifts, the $3D$ Lagrangian must obviously
satisfy a number of restrictions. As already explained, under
dimensional reduction as carried out in this paper, the vector
multiplet Lagrangian is of a restricted type. This implies that the
alternative uplift to four space-time dimensions is only possible when
also the tensor multiplet Lagrangian belongs to this restricted
class. A similar argument applies to the hypermultiplets. Since the
hypermultiplet Lagrangian associated with the hyperk\"ahler potential
$\tilde\chi_\mathrm{hyper}$ cannot be obtained directly by dimensional
reduction from $4D$ hypermultiplets, it can only emerge via
vector-scalar duality from the vector sector. Hence to have two
alternative uplifts to $4D$ the two hyperk\"ahler Lagrangians should
both be such that they can be obtained from scalar-vector duality from
a restricted $3D$ vector Lagrangian. When dualizing $n+2$ vector
multiplets one obtains a hyperk\"ahler cone of quaternionic dimension
$n+2$ with $2n+3$ tri-holomorphic abelian isometries.\footnote{ %%%%%%
  Note that for the on-shell theory the corresponding
  quaternion-K\"ahler manifold of quaternionic dimension $n+1$ has
  only $n+2$ commuting quaternionic abelian isometries
  \cite{deWit:1992wf}.  } %%%%%%%%%%%%%%%%%%%%%%%%%%%%

If one of the inequivalent $4D$ Lagrangians has $n_\mathrm{v}$
(off-shell) vector multiplets and $n_\mathrm{t}$ (off-shell) tensor
multiplets (ignoring the hypermultiplets for convenience), then the
other uplift should have $n_\mathrm{t}-1$ (off-shell) vector and
$n_\mathrm{v}+1$ (off-shell) tensor multiplets (so that the total
number of off-shell vector and tensor multiplets in $3D$ equals
$n_\mathrm{v} +n_\mathrm{t}+1$).  Obviously we have the condition that
there must at least be one off-shell tensor supermultiplet in either
one of the two inequivalent $4D$ Lagrangians!  The map between these
two inequivalent $4D$ theories is known as the c-map. From the
perspective of the $10D$ IIA and IIB supergravities compactified on
the circle $S^1$ times a six-dimensional internal manifold that
preserves eight supersymmetries, the resulting reduction to $3D$ leads
to a Lagrangian that can be uplifted to $4D$ in two different
ways. Those will then correspond to the compactified IIA and the
compactified IIB theories. In string theory these two theories emerge
in the compactification of type-II string theory on a circle in its
two decompactification limits, where either the momentum modes or the
winding modes become massless. Hence this property of
lower-dimensional matter-coupled supergravities can thus be seen as a
consequence of T-duality for type-II strings \cite{Cecotti:1988qn}.

%%%%%%%%%%%%%%%%%%%%%%%%%%%%%%%%%%%%%%%%%%%%%%%%%%%%%%%%
\section{The c-map for higher-derivative couplings}
\label{sec:high-deriv-coupl}
\setcounter{equation}{0}
%%%%%%%%%%%%%%%%%%%%%%%%%%%%%%%%%%%%%%%%%%%%%%%%%%%%%%%%%%%%%
The off-shell reduction scheme introduced in this paper can be
straightforwardly applied to higher-derivative
Lagrangians. Higher-derivative couplings in $4D$ can be generated by
coupling a number of vector multiplets to the Weyl multiplet (its
covariant quantities constitute a chiral multiplet with the
anti-selfdual tensor $T_{ab}{}^{ij}$ as its lowest component, in the
same way as the covariant quantities of the vector multiplet define a
chiral multiplet with the holomorphic scalar $X$ as its lowest
component), by means of a chiral invariant
\cite{Bergshoeff:1980is}. To consider a similar coupling on the tensor
multiplet side is, however, more complicated, although this can be
handled by the standard technique of making use of composite
multiplets.  For instance, one can write an off-shell vector multiplet
in terms of off-shell tensor multiplets \cite{deWit:2006gn}, or an
off-shell tensor multiplet in terms of vector multiplets. Since these
composite multiplets contain two derivatives, their substitution into
a standard two-derivative Lagrangian will lead to four space-time
derivatives.  Another way to generalize higher-derivative couplings is
by making use of the so-called `kinetic multiplet', which leads in
principle to non-chiral invariants
\cite{deWit:2010za,Butter:2013lta,Katmadas:2013mma}.

For simplicity, we will first consider the Lagrangians derived in the
previous section and replace some of the elementary vector and/or
tensor multiplets by composite ones. In this way we will naturally 
obtain higher-derivative actions that can be uplifted to two
different $4D$ theories, which are thus related by the c-map. In the
next subsection we will first introduce the key formulae for these
composite multiplets. In the last subsection we will briefly consider the
coupling to a composite chiral multiplet consisting of the square of
the Weyl multiplet.

%%%%%%%%%%%%%%%%%%%%%%%%%%%%%%%%%%%%%%%%
\subsection{Higher derivative couplings through composite matter multiplets}
\label{sec:high-matter}
%%%%%%%%%%%%%%%%%%%%%%%%%%%%%%%%%%%%%%%%
In order to discuss higher-derivative actions for matter multiplets it
is convenient to introduce some elements of the multiplet calculus
known in four dimensions, which can be straightforwardly reduced to
three dimensions, using the formulae in section
\ref{sec:4d-fields-terms}.

In four dimensions, one may construct composite vector multiplets
out of a set of tensor multiplets \cite{deWit:2006gn}. The starting
point is the lowest-weight components of the vector multiplets, the
complex scalars $X^\mathrm{comp}$, which take the form
\begin{equation}
  \label{eq:comp-X}
  X^\mathrm{comp}=  f(L)_{I} \,\bar G^I  
  +f(L)_{IJ}{}^{ij} \,\bar\varphi^I_i \varphi^J_j\,,
\end{equation}
where the $f(L)_{I}$ are functions of the tensor multiplet scalars
$L_{ij}{}^{I}$, which are homogeneous of degree $-1$. The
$f(L)_{IJ}{}^{ij}$ then denote their derivatives with respect to
$L_{ij}{}^J$, and $G^I$ and $\varphi^I$ denote the auxiliary fields
and the spinor fields of the $4D$ tensor multiplets. The functions
$f_I$ are subject to two additional constraints, namely
\begin{equation}
  \label{eq:chiral-constraints}
  f_{IJ}{}^{ij}= f_{JI}{}^{ij}\,,\qquad 
   \varepsilon^{jk}\; \frac{\partial f_{IJij}}{\partial
     L^{kl K}} =0\,.
\end{equation}
These constraints are similar to the ones noted in subsection
\ref{subsubsection:tensor-multiplet-lagr} for the function $F_{IJ}$
that appears in the $4D$ tensor multiplet action. In components their
solution takes the form \eqref{eq:F-second-derivatives} upon
suppressing the first index $I$.  The functions $f(L)_I$ must be
invariant under the $4D$ $\mathrm{SU}(2)$ R-symmetry group, so that
the composite scalar $X^\mathrm{comp}$ transforms as as a proper $4D$
chiral multiplet scalar.

The remaining components of the composite multiplet are then
identified straightforwardly upon considering consecutive
supersymmetry variations. As an example we present the expression for
the composite spinor associated with the composite vector multiplet,
\begin{align}
  \label{eq:vector-spinor-comp}
  \Omega_{i}{\!}^\mathrm{comp} =&\, -2 \,f_{I} \big[ \Slash{D}
  \varphi_i{}^I + 3 \,L_{ij}{\!}^I\chi^j + \tfrac18 \varepsilon_{ij} \, T_{AB}{}^{jk}\,
  \gamma^{AB}  \varepsilon_{kl} \,\varphi^{l\,I} \big] \nonumber\\
  &\,+ 2\, {f}_{IJij} \, \bar G^I\, \varphi^{jJ}  -  2\,
    {f}_{IJ}{}^{kl} \,\big(\Slash{D} L_{ik}{}^I -\varepsilon_{ik}\,
    \Slash{E}^I \big) \varphi_l{}^J   
    + 2\, {f}_{IJKij}{}^{kl}  \;\varphi^{jK}
    \,\bar\varphi_k{}^I\varphi_l{}^J  \,,
\end{align} 
where $f_{IJKij}{}^{kl} = \partial^2 f_I/ \partial L^{ij J}\,\partial L_{kl}{}^K$. 

Also the reverse situation is possible, and one may construct a
four-dimensional composite tensor multiplets out of a set of vector
multiplets. In this case, the lowest-weight component is an $\mathrm{SU}(2)$
triplet of scalars $L_{ij}{\!\!}^\mathrm{comp}$, which is given by
\begin{align}  
  \label{eq:comp-L}
  L_{ij}{\!\!}^\mathrm{comp} =&\,  g (X)_{\Lambda}\, Y_{ij}{}^\Lambda -
  \tfrac12 g(X)_{\Lambda \Sigma}\, \bar{\Omega}_{(i}{\!}^\Lambda
  \Omega_{j)}{\!}^\Sigma \nonumber\\
  &\, + \varepsilon_{ik} \varepsilon_{jl}\, \big[\bar g_{\Lambda}(\bar X)\,
  Y^{kl\,\Lambda}  - 
  \tfrac12 \bar g(\bar X)_{\Lambda \Sigma}\,
  \bar{\Omega}^{(k\,\Lambda} \Omega^{l)\,\Sigma} \big] \,,  
\end{align}
where the $g_{\Lambda}(X)$ are holomorphic functions of the vector
multiplet scalars $X^{\Lambda}$ which are homogeneous of zeroth
degree. The $g_{\Lambda\Sigma}$ denote the derivative of $g_\Lambda$
with respect to $X^\Sigma$. Again there is a constraint on the
derivatives of the functions $g_{\Lambda}$,
\begin{equation}
  \label{eq:g-symm}
  g_{\Lambda\Sigma} = g_{\Sigma\Lambda}\,, 
\end{equation}
which implies that the $g_\Lambda$ can be expressed in terms of a
derivative of a holomorphic function, $g_\Lambda= \partial g/\partial
X^\Lambda$. Just as before the remaining components of this multiplet
will follow from applying consecutive supersymmetry variations of
\eqref{eq:comp-L}. As an example we present the corresponding
expression for the composite spinor,
\begin{align}
  \label{eq:varphi-comp}
    \varphi_i{\!}^\mathrm{comp} = &\,(g_\Lambda+\bar g_\Lambda) \, \Slash{D}
    \Omega_i{}^\Lambda  - \Slash{D} g_{\Lambda} \,
    \Omega_i{}^{\Lambda} \nonumber\\
    &\, + \tfrac12\,\bar g_{\Lambda\Sigma} \,  Y_{ij}{}^\Lambda \,\Omega^{j\,\Sigma} 
      - \tfrac1{4} \varepsilon_{ij} \,\bar g_{\Lambda\Sigma}\big(  F_{AB}{}^\Lambda
      -\tfrac14 X\,T_{ABkl}\,\varepsilon^{kl} \big) \gamma^{AB} 
      \Omega^j{}^\Sigma  \nonumber \\
  & + \tfrac1{64} \bar g_{\Lambda\Sigma\Xi} \; \varepsilon_{ij}
  \,\gamma_{AB}  \Omega^{j\Lambda}\,  \varepsilon_{kl} \, \bar{\Omega}^{k\Sigma}
  \gamma^{AB}  \Omega^{l\Xi}  \,.
\end{align}
The underlying reason why this construction works is related to the
fact that the equations of motion associated with a (two-derivative)
vector multiplet Lagrangian transform as a tensor multiplet, and vice
versa. This remains true in a superconformal background. The
conditions \eqref{eq:chiral-constraints} and \eqref{eq:g-symm} can be
understood in this perspective: when these conditions hold one can
construct invariant Lagrangians based on such functions.

With the above results one can in principle obtain the corresponding
$3D$ composite multiplets by applying the dictionary given in section
\ref{sec:4d-fields-terms} to all the bosonic components of the
composite multiplets. 

Starting from the composite vector multiplet defined by
\eqref{eq:comp-X}, we write both sides of the equations in terms of
the corresponding components of the $3D$ multiplets. The relevant
functions $f(L)_I$ and $g_\Lambda$ are then written in terms of the
proper $3D$ fields. The $f_I$ are written in terms of the $3D$ of
the $L^i{}_j{}^I$, after extracting a uniform
factor $1/(2\,L^0)$,
\begin{equation}
  \label{eq:f-to-f}
  f(L^{4D})_I \longrightarrow  \frac1{2\,L^0}\, f(L^{3D})_I  \,. 
\end{equation}
The new function thus remains $\mathrm{SU}(2)$ invariant and homogeneous
of degree $-1$ in the $3D$ scalars.  This is all dictated by the
off-shell dictionary (see, in particular, \eqref{eq:D4to3vector} and
\eqref{eq:D4to3tensor}). It is then straightforward to obtain the
following results for the bosonic composite vector multiplet
components (suppressing their fermionic contributions),
\begin{align}
  \label{eq:comp-tens-3D}
  L^p{}_q{\!}^\mathrm{comp}  =&\,  f(L)_I \,Y^p{}_q{\!}^I \,,  \nonumber \\[2mm]
   F(W)_{\mu\nu}{\!}^\mathrm{comp}  =&\,  f_{IJ}{}_i{}^j \,
   \mathcal{D}_{[\mu}L^i{}_{k}{\!}^I \, 
   \mathcal{D}_{\nu]}L^k{}_{j}{\!}^J \tfrac12 f_I \, L^i{}_j{\!}^I \, 
   R(\mathcal{V})_{\mu\nu}{}^j{}_i   +\partial_{[\mu} \big[
   \mathrm{i}\,e^{-1}\varepsilon_{\nu]\rho\sigma}\, f_I \,
   F(E)^{\rho\sigma I} \big]   \, , \nonumber  \\[2mm]
   Y^i{}_{j}{\!}^\mathrm{comp} =&\, f_I\,\big[\mathcal{D}^{2}
   L^i{}_{j}{\!}^I + \tfrac12(\tfrac12 R + D - C^2) L^i{}_{j}{\!}^I
   \big] + f_{IJ}{}^k{}_l \,\mathcal{D}_\mu L^l{}_k{\!}^I
   \,\mathcal{D}^\mu L^i{}_j{\!}^J  \nonumber\\
   &\, +\tfrac12 f_{IJ}{}^i{}_j\,\big[Y^p{}_q{}^I\, Y^q{}_p{}^J +
   F(E)^{abI} F(E)_{ab}{}^J - \mathcal{D}_\mu L^k{}_l{\!}^I
   \,\mathcal{D}^\mu L^l{}_k{\!}^J\big]   \nonumber\\
   &\, + \tfrac12\mathrm{i}\,\varepsilon^{\mu\nu\rho}\, \big[
   f_{IJ}{}^i{}_k \mathcal{D}_\mu L^k{}_{j}{\!}^I - f_{IJ}{}^k{}_j
   \mathcal{D}_\mu L^i{}_{k}{\!}^I\big] F(E)_{\nu\rho}{\!}^J \,.
\end{align}
 
The derivation for the composite tensor multiplet proceeds along
similar lines, except that the Kaluza-Klein vector multiplet will now
also contribute, Hence the sum over the vector multiplets in
\eqref{eq:comp-L} and \eqref{eq:varphi-comp} will now include an
extra vector multiplet. The function $g$ is written in terms of the
fields $L^p{}_q{\!}^A$ (thus including the Kaluza-Klein scalar). The
degree of homogeneity is changed because we have to absorb a factor
$1/L^0$. This is all dictated by the off-shell dictionary (see, in
particular, \eqref{eq:D4to3vector} and \eqref{eq:D4to3tensor}). It is
then straightforward to obtain the conversion to 
\begin{equation}
  \label{eq:g-to-g}
  \big[g(X)+\bar g(\bar X)\big]_\Lambda   \longrightarrow \frac1{L^0}\, g(L^{3D})_A\, ,
\end{equation}
with $A=\Lambda,0$ and where 
\begin{equation}
  \label{eq:3g-A-Lambda}
  g(L^{3D})_A  =  \left\{ \begin{array}{l}  g(L^{3D})_\Lambda \;,\\[2mm]
      \displaystyle g(L^{3D})_\Sigma\, \frac{L^p{}_q{}^\Sigma\, L^q{}_p{\!}^0} {2\,
        (L^0)^2} \;.\end{array}\right. 
\end{equation}
As a consequence of \eqref{eq:3g-A-Lambda} the resulting expressions
will again be invariant under the transformations \eqref{eq:L-L)-invariance}.
We now present the bosonic components of the composite tensor multiplet,
converted to $3D$ and suppressing fermionic contributions,
\begin{align}
  \label{eq:comp-vector-3D}
  L^i{}_j{\!}^\mathrm{comp}  =&\,  g(L)_A \,Y^i{}_j{\!}^A \,,  \nonumber \\[2mm]
   F(E)_{\mu\nu}{\!}^\mathrm{comp}  =&\,  g_{AB}{}_p{}^q \, \mathcal{D}_{[\mu}L^p{}_{r}{\!}^A \,
   \mathcal{D}_{\nu]}L^r{}_{q}{\!}^B -\tfrac12 g_A \, L^p{}_q{\!}^A \, 
   R(\mathcal{A})_{\mu\nu}{}^q{}_p   +\partial_{[\mu} \big[
   \mathrm{i}\,e^{-1}\varepsilon_{\nu]\rho\sigma}\, g_A \,
   F(W)^{\rho\sigma A} \big]   \, , \nonumber  \\[2mm]
   Y^p{}_{q}{\!}^\mathrm{comp} =&\, g_A\,\big[\mathcal{D}^{2}
   L^p{}_{q}{\!}^A + \tfrac12(\tfrac12 R - D - C^2) L^p{}_{q}{\!}^A
   \big] + g_{AB}{}^r{}_s \,\mathcal{D}_\mu L^s{}_r{\!}^A
   \,\mathcal{D}^\mu L^p{}_q{\!}^B  \nonumber\\
   &\, +\tfrac12 g_{AB}{}^p{}_q\,\big[Y^i{}_j{}^A\, Y^j{}_i{}^B +
   F(W)^{abA} F(W)_{ab}{}^B - \mathcal{D}_\mu L^r{}_s{\!}^A
   \,\mathcal{D}^\mu L^s{}_r{\!}^B\big]   \nonumber\\
   &\, + \tfrac12\mathrm{i}\,\varepsilon^{\mu\nu\rho}\, \big[
   g_{AB}{}^p{}_r \mathcal{D}_\mu L^r{}_{q}{\!}^A - g_{AB}{}^r{}_q
   \mathcal{D}_\mu L^p{}_{r}{\!}^A\big] F(W)_{\nu\rho}{\!}^B \,.
\end{align}

The components of the composite vector and tensor multiplets clearly
share a common structure. Apart from the fact that the indices are
different (because they transform under a different $\mathrm{SU}(2)$
factor of the R-symmetry group) and that an additional Kaluza-Klein
vector multiplet has emerged in the composite tensor multiplet, the
only obvious difference is that the field $D$ appears with opposite
signs in \eqref{eq:comp-tens-3D} and \eqref{eq:comp-vector-3D}, which
is consistent with the reflection symmetry noted in
\eqref{eq:su2-interchange}. However, there is also another, more
implicit, difference associated with the field strengths
$F(W)_{\mu\nu}{\!}^\mathrm{comp}$ and
$F(E)_{\mu\nu}{\!}^\mathrm{comp}$. One can show that both of them
satisfy a Bianchi identity, which implies that there should exist
explicit expressions for the corresponding composite gauge fields
$W_\mu {\!}^\mathrm{comp}$ and $E_\mu{\!}^\mathrm{comp}$.  However, as
we have already noted when discussing the Lagrangians with two
derivatives in subsection \ref{sec:lagrangians}, the expression for
$W_\mu{\!}^\mathrm{comp}$ is in general {\it not} invariant under the
relevant $\mathrm{SU}(2)$ R-symmetry, whereas the expression for
$E_\mu {\!}^\mathrm{comp}$ will be manifestly invariant under the
relevant $\mathrm{SU}(2)$. This should not come as a surprise in view
of the fact that the composites can be associated with the field
equations belonging to some appropriately chosen Lagrangian. Since
$F(E)_{\mu\nu}{\!}^\mathrm{comp}$ is therefore a field equation
belonging to a vector multiplet Lagrangian,
$E_{\mu}{\!}^\mathrm{comp}$ will thus be manifestly
$\mathrm{SU}(2)$ invariant. For $W_\mu{\!}^\mathrm{comp}$ the
situation is different and it will not necessarily be $\mathrm{SU}(2)$
invariant. Whether or not this is the case will depend on the
functions $f(L)_I$ that one intends to use.

There is also another feature that is relevant, namely, as was already
alluded to above, the composite tensor multiplet components
\eqref{eq:comp-tens-3D} will necessarily be invariant under the
transformations \eqref{eq:L-L)-invariance}, whereas the vector
multiplet components will in general not be subject to such a
symmetry. Hence consistency with the c-map will requires that the
functions $f(L)_I$ will satisfy such a symmetry as well. Provided that
this is the case, one may construct the actions for vectors and
tensors by including both the elementary and a number of composite
multiplets in the way that was described in section
\ref{sec:lagrangians}, because from this construction there is no
difference between elementary and composite multiplets. One starts
from a $4D$ Lagrangian describing $n_\mathrm{v}$ elementary and
$\tilde n_\mathrm{v}$ composite vector multiplets (the latter
described in terms of $n_\mathrm{t}$ elementary tensor multiplets), and
a second Lagrangian describing $n_\mathrm{t}$ elementary tensor and
$\tilde n_\mathrm{t}$ composite tensor multiplets (the latter
expressed in terms of the $n_\mathrm{v}$ elementary vector
multiplets). This then leads to a $3D$ action which, under the
conditions described above, can then also be uplifted to the sum of
two $4D$ Lagrangians, one describing $n_\mathrm{t}-1$ elementary and
$\tilde n_\mathrm{t}$ composite vector multiplets (the latter
described in terms of $n_\mathrm{v}+1$ elementary tensor multiplets),
and a second one based on $n_\mathrm{v}+1$ elementary and
$\tilde n_\mathrm{v}$ composite tensor multiplets (the latter
expressed in terms of the $n_\mathrm{t}-1$ elementary vector multiplets).

We refrain from working out some of these theories in detail and leave
this to later work. It is clear that, by considering composite
multiplets that themselves depend on both elementary and composite
multiplets, one can successively construct interactions that will
involve even higher-order derivatives. In our next and last subsection
we will briefly discuss other higher-derivative Lagrangians and
describe some details about their reduction to three dimensions.

%%%%%%%%%%%%%%%%%%%%%%%%%%%%%%%%%%%%%%%%%%%%%%
\subsection{More higher-derivative couplings}
\label{sec:high-weyl}
%%%%%%%%%%%%%%%%%%%%%%%%%%%%%%%%%%%%%%%%%%%%%%
As is well-known there exists a larger class of $4D$ higher-derivative
actions for vector supermultiplets, possibly involving the Weyl
multiplet \cite{Bergshoeff:1980is, deWit:2010za,Butter:2013lta}. The
latter is a reduced chiral tensor multiplet, whose lowest-weight
component equals $\varepsilon_{ij}\,T_{AB}{}^{ij}$. In all cases it is
the square of the Weyl multiplet that enters, so that the resulting
multiplet is a composite chiral multiplet whose lowest-weight
component equals the composite scalar
$\hat{A}=(\varepsilon_{ij}\,T_{AB}{}^{ij})^2$. For the subsequent
discussion we also present the bosonic contributions to the
highest-weight component of this multiplet, which is denoted by $\hat
C$,
\begin{align}
  \label{eq:Weyl-back}
  \hat C =&\,  64\, {\cal R}(M)^{-CD}{\!}_{\!AB}\, {\cal
    R}(M)^-_{CD}{\!}^{\!AB}  + 32\, R({\cal V})^{-AB\,i}{}_j \,
   R({\cal V})^-_{AB}{}^{j}{}_i  \nonumber\\
   &\,- 32\, T^{AB\,ij} \, D_A \,D^CT_{CB\,ij} \,,
\end{align}
where ${\cal R}(M)^{-CD}{\!}_{AB}$ is a generalization of the
(anti-selfdual component) of the Weyl tensor. Since this composite
multiplet is a scalar chiral multiplet, it can be directly coupled to
vector multiplets as well as to (composite) tensor multiplets. A full
discussion of these couplings is outside the scope of the present
paper, and here we will mainly confine ourselves to a partial analysis
of square of the the Weyl multiplet upon its reduction to three
dimensions.

Using the dictionary in subsection \ref{sec:4d-fields-terms} we can
express the components $\hat A$ and $\hat C$ in terms of $3D$ fields
(suppressing fermionic contributions),
\begin{align}
  \label{eq:Weyl-back-red}
  \hat A   =&\, 
    -\frac4{(L^0)^4} \Big[ (\bar\upsilon^0
    \stackrel{\leftrightarrow}{\mathcal{D}}_a  x^0 ) -
           \displaystyle{ \frac{\bar\upsilon^0 } { L^0+\tfrac12 x^0 }}
           \,(\upsilon^0 \stackrel{\leftrightarrow}{\mathcal{D}}_a
           \bar\upsilon^0) \Big]^2 \,,
 \nonumber \\[2mm]
  \hat C =&\, 32\, \big[R^{\mu\nu}R_{\mu\nu}-\tfrac38\,R^2\big]
   +64\, (\mathcal{D}_\mu C)^2 + 48\, D^2
  \nonumber\\[2mm]
   &\,
   + 16\,\big[ 
       R({\cal V})^{\mu\nu\,i}{}_j^{~} \,  {R({\cal V})}_{\mu\nu}{}^{j}{}_i^{~}
   +2\,R({\cal A})^{\mu\nu\,p}{}_q^{~} \,  {R(\mathcal{A})}_{\mu\nu}{}^{q}{}_p^{~}   \big]
  \nonumber\\[1mm]
   &\,
   +\frac{32}{(L^0)^{2}}\, \big[{\cal D}^{\mu} {F^{\nu\rho\,0 }}\,
       {\cal D}_{\mu} F_{\nu\rho}{\!}^0  
   - 2\,{\cal D}_{\mu} F^{\mu\nu\,0} \,{\cal D}^{\rho} F_{\rho\nu}{\!}^0  \big]
   \nonumber\\
   &\,
   + \frac{32}{(L^0)^{2}}\,\big[ \mathcal{D}_\mu \,Y^{i}{}_j{\!}^{0}
   \,  \mathcal{D}^\mu Y^{j}{}_i{\!}^{0} 
   + \mathrm{i}e^{-1} \,\varepsilon^{\mu\nu\rho}\,L^0\,
   R(\mathcal{V}){}_{\mu\nu}{}^i{}_j \, \mathcal{D}_\rho Y^{j}{}_i{}^{0}
   \big]  + \cdots\,,
\end{align}
where in $\hat C$ we restricted ourselves to only some characteristic terms.

To elucidate this result let us consider the bosonic terms of the $4D$
superconformal action, which can be written as
\begin{equation}
  \label{eq:conformal-4D}
  \mathcal{L}^\mathrm{s.c.}\big\vert_{4D} = E\, \big[\hat C -\ft1{16}\, \hat
  A\,(T_{ABij}\,\varepsilon^{ij})^2\big]  +\mathrm{h.c.} \,,
\end{equation}
where the second term represents the bosonic contribution of the
chiral superspace measure. Upon its reduction to three dimensions, we
write the result as a linear combination of two terms,
\begin{equation}
  \label{eq:conformal-3D}
  e^{-1} \mathcal{L}^\mathrm{s.c.}\big\vert_{3D} = e^{-1}\mathcal{L}_1
  +e^{-1} \mathcal{L}_2 , 
\end{equation}
with 
\begin{align}
   \label{eq:L1-3D}
   e^{-1}{\cal L}_1 =  \frac{64}{L^0} \Big[& 
   R^{\mu\nu}R_{\mu\nu}-\tfrac38\,R^2 
   +\tfrac12 R({\cal V})^{\mu\nu\,i}{}_j  \,  R({\cal V})_{\mu\nu}{}^{\!j}{}_i
   + R({\cal A})^{\mu\nu\,p}{}_q \,  R({\cal A}){}_{\mu\nu}{}^{q}{}_p  
   \nonumber \\
   &\,
   +2\, (\mathcal{D}_a C)^2 + \tfrac32\,(D-C^2)^2 \Big] \,,
\end{align}
and
\begin{align} 
    \label{eq:L2-3D}
    e^{-1}{\cal L}_2 = 
     \frac{64}{(L^0)^{3}}\, \Big[&\mathcal{D}^{\mu} {F^{\nu\rho\,0 }}\,
       {\cal D}_{\mu} F_{\nu\rho}{\!}^0  
   - 2\,{\cal D}_{\mu} F^{\mu\nu\,0} \,{\cal D}^{\rho} F_{\rho\nu}{\!}^0  
   \nonumber\\[-1mm]
   & 
   +  \mathcal{D}_\mu \,Y^{i}{}_j{\!}^{0}
   \,  \mathcal{D}^\mu Y^{j}{}_i{\!}^{0} 
   + \mathrm{i}e^{-1}\,\varepsilon^{\mu\nu\rho} \,L^0\,
   R(\mathcal{V}){}_{\mu\nu}{}^i{}_j \, \mathcal{D}_\rho Y^{j}{}_i{}^{0}
   \nonumber\\[1mm] 
   & 
   + \tfrac14\,( C^2-D )\, ( 3\,Y^i{}_j{\!}^0 \,Y^{j}{}_i{\!}^0  
              +(F_{\mu\nu}{\!}^0)^2)   \Big]   \nonumber\\[1mm]
    + \frac{4}{(L^0)^5} &\Big[ 3\,(F_{\mu\nu}{\!}^0)^2\,
    (F_{\rho\sigma}{\!}^0)^2  +3\,  (Y^{i}{}_j{\!}^0 \,Y^{j}{}_i{\!}^0)^2 
    +2\,Y^i{}_j{\!}^0\,Y^{j}{}_i{\!}^0 \, (F_{\mu\nu}{\!}^0)^2 \Big]
    +\cdots \,.
\end{align}
We should emphasize that the above expressions concern only a subset
of the terms generated by the reduction and are thus incomplete. As we
have already seen in section
\ref{subsubsection:vector-multiplet-lagr}, where we evaluated the $3D$
Lagrangian for vector multiplets, a full evaluation of the $3D$
results can be rather tedious and this is particularly the case for
Lagrangians with higher-derivative couplings. Nevertheless the above
results already show a number of noteworthy features that will be
present in the final result. Those will be briefly discussed below.

First of all this Lagrangian depends on both the fields of the $3D$
Weyl multiplet and of the Kaluza-Klein vector multiplet. Clearly it is
homogeneous of degree $-1$ in the latter fields, and the
super conformal fields appear in a non-linear fashion. This can be
understood on more general grounds, just as it was clear from the
start that the Lagrangian should contain fourth-order space-time
derivatives.

The Lagrangian $\mathcal{L}_1$ contains terms that are familiar from
previous work on higher-derivative Lagrangians for $3D$
(super)gravity, multiplied by a compensating $(L^0)^{-1}$ factor that
is required by conformal invariance. The linearized result for the
corresponding supergravity invariant was given in \cite{Bergshoeff:2010ui}
and exhibits all the quadratic bosonic terms present in
\eqref{eq:L1-3D}. However, there are some notable differences in the
coefficients. One is that the squares of the two $\mathrm{SU}(2)$
curvatures appear with different coefficients, unlike in
\cite{Bergshoeff:2010ui} where the coefficients are the same. The
other one concerns the coefficient of the kinetic term of the field
$C$, which is positive.  This discrepancy in the coefficients is no
reason for concern: additional curvature terms may arise by
commutators of covariant derivatives, and the scalar kinetic terms are
effected by the presence of additional terms, for instance
proportional to $\mathcal{D}_\mu L^0\,\mathcal{D}^\mu C$, that we have
not extracted but that will change the coefficient of the
$(\mathcal{D}_\mu C)^2$. Obviously the terms shown in \eqref{eq:L2-3D}
have no bearing on the expression in \cite{Bergshoeff:2010ui}, because
the presence of the components of the Kaluza-Klein vector
multiplet is even more crucial here.  It should be of interest to
evaluate the full $3D$ superconformal invariant, either from the
off-shell dimensional reduction or directly in three space-time
dimensions. The latter can be done by utilizing the off-shell multiplet calculus
obtained in \cite{Butter:2013goa,Butter:2013rba}. 

The $4D$ Weyl multiplet can easily be coupled to vector
multiplets. Schematically one has a function $F(X,\hat A)$, which can
for instance be expanded in positive powers of $\hat A$ according to
\begin{equation}
  \label{eq:F-expansion}
     F(X,\hat A) = \sum_{g}  F_g(X)\, \hat A^g \,,
\end{equation}
where each holomorphic function, $F_g(X)$ is of appropriate weight to
ensure consistency with respect to conformal invariance.  Comparing
with \eqref{eq:Weyl-back-red}, where $\hat A$ is expressed in terms of
derivatives of the compensating scalars, it is clear that each term in
\eqref{eq:F-expansion} contributes $2\,g$ first-order derivatives on the
scalars. This can be compared to the situation in $4D$, where the
off-shell Lagrangian contains only four-derivative interactions, while
a similar series of ever increasing derivatives appears when solving
for the auxiliary tensor $T_{AB}{}^{ij}$. 

Finally we emphasize that we have only briefly considered the coupling of the
Weyl multiplet to vector multiplets in this section. There also exist
couplings that involve tensor multiplets. Those will of course be
relevant for establishing consistency with the c-map. Assuming that
this can be achieved, it may further clarify the effective action
description for topological amplitudes involving tensor multiplets or
hypermultipets \cite{Antoniadis:1993ze}.

%%%%%%%%%%%%%%%%%%%%%%%%%%%%%%%%%%%%%%%%%%%%%%%%%%%%%%%%%%%%
\section*{Acknowledgement}
%%%%%%%%%%%%%%%%%%%%%%%%%%%%%%%%%%%%%%%%%%%%%%%%%%%%%%%%%%%%
We thank Nathan Berkovits, Daniel Butter, Martin Ro\v{c}ek, Warren
Siegel and Stefan Vandoren for valuable discussions. N.B. acknowledges
the hospitality extended to her at Nikhef where her work was supported
by a Veni grant of the `Nederlandse Organisatie voor Wetenschappelijk
Onderzoek (NWO)'. At present her work is partially supported by a
Ramanujan Fellowship, DST, Government of India. B.d.W. is supported by the ERC
Advanced Grant no. 246974, {\it``Supersymmetry: a window to
  non-perturbative physics''}.  The work of S.K. is supported by the
ERC Starting Grant no. 307286, {\it ``The structure of the extra
  dimensions of string theory''} and by the INFN.
%%%%%%%%%%%%%%%%%%%%%%%%%%%%%%%%%%%%%%%%%%%%%%%%%%%%%%%%%%%%%%%%

%%%%%%%%%%%%%%%%%%%%%%%%%%%%%%%%%%%%%%%%%%%%%%%%%%%%%%%%%%%%%%%%
\begin{appendix}
%
%%%%%%%%%%%%%%%%%%%%%%%%%%%%%%%%%%%%%%%%%%%%%%%%%%%%%%%%%%%%%%%%%%%%
\section{Relations between 4D and 3D Riemann curvatures}
\label{App:4-3D-Riemann-curv}
\setcounter{equation}{0}
%%%%%%%%%%%%%%%%%%%%%%%%%%%%%%%%%%%%%%%%%%%%%%%%%%%%%%%%%%%%%%%%%%%%
Based on \eqref{eq:kk-ansatz} one can evaluate the relation between
$4D$ and $3D$ curvature components. In the equations below,
derivatives $\mathcal{D}_a$ are covariant with respect to $3D$ local
Lorentz transformations and dilatations. The results are as follows
(in this appendix the $4D$ curvature components are consistently
denoted by $\hat R$),
\begin{align}
  \label{eq:curvatures-world}
  \hat R_{\mu\nu}{}^{ab} =& \, R_{\mu\nu}{}^{ab} +\tfrac12 \phi^{-2} \Big[
  F(B)_{\mu}{}^{[a} \,F(B)_\nu{}^{b]} + F(B)_{\mu\nu} F(B)^{ab}
  \Big]\nonumber\\
  &\, - B_{[\mu} \Big[2\, \phi^{-3} F(B)_{\nu]}{}^{[a}
  \, \mathcal{D}^{b]} \phi + \mathcal{D}_{\nu]}[\phi^{-2} F(B)^{ab}]
  \Big]\,, \nonumber\\[.4ex]
  \hat R_{\mu\nu}{}^{a4} =&\,
   - \mathcal{D}_{[\mu}[\phi^{-1} F(B)_{\nu]}{}^a] -
  \phi^{-2} \mathcal{D}^a\phi\, F(B)_{\mu\nu}\nonumber\\
  &\,
  +B_{[\mu}\Big[2\,\mathcal{D}_{\nu]} [\phi^{-2}\,\mathcal{D}^a\phi] + \tfrac12
  \phi^{-3}\,F(B)_{\nu]b} \, F(B)^{ab}\Big]  \,,  \nonumber\\[.4ex]
  \hat R_{\mu\hat 4}{}^{ab} =&\,\tfrac12 \mathcal{D}_\mu [\phi^{-2}
  F(B)^{ab} ] + \phi^{-3} F(B)_\mu{}^{[a} \,\mathcal{D}^{b]}\phi\,,
  \nonumber\\[.4ex]
  \hat R_{\mu\hat 4}{}^{a4} =&\,- \mathcal{D}_\mu
  [\phi^{-2} \mathcal{D}^a\phi] - \tfrac14 \phi^{-3} \,F(B)_{\mu
    b}F(B)^{ab}  \,.
\end{align}
With tangent-space indices, $\hat R_{CD}{}^{AB}$ takes the form,
\begin{align}
  \label{eq:curvatures-tangent}
  \hat R_{cd}{}^{ab} =& \, R_{cd}{}^{ab} +\tfrac12 \phi^{-2} \Big[
  F(B)_{c}{}^{[a} \,F(B)_d{}^{b]} + F(B)_{cd} F(B)^{ab}
  \Big]\,, \nonumber\\[.4ex]
  \hat R_{cd}{}^{a4} =&\,
  %- \mathcal{D}_{[c} [\phi^{-1} F(B)_{d]}{}^a] -
  %\phi^{-2} \mathcal{D}^a\phi\, F(B)_{cd}\nonumber \\
  %=&\,
  %-\phi^{-1} \,\mathcal{D}_{[c} F(B)_{d]}{}^a +
   %\phi^{-2} \Big[ \mathcal{D}_{[c}\phi \,F(B)_{d]}{}^a -
  %\mathcal{D}^a\phi\, F(B)_{cd}\Big]  \nonumber\\
  %=&\,
  \tfrac12 \phi^{-1} \,\mathcal{D}^aF(B)_{cd}  -
  \phi^{-2} \Big[\mathcal{D}^a\phi\, F(B)_{cd} -
  \,F(B)^a{}_{[c} \mathcal{D}_{d]}\phi \Big]   \,,  \nonumber\\[.4ex]
  \hat R_{c 4}{}^{ab} =&\,
  \tfrac12\phi^{-1} \, \mathcal{D}_c F(B)^{ab} -
  \phi^{-2}\Big[F(B)^{ab}\, \mathcal{D}_{c}\phi - F(B)_c{}^{[a}
  \,\mathcal{D}^{b]}\phi \Big]\,,   \nonumber\\[.4ex]
  \hat R_{c4}{}^{a4} =&\,-\phi\, D_c(\omega)
  [\phi^{-2} \mathcal{D}^a\phi] - \tfrac14 \phi^{-2} \,F(B)_{cb}F(B)^{ab}
  \,.
\end{align}
Note that these components satisfy the pair-exchange property of the
Riemann tensor.  Contracted versions of the Riemann tensor take the
form,
\begin{align}
  \label{eq:contracted-R}
  \hat R_{cB}{}^{aB} =&\, R_{cb}{}^{ab} + \tfrac12 \phi^{-2}
  F(B)_{cb}F(B)^{ab} -\phi\, \mathcal{D}_c [\phi^{-2}\mathcal{D}^a\phi] \,,
  \nonumber\\
  \hat R_{A4}{}^{Ab} =&\, \tfrac12\phi^{-1} \, \mathcal{D}_a F(B)^{ab}
  -\tfrac32 \phi^{-2}\,F(B)^{ab}\, \mathcal{D}_{a}\phi\,,\nonumber\\
    \hat R_{A4}{}^{A4}=&\, -\phi\, \mathcal{D}_a(\omega)
  [\phi^{-2} \mathcal{D}^a\phi] - \tfrac14 \phi^{-2}
  \,F(B)_{ab}F(B)^{ab}\,, \nonumber\\
  \hat R_{AB}{}^{AB} =&\, R_{ab}{}^{ab} - 2\, \phi\,
  \mathcal{D}_a [\phi^{-2} \mathcal{D}^a\phi]
  +\tfrac14 \phi^{-2} \,F(B)_{ab} F(B)^{ab}  \,.
\end{align}
Furthermore one may consider the components of $\hat R_{[AB}{}^{EF} \,\hat
R_{CD]EF}$,
\begin{align}
  \label{eq:RwedgeR}
  \hat R_{[ab}{}^{EF}\,\hat R_{cd]EF} =&\, 0\,,
  \nonumber\\[.4ex]
%%%%%%%%%%%%%%%%%%%
   \hat R_{\hat 4[a}{}^{EF}\,\hat R_{cd]EF} =&\,- \phi\,\mathcal{D}_{[a}\Big[
   \tfrac12\phi^{-2} R_{cd]}{}^{ef} F(B)_{ef} \nonumber\\
   &\qquad\qquad +\tfrac18\phi^{-4} \big[ F(B)^2
   F(B)_{cd]} + 2\, F(B)^{ef} F(B)_{ce}F(B)_{df}\big]  \nonumber\\
   &\qquad\qquad -2\, \phi^{-1}F(B)_{c}{}^{e}
   \,\mathcal{D}_{d]}(\mathcal{D}_e\phi^{-1})
   + F(B)_{cd]}  (\mathcal{D}\phi^{-1})^2 \Big] \,.
\end{align}
where we made use of the Bianchi identity on $F(B)$ in the $3D$ Riemann
tensor.

%%%%%%%%%%%%%%%%%%%%%%%%%%%%%%%%%%%%%%%%%%%%%%%%%%%%%%%%%%%%%%%%
%%%%%%%%%%%%%%%%%%%%%%%%%%%%%%%%%%%%%%%%%%%%%%%%%%%%%%%%%%%%%%%%
\section{The conversion of \texorpdfstring{$\boldsymbol{4D}$}{4D}
  chiral to  \texorpdfstring{$\boldsymbol{\mathrm{SU}(2)\!\times
      \!\mathrm{SU}(2)}$}{SU(2) x SU(2)}
  covariant \texorpdfstring{$\boldsymbol{3D}$}{3D} spinors}
%%%%%%%%%%%%%%%%%%%%%%%%%%%%%%%%%%%%%%%%%%%%%%%%%%%%%%%%%%%%%%%%
\label{sec:conv-spin-basis}
\setcounter{equation}{0}
%%%%%%%%%%%%%%%%%%%%%%%%%%%%%%%%%%%%%%%%%%%%%%%%%%%%%%%%%%%%%%%%
The original $4D$ theory contains doublets of Majorana spinors that
transform under the chiral R-symmetry group
$\mathrm{SU}(2)\times\mathrm{U}(1)$. Hence they transform irreducibly
according to the pseudo-real $(4,2)_\mathrm{c}$ representation of
$\mathrm{Spin}(3,1)\times \mathrm{SU}(2)\times\mathrm{U}(1)$, where
the subscript denotes the chiral $\mathrm{U}(1)$ charge. When reducing
to three dimensions, a $4D$ spinor decomposes into two real $3D$
spinors and, as we shall demonstrate, the $\mathrm{U}(1)$ component of
the R-symmetry will then extend to a second $\mathrm{SU}(2)$ group, so
that we obtain a pseudoreal irreducible representation $(2,2,2)$ of
$\mathrm{Spin}(2,1)\times \mathrm{SU}(2)\times\mathrm{SU}(2)$. To
obtain the spinor fields in their $3D$ form, we must convert the $4D$
spinors such that the $3D$ symmetry assignments become manifest. This
conversion is the topic of this section where we will base ourselves
on previous results presented in \cite{DeJaegher:1997ka,deWit:2006gn}

The analysis starts from the underlying Clifford algebra for the $4D$
gamma matrices, which has to be defined such that they act reducibly
on the original spinor. We remind the reader that the reduction
amounts to compactifying the fourth coordinate $\hat x^4$ on a circle
which is subsequently shrunk to zero size. The proper $3D$ gamma
matrices are now defined in terms of the $4D$ gamma matrices by
\begin{equation}
  \label{eq:3-gamma}
  \hat\gamma^a  = \gamma^a \tilde\gamma\,, \quad \mbox{where} \quad\tilde
  \gamma= -\mathrm{i}\gamma_4\gamma_5\,.
\end{equation}
The hermitian matrices $\tilde\gamma$, $\gamma_4$ and $\gamma_5$ are
mutually anti-commuting, and square to the unit matrix. Furthermore
they commute with the $\hat\gamma^a$. Hence we have obtained two
mutually commuting three-dimensional Clifford algebras, generated by
the $\hat\gamma^a$ and by $(\tilde\gamma,\gamma_4,\gamma_5)$,
respectively. Observe that we have the identity,
\begin{equation}
  \label{eq:3-gamma-1}
  \hat\gamma^{[a} \,\hat\gamma^b \,\hat\gamma^{c]} = \mathrm{i}
  \varepsilon^{abc} \, {\bf 1}\,,
\end{equation}
showing that the two separate $3D$ Lorentz spinors into which a
generic $4D$ spinor decomposes transform according in the same
Clifford algebra representation. Starting from a single $4D$ spinor one
thus obtains a doublet of $3D$ spinors transforming under an extended
R-symmetry group $\mathrm{SU}(2)$ with generators
$(\tilde\gamma,\gamma_4,\gamma_5)$, subject to
$\tilde\gamma\,\gamma_4\,\gamma_5= \mathrm{i} {\bf 1}$. Obviously the
generator proportional to $\gamma_5$ corresponds to the generator of
chiral $\mathrm{U}(1)$ R-symmetry that is already present in $4D$.

As a result of the redefinition of the $3D$ gamma matrices, the
definition of the Dirac conjugate will change, and consequently also
the $4D$ charge conjugation matrix must be redefined. The new Dirac
conjugate and the new charge conjugation matrix read,
\begin{equation}
  \label{eq:3D-charge-conj}
  \hat C = C\,\tilde\gamma\,,\qquad \hat{\bar\psi} =
  \bar\psi\,\tilde\gamma\,.
\end{equation}
Note, however, that it is still possible to further modify the charge
conjugation matrix. Indeed, the $\mathrm{SU}(2)\times \mathrm{SU}(2)$
covariant $3D$ spinor basis that we are about to construct will
require such a modification. Based on the present redefinitions one
easily verifies the following equations (using the properties of the
charge conjugation matrix $C$ in $4D$),
\begin{align}
  \label{eq:C3d}
  \hat C \hat\gamma^{a} \hat C^{-1} =&\, - \hat\gamma^{a {\rm T}}
  \,, \qquad \hat C^{\rm T} = - \hat C \,, \nonumber\\
  \hat C \gamma_{4} \hat C^{-1} =&\, \gamma_4{}^{\rm T} \,,\qquad \hat
  C \tilde{\gamma} \hat C^{-1} = \tilde{\gamma}^{{\rm T}} \,,\qquad
  \hat C \gamma_{5} \hat C^{-1} = - \gamma_5{}^{\rm T} \,.
\end{align}

Now we have to extend the previous analysis to the case of a doublet
of $4D$ Majorana fermions. It is convenient to still express the
fermions in terms of $4D$ chiral components, because those transform
systematically under the action of the $\mathrm{SU}(2)$ R-symmetry
group that is manifest in $4D$. The new $\mathrm{SU}(2)$ group that
emerges in the reduction to $3D$ as an extension of the $4D$ chiral
$\mathrm{U}(1)$ group, will commute with the original $4D$ chiral
$\mathrm{SU}(2)$. To study the way in which the two $\mathrm{SU}(2)$
factors are realized, let us start from a positive-chirality $4D$
spinor $\psi^i$ with $\mathrm{U}(1)$ charge $+1/2$, which we combine
with the negative-chirality conjugate $\psi_i$, which is provided by
the $4D$ Majorana condition. The latter spinor thus has
$\mathrm{U}(1)$ charge equal to $-1/2$, respectively. Since the
spinors transform uniformly under the $3D$ Lorentz transformations we
will only be concerned with the possible R-symmetry
transformations. Observing that the symmetry enhancement of the
R-symmetry group will be based on the generators
$(\tilde\gamma,\gamma_4,\gamma_5)$ identified before, one expects that
the extended symmetry involves the following infinitesimal variation,
\begin{equation}
  \label{eq:R-symm-var}
  \delta\begin{pmatrix} \psi^i\\[4mm] \psi_i \end{pmatrix} = \frac12
  \begin{pmatrix} \Lambda^i{}_j +\mathrm{i}\alpha\,\delta^i_j &
    -\beta \,\gamma_4\,\varepsilon^{ij} \\[4mm]
    -\bar\beta\,\gamma_4\,\varepsilon_{ij}&\Lambda_i{}^j -
    \mathrm{i}\alpha\,\delta_i{}^j  \end{pmatrix}
  \begin{pmatrix} \psi^j\\[4mm] \psi_j \end{pmatrix} \;,
\end{equation}
where $\Lambda^i{}_j$ is an anti-hermitian traceless matrix, i.e.
it satisfies the relations,
\begin{equation}
  \label{eq:properties-Lambda}
  \Lambda^i{}_i=0\, \quad
  \Lambda^i{}_k\,\varepsilon^{kj}+ \Lambda^j{}_k\,\varepsilon^{ik}=
  0\,, \quad \Lambda_i{}^j\equiv (\Lambda^i{}_j)^\ast\,,
\end{equation}
and $\alpha$, $\beta$ and $\bar \beta$ are the transformation
parameters of the new $\mathrm{SU}(2)$. The normalization of
these parameters is of no concern at this point. The reader can
directly verify that these transformations form a group and that
the new $\mathrm{SU}(2)$ group commutes with the original one
generated by the matrix $\Lambda^i{}_j$.

The representation \eqref{eq:R-symm-var} has the disadvantage that it
involves spinor components of opposite chirality. However, since we
have reduced the space-time dimension, it is possible to apply a
further redefinition,
\begin{equation}
  \label{eq:equal-chir-spinors}
  \psi^{i+} = \psi^i\,,\qquad
  \psi^{i-} = -\varepsilon^{ij}\,\gamma_4 \,\psi_j\,,
\end{equation}
where the superscripts $\pm$ denote the sign of the $\mathrm{U}(1)$
charge. Because of the presence of the matrix $\gamma_4$, the spinors
are defined in the same eigenspace of $\gamma_5$ and we choose a
positive eigenvalue, i.e.,
\begin{equation}
  \label{eq:projection-spinor}
  (\gamma_5-{\bf 1}) \psi^{i\pm}= 0\,,
\end{equation}
so that in the new basis we have replaced the doublets $\psi^i$ and
$\psi_i$ of opposite chirality by four equal-chirality spinors
$\psi^{i\pm}$. For the Dirac conjugate spinors, the corresponding
relations follow from \eqref{eq:3D-charge-conj},
\begin{equation}
  \label{eq:conj-chiral-spinors}
  \hat{\bar\psi}_{i+} = -\mathrm{i} \bar\psi_i\,\gamma_4 \,,\qquad
 \hat{\bar\psi}_{i-} = -\mathrm{i} \varepsilon_{ij} \,  \bar\psi^j \,,
\end{equation}
where on the left-hand side we have the $3D$ Dirac conjugate spinors
and on the right-hand side the $4D$ conjugate spinors.  Note that we
have $\bar\psi_{i\pm} (\gamma_5 -{\bf 1}) = 0$.  In this basis the
transformation rule \eqref{eq:R-symm-var} takes the form ($p,q=+,-$),
\begin{equation}
  \label{eq:fourplet-spinor}
  \psi^{ip} \to U^i{}_j\,V^p{}_q \,\psi^{j q}\,,
\end{equation}
where $U$ denotes the chiral $\mathrm{SU}(2)$ transformation that was
originally present in $4D$, and $V$ the new $\mathrm{SU}(2)$
transformation that has emerged in $3D$. In terms of the parameters in
\eqref{eq:R-symm-var}, we have
\begin{equation}
  \label{eq:V}
  V\approx {\bf 1}+ \tfrac12\begin{pmatrix} \mathrm{i}\alpha & \beta\\[2mm]
   - \bar \beta & -\mathrm{i}\alpha \end{pmatrix} \,.
\end{equation}

The next topic is to derive the consequences of the Majorana property
of the spinors. For chiral $4D$ Majorana spinors the
constraint on the chiral components is given by,
\begin{align}
  \label{eq:majorana}
  C^{-1} \bar\psi_i{}^\mathrm{T}= \hat C^{-1}
  \hat{\bar\psi}_i{}^\mathrm{T}= \psi_i\,,\nonumber\\
    C^{-1} \bar\psi^i{}^\mathrm{T}= \hat C^{-1}
  \hat{\bar\psi}^i{}^\mathrm{T}= \psi^i\,,
\end{align}
where the left-hand side contains the Dirac conjugate according to the
$4D$ and $3D$ definition, respectively, where the indices are lowered
or raised as a result of complex conjugation. From these constraints,
one straightforwardly derives,
\begin{equation}
  \label{eq:majorana-2}
  \hat C^{-1} \hat{\bar\psi}_{i+}{}^\mathrm{T} =  \varepsilon_{ij}
  \gamma_4\, \psi^{j-}\,,\qquad \hat C^{-1}
  \hat{\bar\psi}_{i-}{}^\mathrm{T}  = -\varepsilon_{ij}  \gamma_4\,
  \psi^{j+} \,.
\end{equation}
Upon absorbing $\gamma_4$ into the definition of the charge conjugation
matrix $\hat C$, one then proves the pseudo-reality relation
\begin{equation}
  \label{eq:Majorana-3}
  C^{-1} \,\bar\psi_{i,p}{}^\mathrm{T} =
  \varepsilon_{ij}\,\varepsilon_{pq}\; \psi^{j,q}\,.
\end{equation}
Hence the appropriate charge conjugation matrix in the covariant
$\mathrm{SU}(2) \times \mathrm{SU}(2)$ basis is given by,
\begin{equation}
  \label{eq:cov-C}
  C=\hat C\gamma_4\,,
\end{equation}
satisfying $C\gamma^a C^{-1}= -\gamma^a{}^\mathrm{T}$ with
$C^\mathrm{T}=-C$. In \eqref{eq:Majorana-3} and \eqref{eq:cov-C} and
henceforth we suppress the caret on $3D$ quantities. The indices
$p,q=+,-$ refer to the spinor components with positive and negative
$\mathrm{U}(1)$ charge respectively.  With these results we derive the
Majorana re-ordering for fermionic bilinears,
\begin{equation}
  \label{eq:3-Maj-reordering}
  \bar\psi_{i,p} \Gamma \psi^{j,q} = \pm \varepsilon_{ik}\,
  \varepsilon^{jl}\, \varepsilon_{pr}\, \varepsilon^{qs}
  \,\bar\psi_{l,s} \Gamma \psi^{k,r}\,,
\end{equation}
where the plus and the minus sign refer to $\Gamma=\bf{1}$ and
$\Gamma=\gamma^\mu$, respectively.

Finally we redefine the $4D$ spinors such that the previous
redefinitions can be applied uniformly. This is done by choosing a
chiral Majorana spinor and modify it such that we obtain a field
$\psi^i$ of positive chirality and positive $\mathrm{U}(1)$
charge. The field $\psi_i$ then follows from applying the $4D$
Majorana condition. However, this only determines $\psi^i$ and
$\psi_i$ up to a phase factor which implies that the $\mathrm{SU}(2)$
transformations induced by \eqref{eq:R-symm-var} on the underlying
fields, are also determined up to phase factors. Insisting that the
$3D$ supersymmetry transformations are manifestly covariant with
respect to the additional $\mathrm{SU}(2)$ R-symmetry component will
fix these relative phase factors.

As an example let us start with the supersymmetry parameter
$\epsilon_i$, which has positive $\mathrm{U}(1)$ charge and negative
chirality. This identifies corresponding fields $(\psi^i, \psi_i)$ up
to a phase factor $z$,
\begin{equation}
  \label{eq:psi-epsilon}
  \psi^i(\epsilon)= z\varepsilon^{ij}\gamma_4\,\epsilon_j \,,\qquad
  \psi_i(\epsilon) =\bar
  z \varepsilon_{ij}\gamma_4 \,\epsilon^j\,.
\end{equation}
As long as we consider a single field, we are free to fix the phase
factor, so we will eventually choose $z=1$. However, for the remaining spinors we
should then leave the phase factor arbitrary. Hence for the remaining
independent spinors we choose,
\begin{equation}
  \label{eq:psi-remaining}
  \begin{array}{rcl} 
    \psi^i(\eta) &\!\!\!=&\!\!\!z_\eta\varepsilon^{ij} \,\eta_j \,,\\
    \psi^i(\Omega) &\!\!\!=&\!\!\! z_\Omega \gamma_4\,\Omega^i \,, \\
    \psi^i(\varphi) &\!\!\!=&\!\!\! z_\varphi \,\varphi^i\,, 
  \end{array}
  \qquad
  \begin{array}{rcl}
    \psi_i(\eta) &\!\!\!=&\!\!\! \bar z_\eta \varepsilon_{ij} \,\eta^j \,,\\
    \psi_i(\Omega)&\!\!\!=&\!\!\! \bar z_\Omega \gamma_4 \,\Omega_i \,,\\
    \psi_i(\varphi) &\!\!\!=&\!\!\!\bar z_\varphi \,\varphi_i\,.
  \end{array}
\end{equation}
The assignments of the conformal gauge fields $\psi_\mu{}^i$ and
$\phi_\mu{}^i$ are the same as those of the transformation parameters
$\epsilon^i$ and $\eta^i$, respectively. These ans\"atze now lead to
the corresponding definitions of the quantities $\epsilon^{ip}$,
$\eta^{ip}$, $\Omega^{ip}$ and $\varphi^{ip}$ which are all subject to
the Majorana condition \eqref{eq:Majorana-3}. They are summarized as
follows,
\begin{equation}
  \label{eq:spinors-4-3}
  \begin{array}{rcl} 
    \epsilon^{i+}\big\vert_{3D} &\!\!\!=&\!\!\!
    z\varepsilon^{ij}\gamma_4\,\epsilon_j\big\vert_{4D} \,,\\ 
    \eta^{i+}\big\vert_{3D} &\!\!\!=&\!\!\! z_\eta\varepsilon^{ij}
    \,\eta_j\big\vert_{4D}  \,,\\ 
    \Omega^{i+}\big\vert_{3D} &\!\!\!=&\!\!\! z_\Omega\,
    \gamma_4\,\Omega^i\big\vert_{4D}   \,, \\
    \varphi^{i+}\big\vert_{3D}  &\!\!\!=&\!\!\! z_\varphi
    \,\varphi^i\big\vert_{4D} \,, \\
    \chi^{i+}\big\vert_{3D} &\!\!\!=&\!\!\!
    z_\chi \varepsilon^{ij}\gamma_4\,\chi_j\big\vert_{4D} \,,
  \end{array}
  \qquad
  \begin{array}{rcl}
    \epsilon^{i-} \big\vert_{3D}&\!\!\!=&\!\!\! \bar z \,\epsilon^i\big\vert_{4D}\,,\\ 
    \eta^{i-} \big\vert_{3D} &\!\!\!=&\!\!\! \bar z_\eta\,
    \gamma_4\,\eta^i\big\vert_{4D} \,,\\ 
    \Omega^{i-} \big\vert_{3D}&\!\!\!=&\!\!\! -\bar z_\Omega\,
    \varepsilon^{ij}\,\Omega_j\big\vert_{4D}  \,,\\ 
    \varphi^{i-} \big\vert_{3D} &\!\!\!=&\!\!\! - \bar z_\varphi
    \,\varepsilon^{ij}\gamma_4\, \varphi_j\big\vert_{4D} \,,\\
   \chi^{i-} \big\vert_{3D}&\!\!\!=&\!\!\! \bar z\chi
   \,\chi^i\big\vert_{4D}\,.
  \end{array}
\end{equation}
For the convenience of the reader we also add the expressions for the Dirac conjugate
spinors, 
\begin{equation}
  \label{eq:D-conj-spinors-4-3}
  \begin{array}{rcl} 
    \bar\epsilon_{i+}\big\vert_{3D} &\!\!\!=&\!\!\!  \mathrm{i}
    \bar{z}\,\varepsilon_{ij}\,\bar\epsilon^j\big\vert_{4D} \,,\\  
    \bar\eta_{i+}\big\vert_{3D} &\!\!\!=&\!\!\! -\mathrm{i}\bar{z}_\eta\,\varepsilon_{ij}
    \,\bar\eta^j \gamma_4 \big\vert_{4D}  \,,\\ 
    \bar\Omega_{i+}\big\vert_{3D} &\!\!\!=&\!\!\! \mathrm{i} \bar{z}_\Omega
    \,\bar\Omega_i\big\vert_{4D}   \,, \\ 
    \bar\varphi_{i+}\big\vert_{3D}  &\!\!\!=&\!\!\! -\mathrm{i}
    \bar{z}_\varphi \,\bar\varphi_i \gamma_4 \big\vert_{4D} \,,  \\
    \bar\chi_{i+}\big\vert_{3D} &\!\!\!=&\!\!\!  \mathrm{i}
    \bar{z}\chi\,\varepsilon_{ij}\,\bar\chi^j\big\vert_{4D} \,,
  \end{array}
  \qquad
  \begin{array}{rcl}
    \bar\epsilon_{i-} \big\vert_{3D}&\!\!\!=&\!\!\! -\mathrm{i} z
    \,\bar\epsilon_i\gamma_4 \big\vert_{4D}\,,\\ 
    \bar\eta_{i-} \big\vert_{3D} &\!\!\!=&\!\!\!  \mathrm{i} z_\eta\,
    \bar\eta_i \big\vert_{4D} \,,\\ 
    \bar\Omega_{i-} \big\vert_{3D}&\!\!\!=&\!\!\! \mathrm{i} z_\Omega\,
    \varepsilon_{ij}\,\bar\Omega^j\gamma_4 \big\vert_{4D}  \,,\\ 
    \bar\varphi_{i-} \big\vert_{3D} &\!\!\!=&\!\!\! - \mathrm{i}  z_\varphi
    \,\varepsilon_{ij} \, \bar\varphi^j\big\vert_{4D} \,,\\
    \bar\chi_{i-} \big\vert_{3D}&\!\!\!=&\!\!\! -\mathrm{i} z_\chi
    \,\bar\epsilon_i\gamma_4 \big\vert_{4D}\,.
  \end{array}
\end{equation}
In the main text we have defined the set of phase factors
consistent with supersymmetry and R-symmetry, as
\begin{equation}
  \label{eq:z-factors}
  z=1\,, \quad z_\eta=-\mathrm{i}\,,\quad z_\Omega= 1\,, \quad
  z_\phi= \mathrm{i}\, \quad z_\chi =1 \,.
\end{equation}

%%%%%%%%%%%%%%%%%%%%%%%%%%%%%%%%%%%%%%%%%%%%%%%%%%%%%%%%%%%%%%%%%%%%
\end{appendix}
%%%%%%%%%%%%%%%%%%%%%%%%%%%%%%%%%%%%%%%%%%%%%%%%%%%%%%%%%%%%%%%%
%\begin{thebibliography}{99}
\providecommand{\href}[2]{#2}
%\begingroup\raggedright

%%%%%%%%%%%%%%%%%%%%%%%%%%%%%%%%%%%%%%%%%%%%%%%%%%%%%%%%%%%%%%%%%%
\end{document}